\documentclass[useAMS,usenatbib]{mnras}
\usepackage{amsmath}
\usepackage{graphicx,txfonts,fleqn,enumerate,color}

\renewcommand*{\vec}[1]   {\boldsymbol{#1}}
\newcommand*  {\sub}[2]   {{#1}_{\mathrm{#2}}}
\newcommand*  {\rcirc}    {\sub{r}{circ}}
\newcommand*  {\Msun}     {\mathrm{M}_\odot}

\title[Feeding SMBHs by collisional cascades]{Feeding supermassive black holes by collisional cascades}
\author[Faber \& Dehnen]
       {Christian Faber\thanks{Email: cf197@leicester.ac.uk},
        Walter Dehnen\thanks{Email: wd11@leicester.ac.uk} \\ 
        Department for Physics \& Astronomy, University of Leicester, Leicester LE1 7RH}
\pagerange{\pageref{firstpage}--\pageref{lastpage}}
\pubyear{2018}

\begin{document}
\maketitle
\label{firstpage}
\begin{abstract}
The processes driving gas accretion on to supermassive black holes (SMBHs) are still poorly understood. Angular momentum conservation prevents gas within $\sim10\,$pc of the black hole from reaching radii $\sim 10^{-3}\,$pc where viscous accretion becomes efficient. Here we present simulations of the collapse of a clumpy shell of swept-up isothermal gas, which is assumed to have formed as a result of feedback from a previous episode of AGN activity. The gas falls towards the SMBH forming clumps and streams, which intersect, collide, and often form a disc. These collisions promote partial cancellations of angular momenta, resulting in further infall and more collisions. This continued collisional cascade generates a tail of gas with sufficiently small angular momenta and provides a viable route for gas inflow to sub-parsec scales. The efficiency of this process hardly depends on details, such as gas temperature, initial virial ratio and power spectrum of the gas distribution, as long as it is not strongly rotating.

Adding star formation to this picture might explain the near-simultaneous formation of the S-stars (from tidally disrupted binaries formed in plunging gas streams) and the sub-parsec young stellar disc around Sgr\,A$^{\!\star}$.

\end{abstract}

\begin{keywords}
	accretion, accretion discs -- black hole physics -- hydrodynamics -- galaxies: supermassive black holes
\end{keywords}

\section{Introduction}
\label{sec:introduction}
It is commonly accepted that most massive galaxies contain a supermassive black hole (SMBH) in their centre \citep[see reviews e.g. by][]{Kormendy95, Kormendy13}. The observed $M_\bullet$-$\sigma$ relation linking the mass $M_\bullet$ of the SMBH to the stellar velocity dispersion $\sigma$ in the bulge of the galactic host \citep[][or \citealt{McConnell13} for a more recent analysis]{Ferrarese00,Gebhardt00} provides a compelling argument that the black hole presents a critical component of the galaxy and a crucial ingredient for its evolution \citep[][for reviews see e.g.\ \citealt{Frank02,Fabian12}]{Haehnelt98,Silk98,King03,Sijacki15}. However, the gravitational influence of the SMBH is negligible, since its mass $M_\bullet\ll\sub{M}{host}$, the mass of the hosting bulge/spheroid, and hence cannot cause the $M_\bullet$-$\sigma$ relation.

A comparison of the total quasar luminosity density to the mass density of SMBHs shows that the dominant mode of SMBH growth is via gas accretion \citep{Soltan82}. The total energy released from an accreting SMBH exceeds the binding energy of the host, e.g. $\eta M_\bullet c^2 \gg \sub{M}{host} \sigma^2$, even if one assumes only a $\eta=10\%$ efficiency for converting gravitational energy of the accreted gas into radiation. This radiation probably drives powerful gas outflows \citep{Silk98, Fabian99, Pounds03, King03, King15}, which are much more efficient at communicating their energy to the host's interstellar medium (ISM) than the original radiation. These outflows can be highly collimated (jets, often associated with low accretion rates), when some form of isotropisation is required to affect most of the host \citep{Quilis01, McNamara05, Sijacki06, Fabian12}. Conversely, the outflows generated by accretion rates close to the \cite{Eddington16} limit are usually associated with near-spherical ionised winds \citep{Halpern84, Reynolds95, McKernan07}. Once the SMBH reaches the $M_\bullet$-$\sigma$ relation, the outflows become efficient in expelling most of the gas from the galaxy, inhibiting further SMBH growth and star formation \citep{King05}. This picture suggests, independently of the \cite{Soltan82} argument, that SMBHs grow predominantly by gas accretion.

This scenario of SMBH growth by gas accretion has been challenged by the observation of a number of black holes with masses $M_\bullet\gtrsim10^{9}\Msun$ at redshifts $z\sim6$ \citep[e.g.][]{Willott03, Riechers09, Mortlock11}, which require an e-folding time of $\lesssim50$\,Myr to grow from stellar-mass seeds. If the SMBH is spinning near the maximum, as initially thought \citep{Volonteri05} based on the assumption that accretion always spins the hole up \citep{Bardeen70, Scheuer96}, the e-folding time exceeds 300\,Myr \citep[e.g.][]{King06}, and therefore requires more massive, non-stellar black-hole seeds \citep{Haehnelt93, Latif16}. However, if the SMBH grows from consecutive accretion discs generated by randomly orientated inflows (`stochastic accretion'), the black-hole spin remains low \citep{King05B,King08,Fanidakis11} implying e-folding times of $\lesssim25$\,Myr. Hence the observed SMBHs at $z\sim6$ are compatible with stellar mass seeds, if the holes maintain an accretion duty cycle of $\gtrsim50\%$. Consequently these high-redshift SMBHs may well originate from the extreme end of the distribution of SMBH growth by gas accretion from stellar-mass seeds.

However, the process(es) responsible for the transportation of gas to the hole are still unclear. The main obstacle is the conservation of angular momentum within the gravitational influence of the SMBH, which prevents the gas from approaching the black hole. Instead the gas is likely to dissipate energy and form a disc at the circularisation radius dictated by its angular momentum content. Once a disc has formed, mass is transported inwards and angular momentum outwards by viscosity \citep[proposed for differentially rotating stars by][]{Goldreich67}. The widely used parameterisation of \cite{Shakura73} for this process gives a viscous accretion time scale of 
\begin{eqnarray}
	\label{eq:t:visc}
	\sub{t}{visc}
	&=& \frac{1}{\alpha}
	\left(\frac{R}{H}\right)^2 \left(\frac{R^3}{GM_\bullet}\right)^{1/2}
	\\[0.5ex] \nonumber
	&\sim&
	3\times10^5\,\mathrm{yr}\,
	\left(\frac{0.1}{\alpha}\right)
	\left(\frac{H/R}{0.002}\right)^{-2}
	\left(\frac{R}{0.002\,\mathrm{pc}}\right)^{3/2}
	\left(\frac{M_\bullet}{10^8\mathrm{M}_\odot}\right)^{-1/2},
\end{eqnarray}
where $R$ and $H$ are the size and vertical extent of the disc respectively. $\alpha$ is a dimensionless viscosity parameter, for which observational evidence gives $\alpha\sim0.1-0.4$ (\citealt{Smak99}; \citealt*{Dubus01}, for review see e.g. \citealt*{King07}). There are two lines of evidence that SMBH accretion discs are restricted to very small scales. First, self-gravity limits the discs \citep{Kolykhalov80,Pringle81,Ladato07} to $R\lesssim10^{-2}\,$pc \citep{KingPringle2007}. Second, since $H/R\lesssim0.002$ for AGN discs \citep[][but see \citealt{Floyd09}]{King08, Poindexter08, Bate08}, $\sub{t}{visc}$ becomes comparable to the observationally inferred duration $\sim10^5\,$yr of AGN phases \citep{Schawinski2015} only at $R\lesssim0.002\,$pc \citep{KingNixon2015}. On the other hand, the region containing $\sim10^8\Msun$ of accretable material is $\sub{r}{acc}\gtrsim10\,$pc in radius (assuming a SMBH host on the $M_\bullet$-$\sigma$ relation and with $M_\bullet=10^8\Msun$). Within this region most gas will have some (mostly random) angular momentum preventing it from reaching the required $10^{-3}\,$pc scale. Therefore some other mechanism is required to bridge the gap of a factor $\gtrsim10^4$ in radius (or $\gtrsim10^2$ in angular momentum).

\cite{Dehnen13} suggested a mechanism for driving gas into the immediate vicinity of the hole. This is based on the picture of stochastic accretion described above, where SMBH growth results from many accretion events \citep{King05B,King08}. These events correspond to quasar-like activity and generate a radiation-driven wind. As long as the hole is still in its infancy, i.e.\ below the $M_\bullet$-$\sigma$ relation, this quasi-spherical outflow is not powerful enough to clear the galaxy of gas, but strong enough to push most of the ambient gas away from the hole and sweep it up into a shell of radius $\sub{r}{shell}\sim1$-10\,pc. The sweeping up of the gas may have caused some cancellation of orbital angular momentum \citep{Zubovas15}, generating a tail of low-angular momentum material in the shell. More importantly, the shell of gas has gained gravitational potential energy by the outflow and, if anything, lost kinetic energy by dissipation. It is thus prevented from falling back only by the ongoing outflow. But as soon as the outflow ends, the gas must fall back in the form of clouds and streams on plunging orbits. The infall of multiple streams from different directions increases the likelihood of collisions near pericentre with the potential of further angular momentum cancellation. This in turn promotes further infall, and results in a cascade of collisions at continually decreasing radii, generating a significant tail of very low angular momentum material, from which eventually an accretion disc forms. In this paper we shall test this idea by using smoothed particle hydrodynamics (SPH) simulations. 

This paper is organised as follows: In Section~\ref{sec:method} we describe the initial conditions and the parameter choices for the different simulations. We choose a suite of simulations for our default choice of the physical parameters in Section~\ref{sec:ref} to outline the general evolution of the infalling gas. We study the effects of varying the physical parameters in Section~\ref{sec:parasweep}. The results are summarised and discussed in Section~\ref{sec:discuss}, while Section~\ref{sec:conclude} concludes.

\section{Modelling approach}\label{sec:method}
\subsection{The hydrodynamical method}\label{sec:sph}
The smoothed-particle hydrodynamics (SPH) computational method, developed by \cite{Lucy77,Gingold77}, uses a Lagrangian description in which the particles follow the flow and serve as interpolation points for the fluid properties. The simulations reported here have been performed with the SPH code \textsc{sphinx} \citep[utilised by e.g.][]{Aly15}, which implements fully con\-serva\-tive SPH with individually adaptive smoothing length \citep[for review e.g.][]{Price12}. \textsc{sphinx} features the widely used method of \cite{Cullen10} for suppressing artificial viscosity away from shocks and employs as smoothing kernel the fourth-order \cite{Wendland95} function as proposed by \cite{Dehnen12} to improve numerical convergence.

The time integration is performed using individually adaptive particle time steps organised in the standard block-step scheme with hierarchically ordered time steps differing by a factor of two \citep{Hayli1967, Makino1991}. Individual steps are done with the second-order accurate leapfrog integrator implemented as a predictor-corrector scheme. In order to resolve cloud-cloud/stream-stream and similar collisions, a wake-up mechanism ensures that the time steps of neighbouring particles differ no more than a factor of 4. 

The implementation uses a ``one-sweep'' algorithm, which requires only a single neighbour search per particle and time step and avoids storing of neighbour lists. The code uses explicit vectorisation and multi-threading for shared-memory hardware.

\subsection{Initial conditions and model setup}\label{sec:inicond}
\begin{center}
    \begin{table*}
        \caption{Summary of the initial conditions and parameters used in the simulations presented. We use model units in which Newton's constant of gravity $G=1$, the mass of the SMBH $M_\bullet=1$, and the initial radius of the gas shell $\sub{r}{shell}=1$. $\sub{M}{shell}$ is the total amount of gas and is modelled by $\sub{N}{shell}=2\times10^6$ SPH particles initially placed in a spherical shell of radius $\sub{r}{shell}$ and Gaussian width $\sub{r}{width}$. The initial gas velocities are set according to equations~(~\ref{eq:v:initial}-\ref{eq:v:rot}) and are controlled by the parameters $\eta$ and $\chi$ as described in the text. $c_s$ is the sound speed of the gas (in units of the circular speed at the shell radius $\sqrt{GM/\sub{r}{shell}}$). $\sub{r}{sink}$ is a computationally motivated boundary region around the SMBH; Particles inside the radius are removed from the simulation. The penultimate column specifies whether gas self-gravity was included, and the final column indicates the Gaussian width of the initial shell.
        }
        \centering
        \begin{tabular}{| l || l | l | l | l | l | l | l | l | l |}
    	\hline
    	 Section         & $\sub{M}{shell}$         &  $\sub{\vec{v}}{turb}$ & power $n$ & $\eta$ & $\chi$ & $c_\mathrm{s}$ & $\sub{r}{sink}$ & $\sub{G}{gas}$ & $\sub{r}{width}$\\
    	\hline\hline
    	\ref{sec:ref}    & 0.01 &   $\vec{\nabla}\cdot\sub{\vec{v}}{turb}\neq 0$ & $-11/3$     & 0.9  & 0.75   & 0.1  & 0.01  & 0 & 0.2  \\ \hline
     	\ref{sec:cs}     & 0.01 &  $\vec{\nabla}\cdot\sub{\vec{v}}{turb}\neq 0$  & $-11/3$     & 0.9 & 0.75  & \textbf{0.05, 0.2}  & 0.01 & 0 & 0.2 \\ \hline
    	\ref{sec:pwr}    & 0.01 &  $\vec{\nabla}\cdot\sub{\vec{v}}{turb}\neq 0$ & \boldmath$-5/2$, $-9/2$  & 0.9 & 0.75 & 0.1  & 0.01 & 0 & 0.2 \\ \hline
        \ref{sec:ratio}  & 0.01 &  $\vec{\nabla}\cdot\sub{\vec{v}}{turb}\neq 0$ & $-11/3$     & 0.9 & \textbf{0.25, 0.5, 1.0} & 0.1 & 0.01 & 0 & 0.2 \\ \hline
    	\ref{sec:sol}    & 0.01 & $\vec{\nabla}\cdot\sub{\vec{\mathbf{v}}}{\mathbf{turb}}\,\mathbf{=\,0}$ & $-11/3$          & 0.9 & 0.75 & 0.1  & 0.01 & 0 & 0.2 \\ \hline
    	\ref{sec:addpar:mass} & \textbf{0.1, 1} &  $\vec{\nabla}\cdot\sub{\vec{v}}{turb}\neq 0$ & $-11/3$  & 0.9 & 0.75 & 0.1   & 0.01 & 0 & 0.2 \\ \hline
        \ref{sec:addpar:width} &
        0.01 &
        $\vec{\nabla}\cdot\sub{\vec{v}}{turb}\neq 0$ &
        $-11/3$ &
        0.9 &
        \textbf{1.0} &
        0.1 &
        0.01 &
        0 &
        \textbf{0.1, 0.3} \\ \hline
       	\ref{sec:addpar:eta} & 0.01 &  $\vec{\nabla}\cdot\sub{\vec{v}}{turb}\neq 0$ & $-11/3$     & \textbf{0.5, 1.1} & 0.75 & 0.1  & 0.01 & 0 & 0.2 \\ \hline
     	\ref{sec:addpar:grav} & \textbf{0.01, 0.1, 1} &  $\vec{\nabla}\cdot\sub{\vec{v}}{turb}\neq 0$ & $-11/3$     & 0.9 & 0.75  & 0.1  & 0.01 & \textbf{1} & 0.2 \\ \hline
        \ref{sec:hstar} &
        0.01 &
        $\vec{\nabla}\cdot\sub{\vec{v}}{turb}\neq 0$ &
        $-11/3$ &
        0.9 &
        \textbf{1.0} &
        0.1 &
        \textbf{0.005, 0.0025, 0.001} &
        0 &
        0.2\\ \hline
    	\end{tabular}
        \label{tab:paroverview}
    \end{table*}
\end{center}
All initial conditions are based on a spherical shell of gas modelled by SPH particles with normally distributed radii of mean $\sub{r}{shell}$ and a standard deviation $0.2r_{\mathrm{shell}}$ centred on a massive sink particle at the origin representing the SMBH (plus unresolved material at $\lesssim \sub{r}{sink}$). We employ units such that $G=1$, $M_\bullet=1$ and $\sub{r}{shell}=1$.

The initial positions of the gas particles are sampled using a quasi-random generator, which provides a low-discrepancy sequence of numbers \citep[for review see e.g.][]{Niederreiter92}. Unlike the common pseudo-random numbers, quasi-random numbers avoid shot noise and therefore create more glass-like initial conditions suitable for SPH simulations. This shell models the gas swept-up by a feedback event, which itself is not modelled. The begin of the simulation coincides with the end of an accretion-driven outflow (responsible for sweeping up the gas) and the begin of subsequent gas infall.

Clearly a smooth and symmetric gas shell is not realistic in view of the turbulent motion and a non-uniform distribution of the swept-up ISM. Hence we add some turbulence to the initial velocity field. In particular, the velocity of each particle is the weighted sum of a turbulent and a purely rotational velocity field, i.e.
\begin{equation}
    \label{eq:v:initial}
    \vec{v}_i = \chi \sub{\vec{v}}{turb}(\vec{r}_i) + (1-\chi) \sub{\vec{v}}{rot}(\vec{r}_i),
\end{equation}
where the mixture parameter $\chi$ equals 0.75 for most of our simulations. The turbulent velocity field is a Gaussian random field with a \cite{Kolmogorov41}\footnote{English translation as \cite{Kolmogorov91}.} power spectrum $P\propto k^{n}$, where $n=-11/3$ for most of our simulations (see Appendix~\ref{sec:turb} for details of how the velocities are generated) and scaled such that
\begin{equation}
    \label{eq:eta:turb}
    \sum_i m_i \sub{\vec{v}}{turb}^2(\vec{r}_i) = \sub{\eta}{turb} \sum_i\frac{Gm_iM_\bullet}{|\vec{r}_i|},
\end{equation}
i.e.\ $\sub{\eta}{turb}$ is the virial ratio in the case $\chi=1$ (when only the turbulent velocity component contributes to $\vec{v}_i$). In general, $\sub{\vec{v}}{turb}$ is not divergence free, though we also study the case where $\sub{\vec{v}}{turb}(\vec{r})$ is constructed to satisfy $\vec{\nabla}\cdot\sub{\vec{v}}{turb}=0$ everywhere. $\sub{\vec{v}}{rot}$ corresponds to solid-body rotation with a fraction $\sub{\eta}{rot}$ of the circular speed at the shell radius, i.e.
\begin{equation}
    \label{eq:v:rot}
    \sub{\vec{v}}{rot}(\vec{r}) = \sub{\eta}{rot} \vec{r}\times\hat{\vec{e}}_z \sqrt{GM_\bullet/\sub{r}{shell}^3}.
\end{equation}
In all our simulations the numerical values for $\sub{\eta}{turb}$ and $\sub{\eta}{rot}$ are identical and set to $\eta=0.9$ for most simulations. This implies that the virial ratio for our initial conditions satisfies
\begin{equation}
    \label{eq:virial:ratio}
    \frac{2\sub{E}{kin}}{-\sub{E}{pot}} = 
    \frac{\sum_im_i\vec{v}_i^2}{\sum_iGm_iM_\bullet/|\vec{r}_i|} \approx
    \chi^2\sub{\eta}{turb} + \tfrac{2}{3}(1-\chi)^2\sub{\eta}{rot}^2
\end{equation}
This relation would be exact, if the shell was infinitely thin and the turbulent and rotational velocities were uncorrelated over the particles such that $\sum_im_i \vec{v}_{\mathrm{turb},i} \cdot \vec{v}_{\mathrm{rot},i}=0$. For our default parameter setting of $\chi=0.75$ and $\eta=0.9$, this evaluates to $-2\sub{E}{kin}/ \sub{E}{pot}\approx0.54$, i.e.\ the system is sub-virial, but not excessively so. By varying both $\chi$ and $\eta$ any combination for the contributions of rotation and turbulence to the velocity field can be obtained, but we only investigate changes in one or the other parameter.

Finally, we assume an isothermal equation of state for the gas with sound speed $c_s$, which for most simulations equals 0.1 times the circular speed at the initial shell radius, $\sqrt{GM_\bullet/\sub{r}{shell}}$.

Particles coming closer to the central sink particle than $\sub{r}{sink}$, which defaults to $0.01\sub{r}{shell}$, are absorbed, i.e.\ their mass, momentum, and angular momentum is added to the sink particle, which carries a spin for this purpose. This effectively implements an inner boundary condition to the model and is necessary to avoid excessively short time steps.

The total gas mass is set to be $\sub{M}{shell}=0.01M_\bullet$ for most of our simulations, but we run a set of simulations with ten or hundred times more gas as well. The simulations presented all contain $\sub{N}{gas}=2\times10^6$ gas particles. We experimented with larger numbers ($4\times10^6$ and $8\times10^6$) and found no significant difference in the results presented below in contrast to simulations with  $\sub{N}{gas}=10^6$ or less.

Although by default we would normally ignore the self-gravity of the gas, it has been included in a few sets of simulations. However, we ignore the gravity from the galactic host, because the gravity in the simulated volume is dominated by the hole (also the dynamics studied are not critically dependent on the Keplerian nature of ballistic trajectories).

The random nature of the initial turbulent velocities implies that details of the simulated flows (e.g. position and angular momentum of the gas clumps) are random, too. Indeed we find variations between simulations when we utilise different random seeds to generate the turbulent velocities, but keep the other parameters identical. In order to assess this variation and the main trends, we ran for each set of physical parameters considered a set of six simulations differing only in the random seed for the turbulent velocities.

Table~\ref{tab:paroverview} gives an overview over all simulations and their parameters presented here.

\begin{figure}
	\includegraphics[width=80mm]{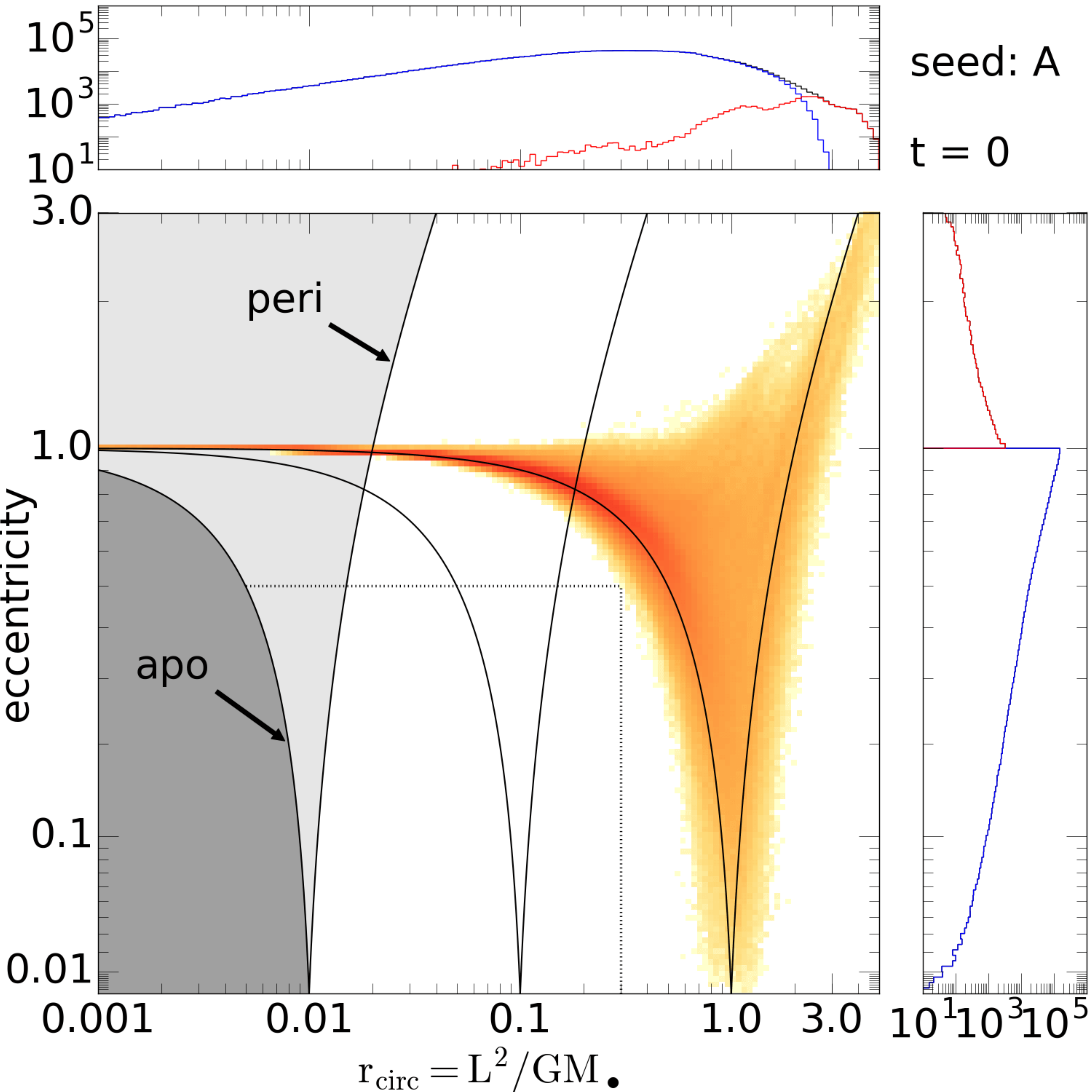} 
	\caption{
	\label{fig:initial2Dhist_ref-seed633}
	Distribution of the initial gas shell (at $r\approx\sub{r}{shell}$) over eccentricity $e$ and circularisation radius $\rcirc=\vec{L}^2 /GM_\bullet$ for the reference simulation (see Section~\ref{sec:ref:detail}). Loci of constant peri/apo-centre, $\sub{r}{apo,peri}=\sub{r}{circ}/(1\mp e)$, are indicated by outward/inward bending curves.	In particular, the dark and light grey region correspond to orbits with, respectively, apo-centre and pericentre within the absorption radius of the central sink particle: orbits in the light grey region cross into the absorption region, whereas the dark grey region is inaccessible to simulated gas. The thin rectangular box indicates orbits classified as `disc' in later figures. In the distributions over $e$ (right) and $\rcirc$ (top), black indicates the total, blue the bound ($e<1$), and red the unbound ($e\ge1$) fraction.
	Most gas is initially in the region $\sub{r}{peri}-\sigma_r\le \sub{r}{shell}<\sub{r}{apo}+\sigma_r$, as expected. The tail at $e\sim1$ and small $\rcirc$ originates from gas with near-zero angular momentum currently near apo-centre $\sub{r}{apo}\approx\sub{r}{shell}$.}
\end{figure}

\section{The reference simulations}
\label{sec:ref}

\begin{figure*}
	\includegraphics[width=43.3mm]{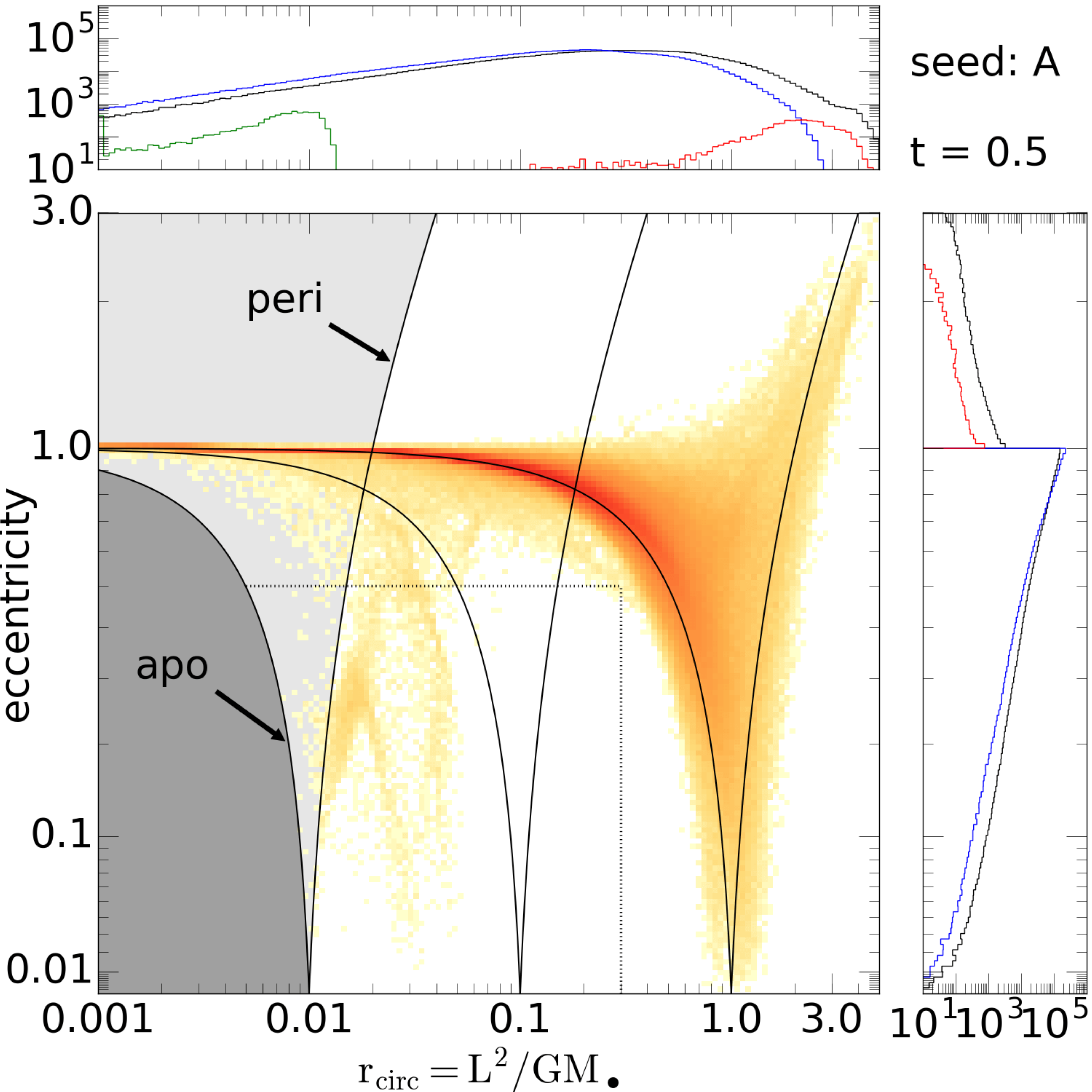}
	\hfill
	\includegraphics[width=43.3mm]{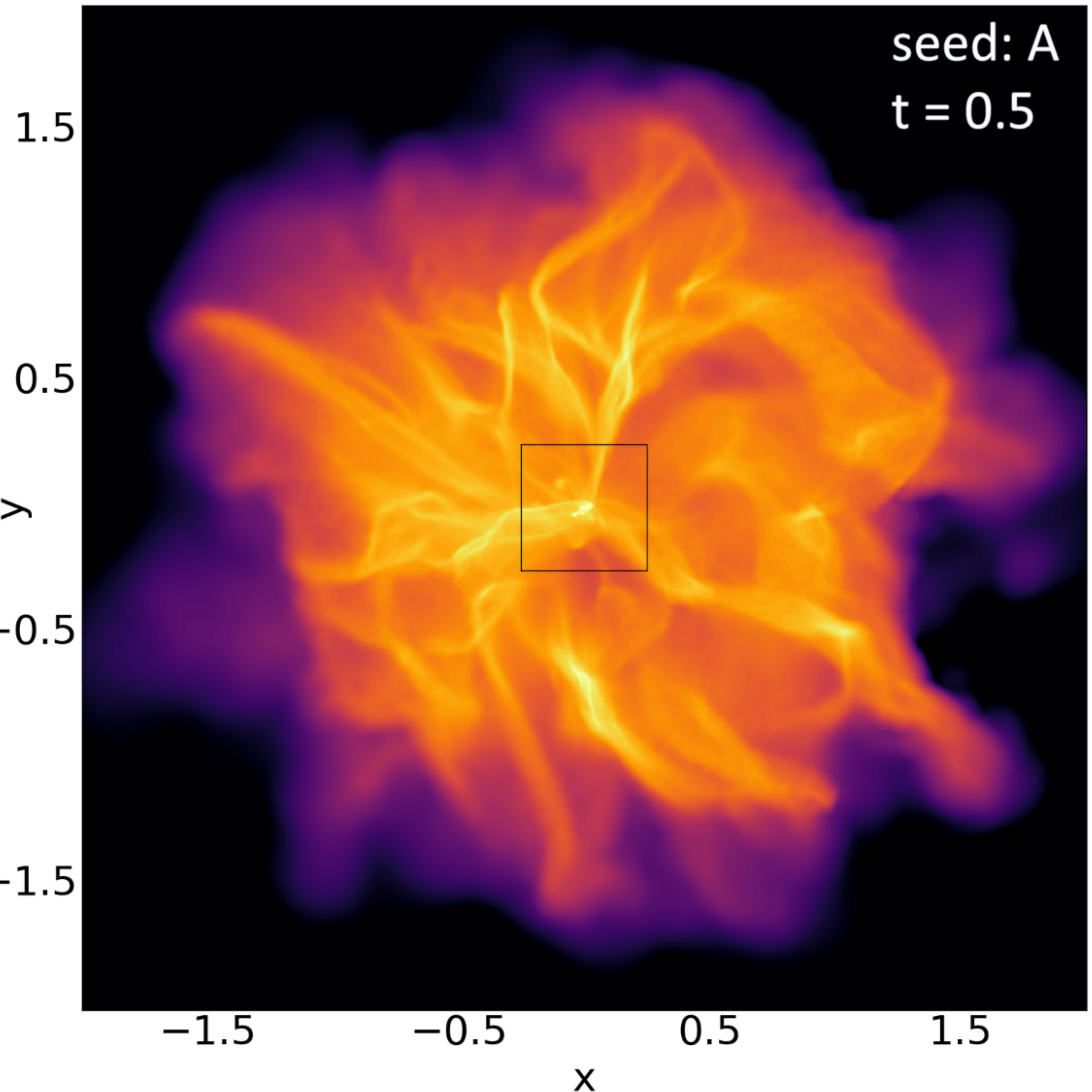}
	\includegraphics[width=43.3mm]{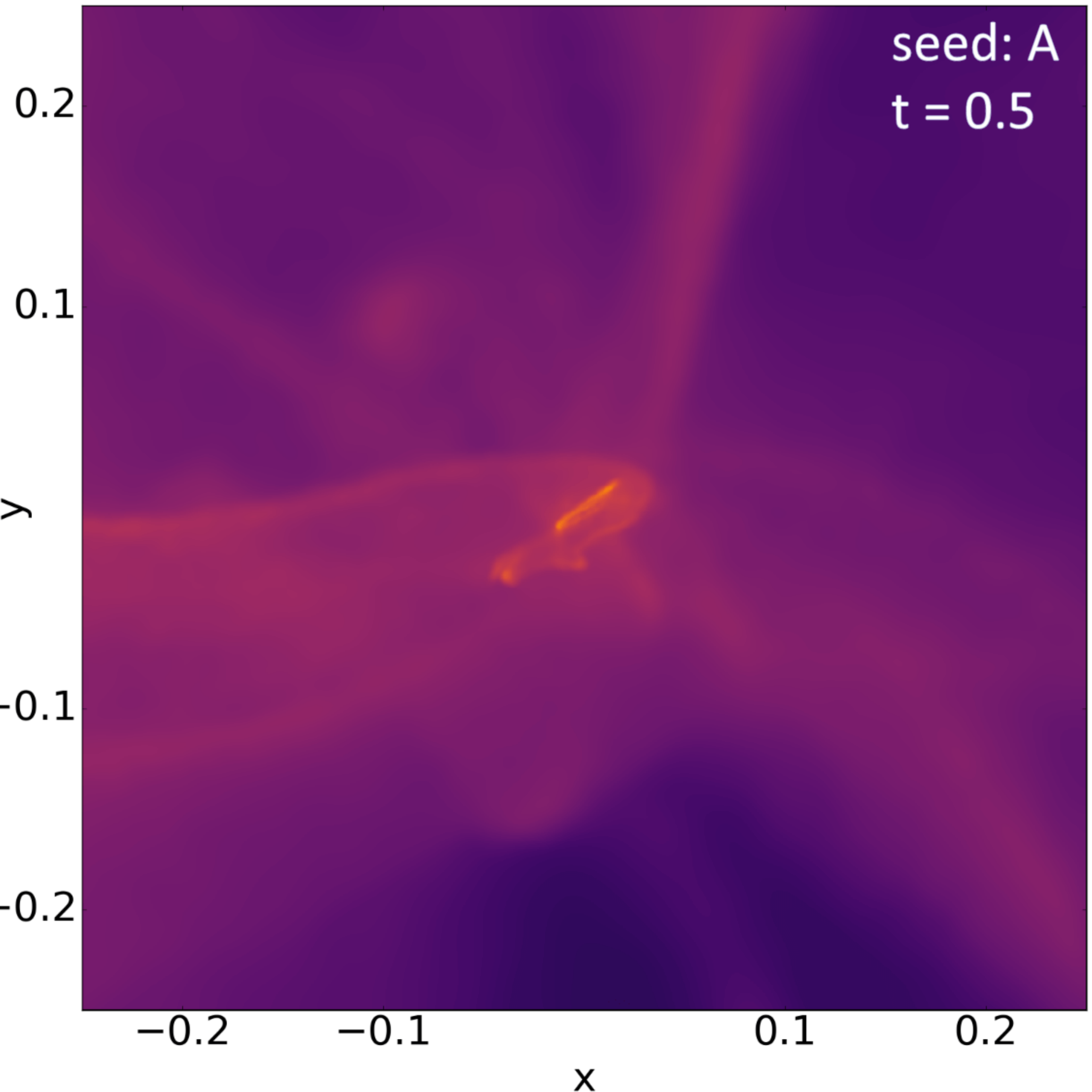}
	\includegraphics[width=43.3mm]{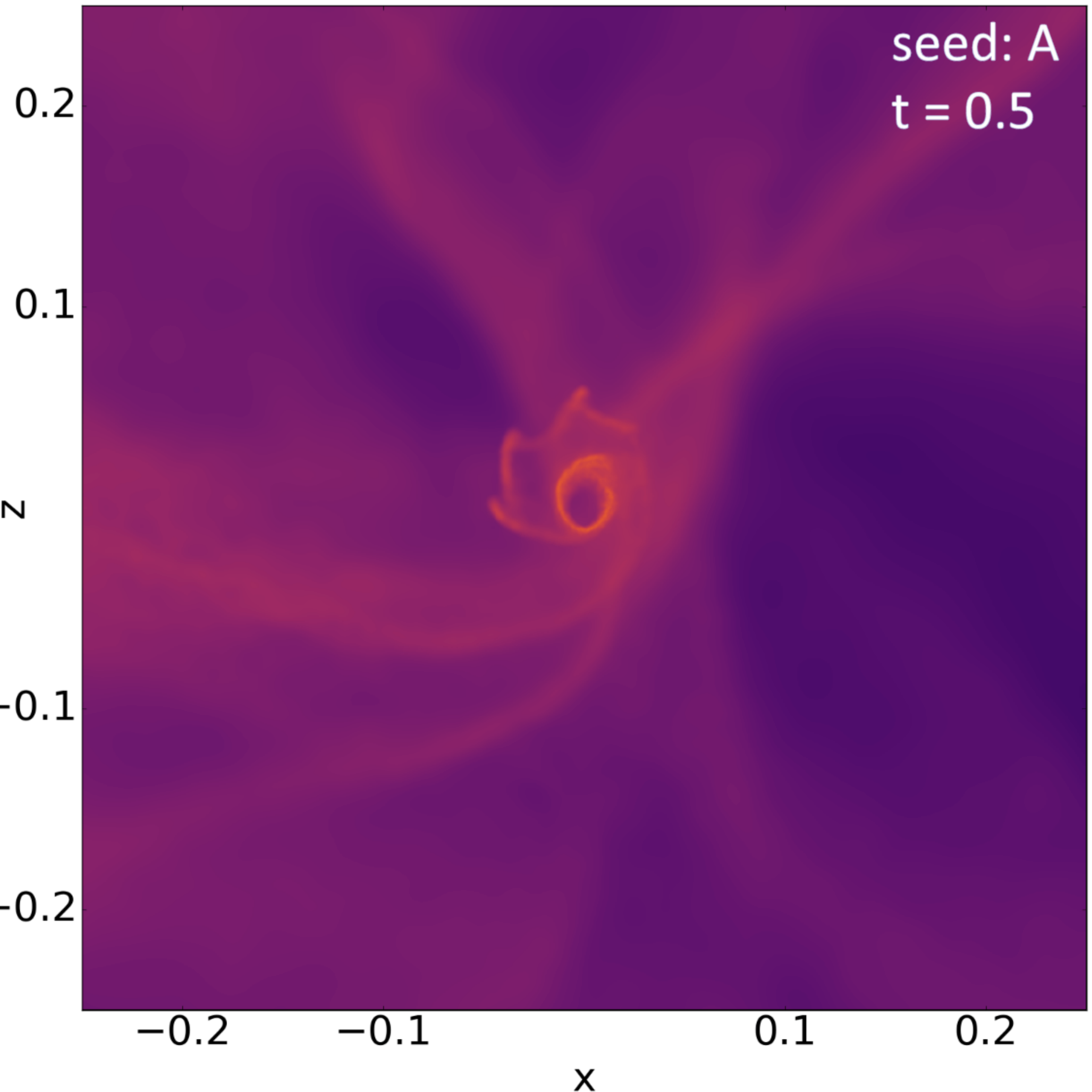}
    
	\includegraphics[width=43.3mm]{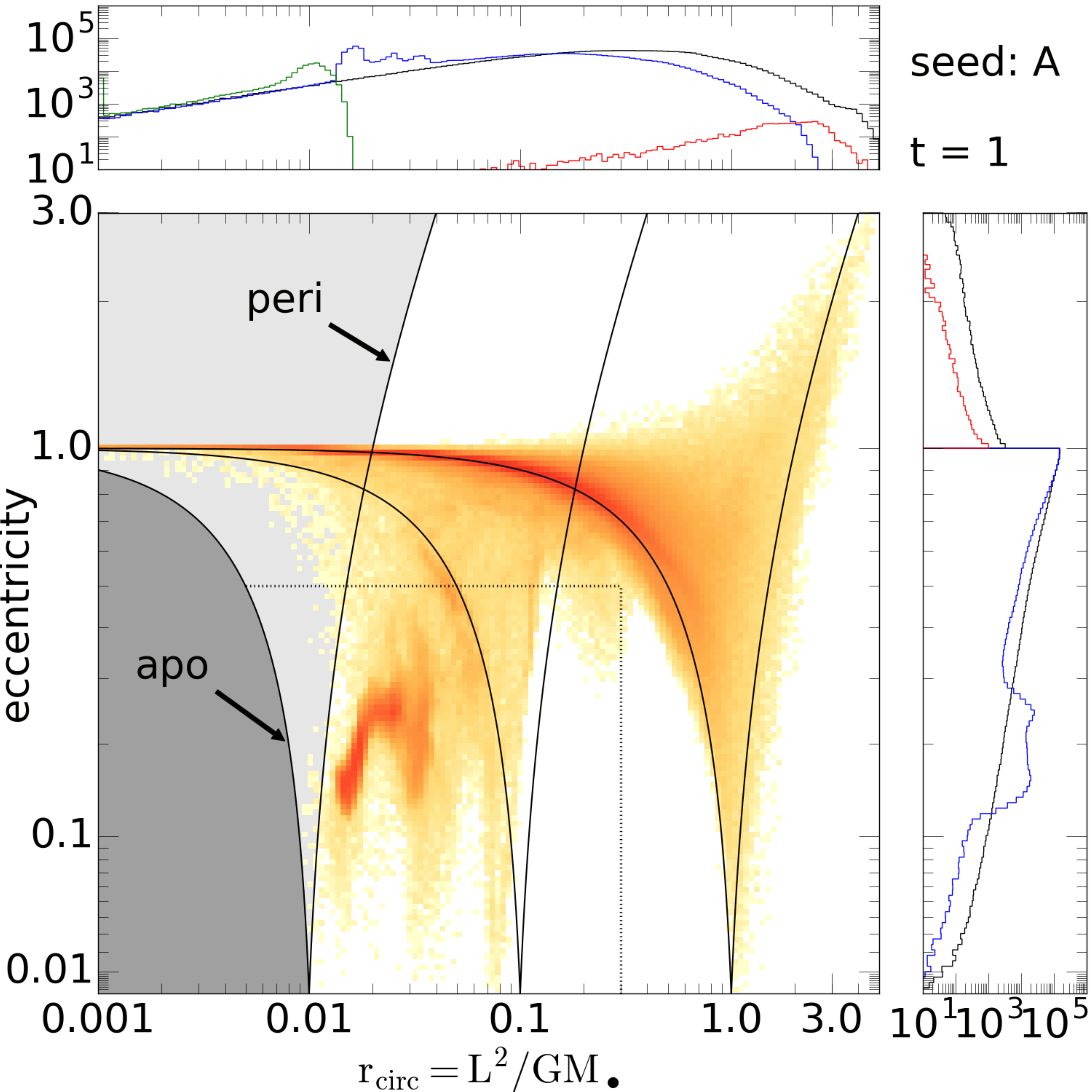}
	\hfill
	\includegraphics[width=43.3mm]{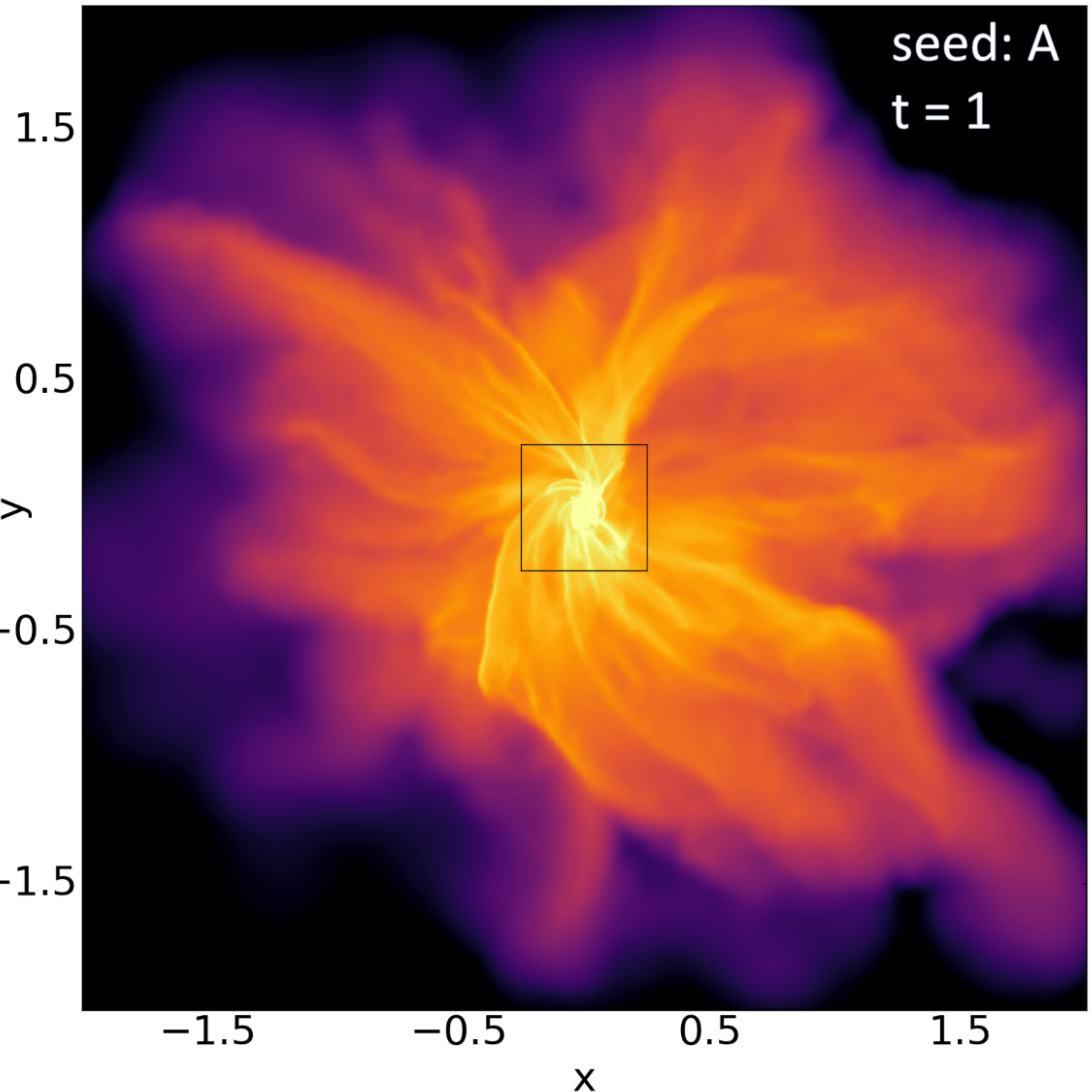}
	\includegraphics[width=43.3mm]{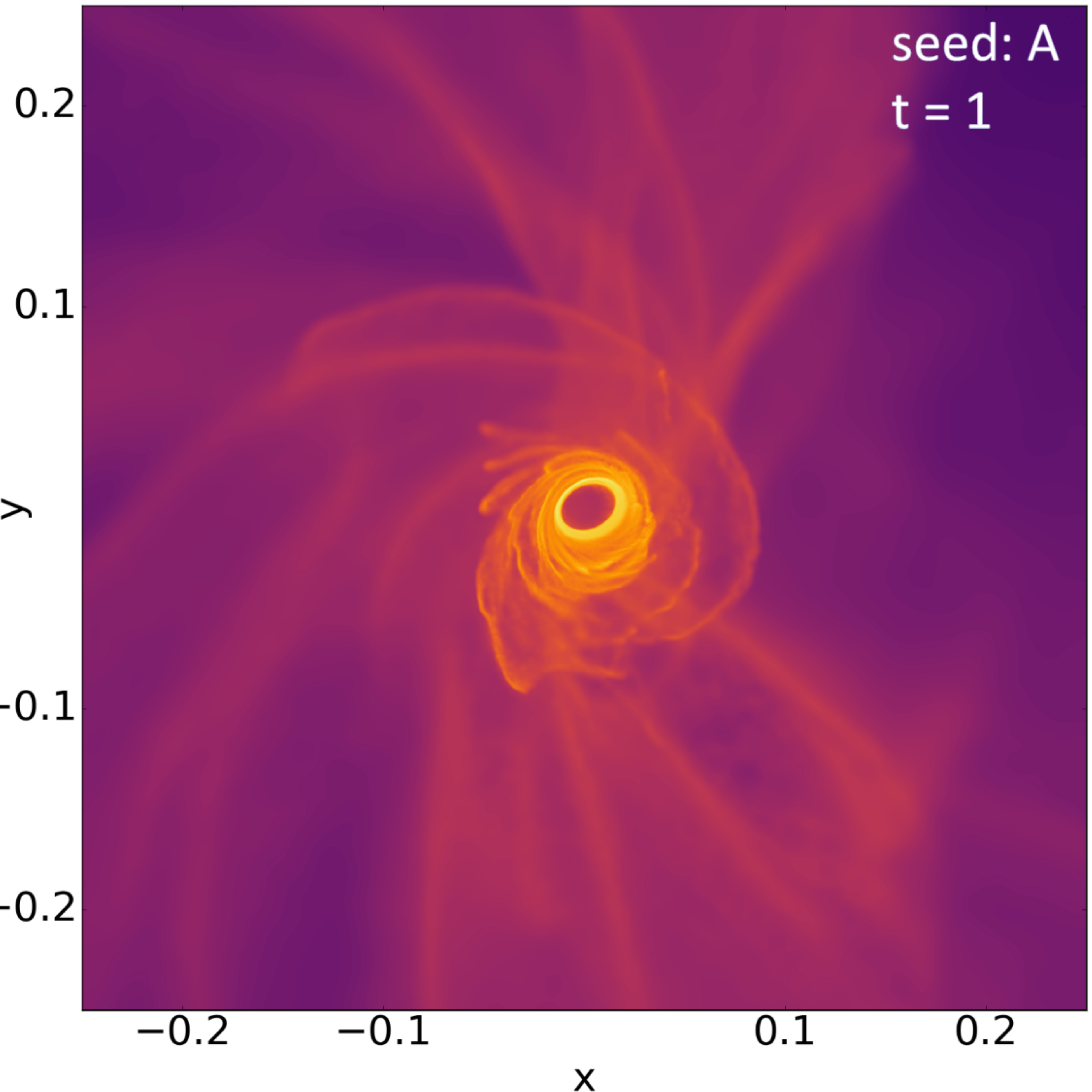}
	\includegraphics[width=43.3mm]{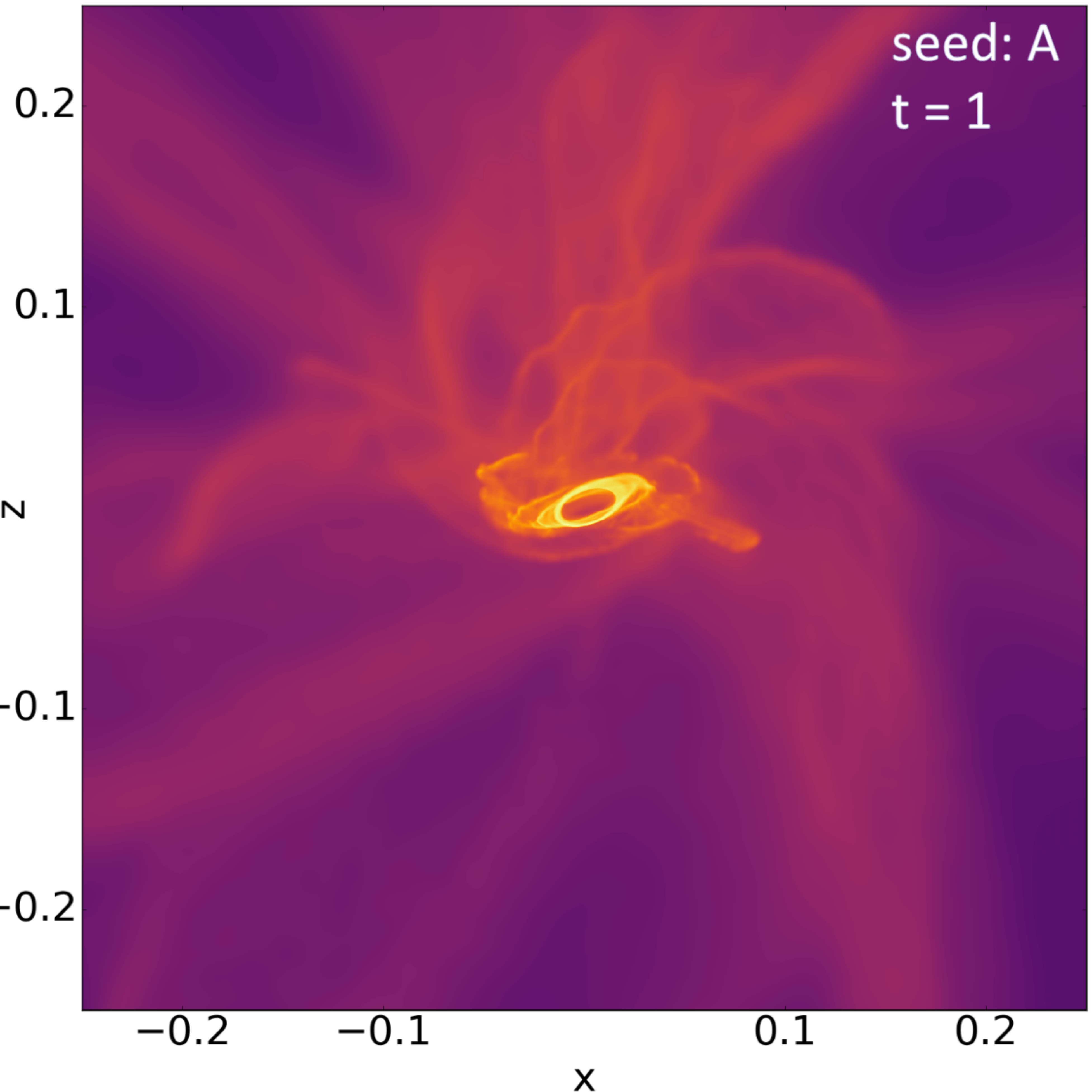}
    
	\includegraphics[width=43.3mm]{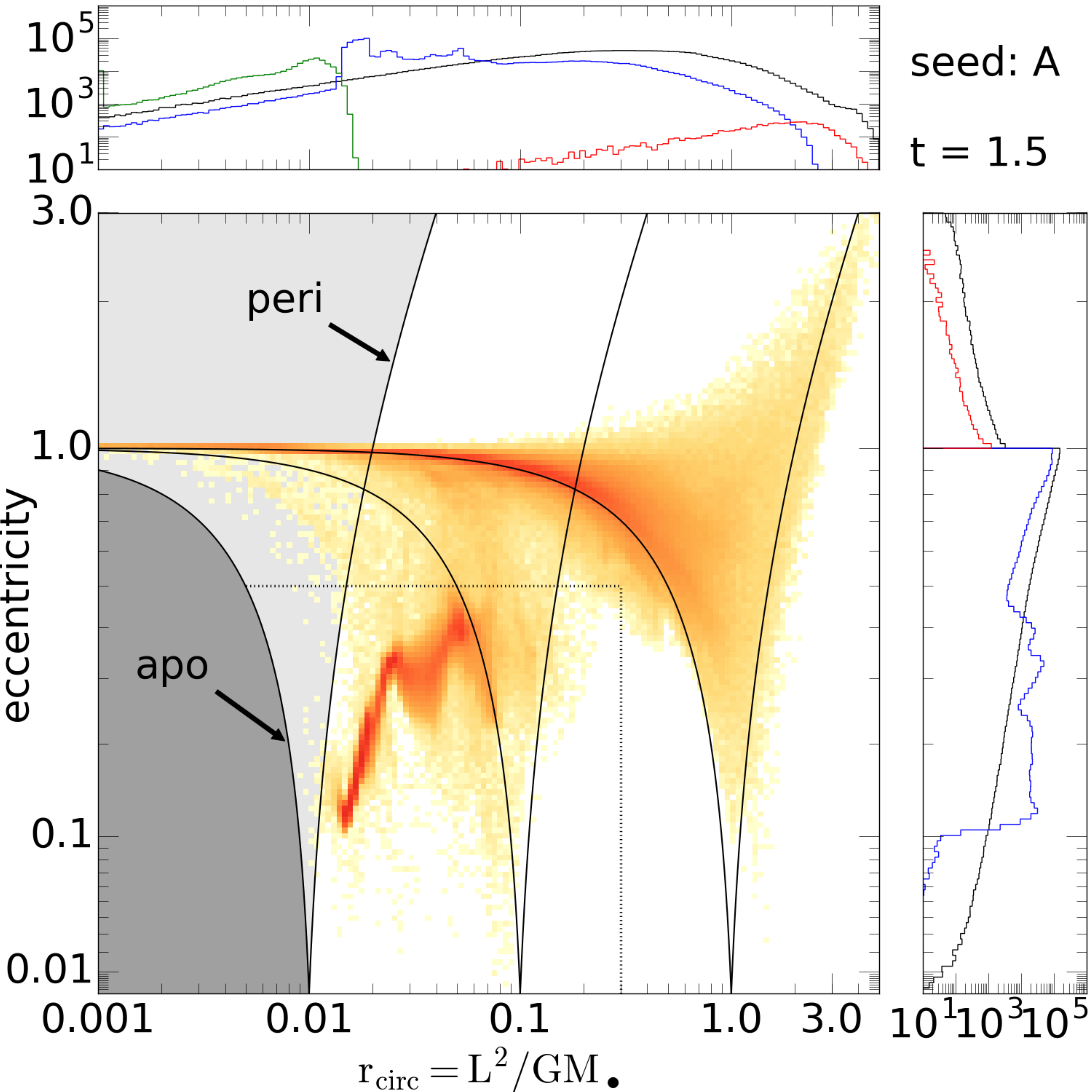}
	\hfill
	\includegraphics[width=43.3mm]{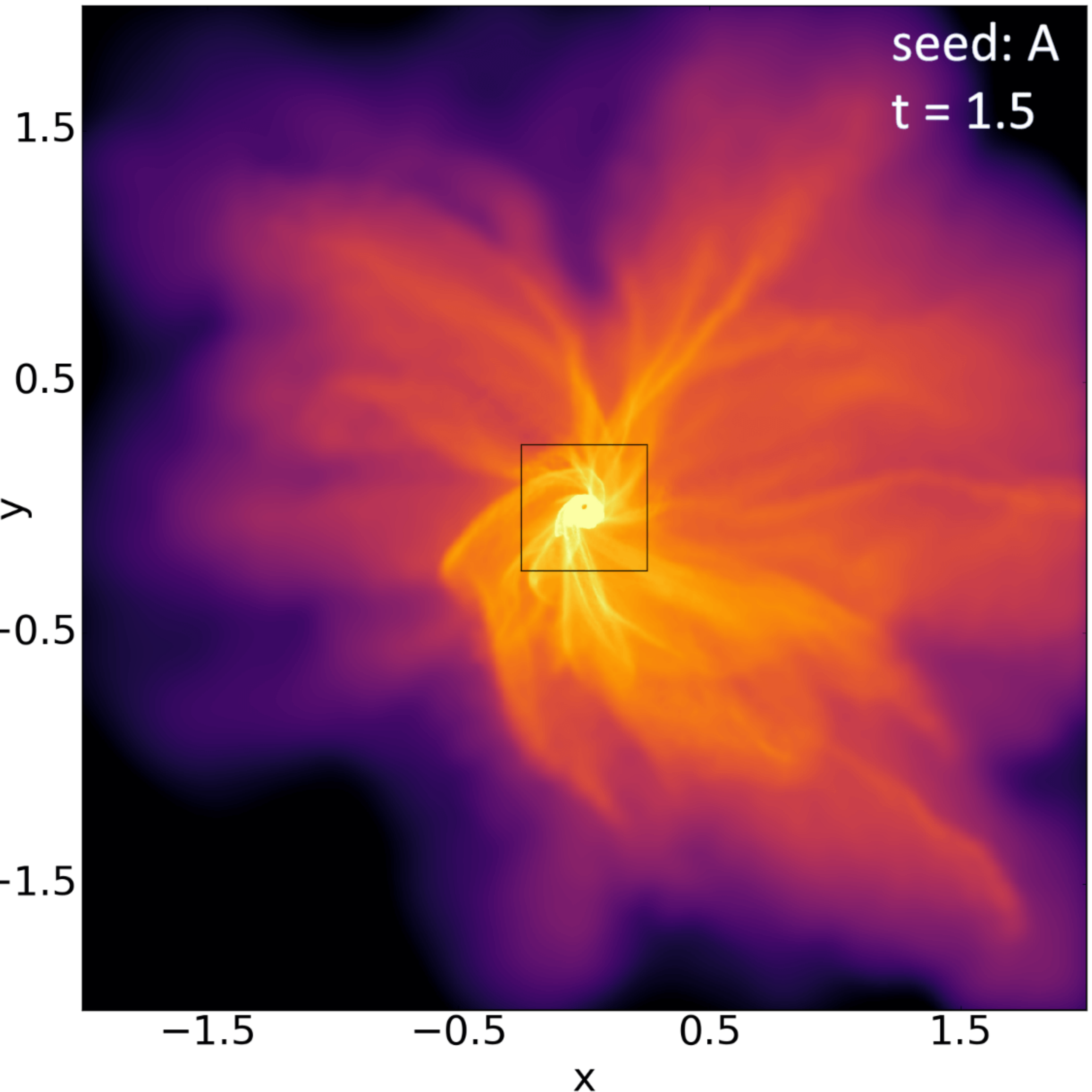}
	\includegraphics[width=43.3mm]{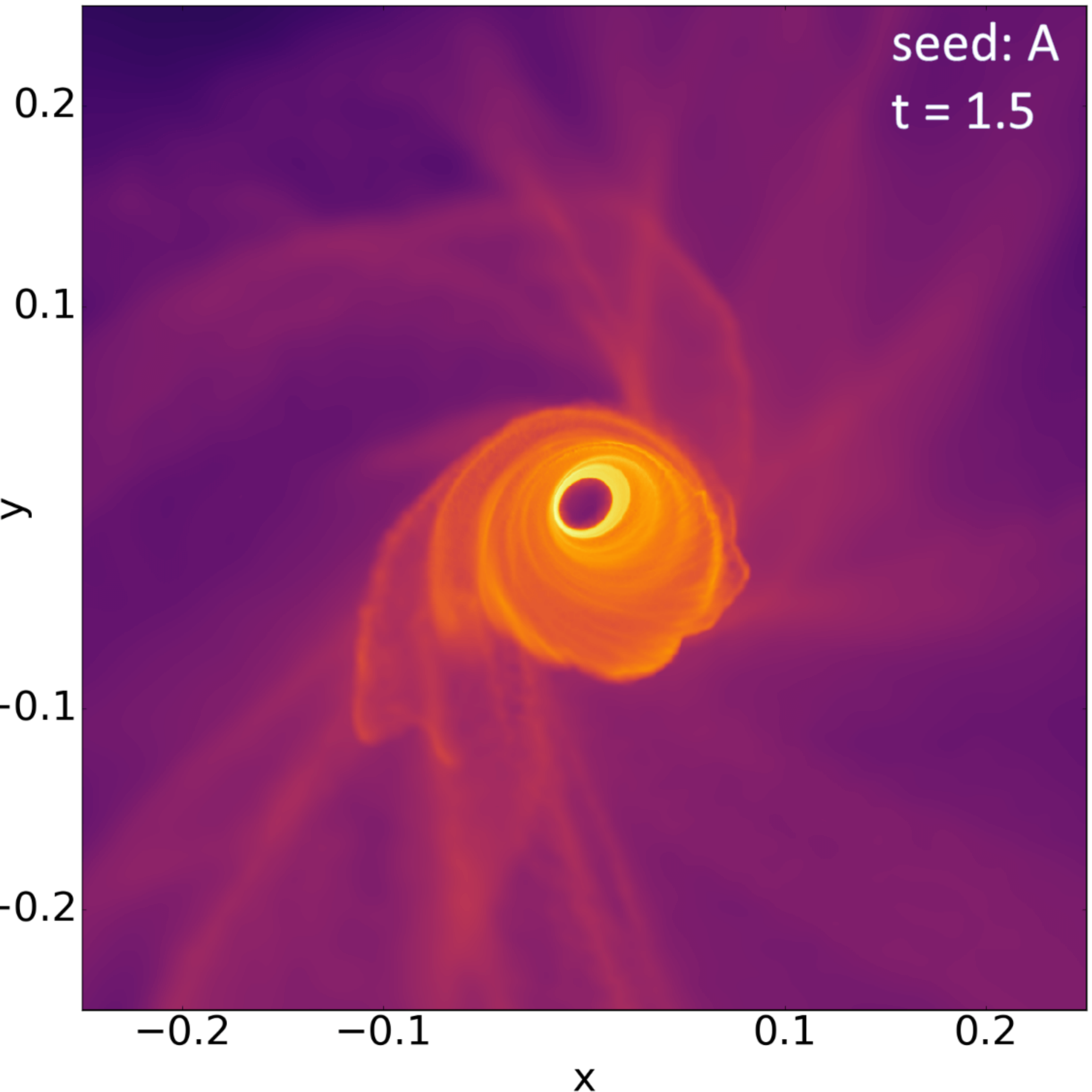}
	\includegraphics[width=43.3mm]{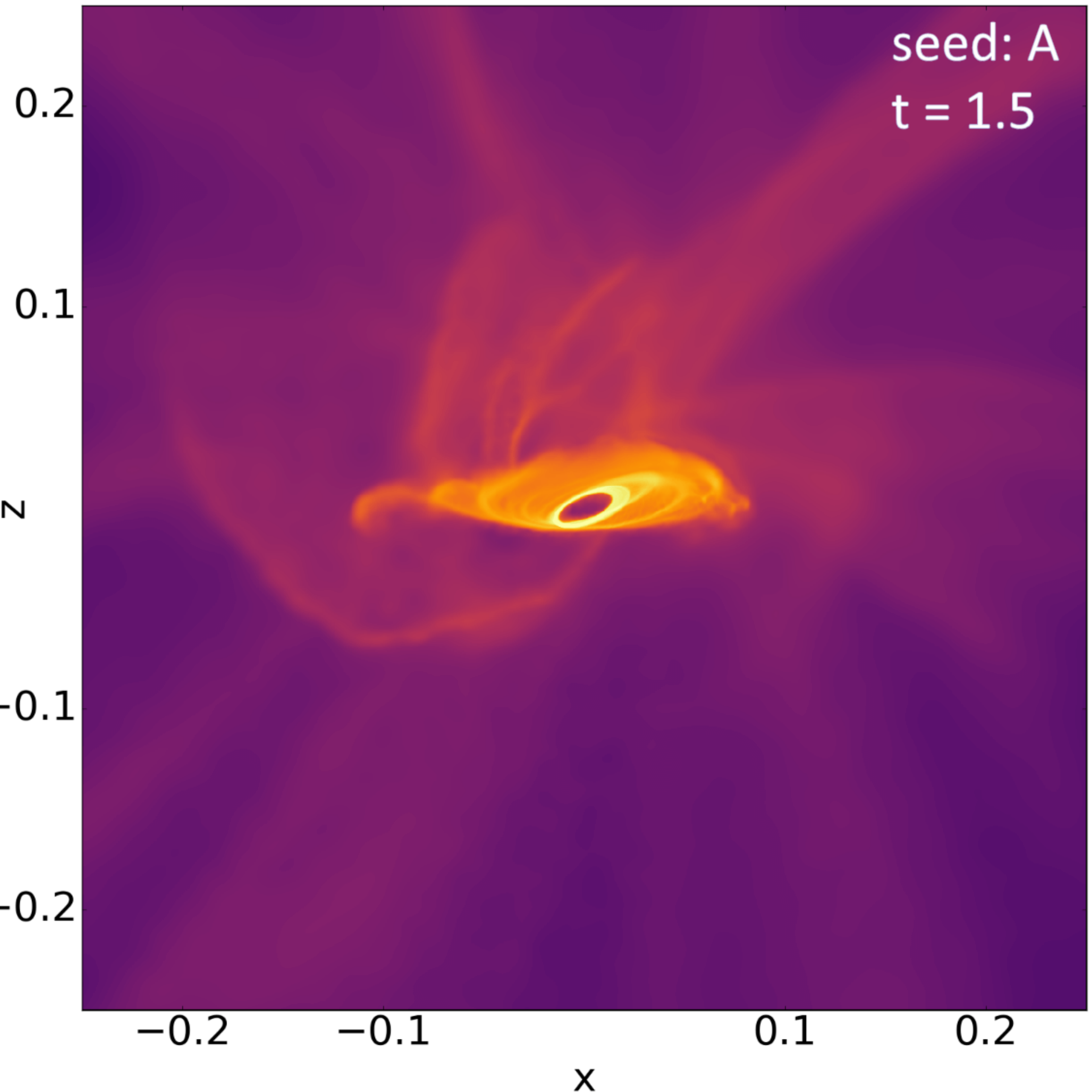}
    
	\includegraphics[width=43.3mm]{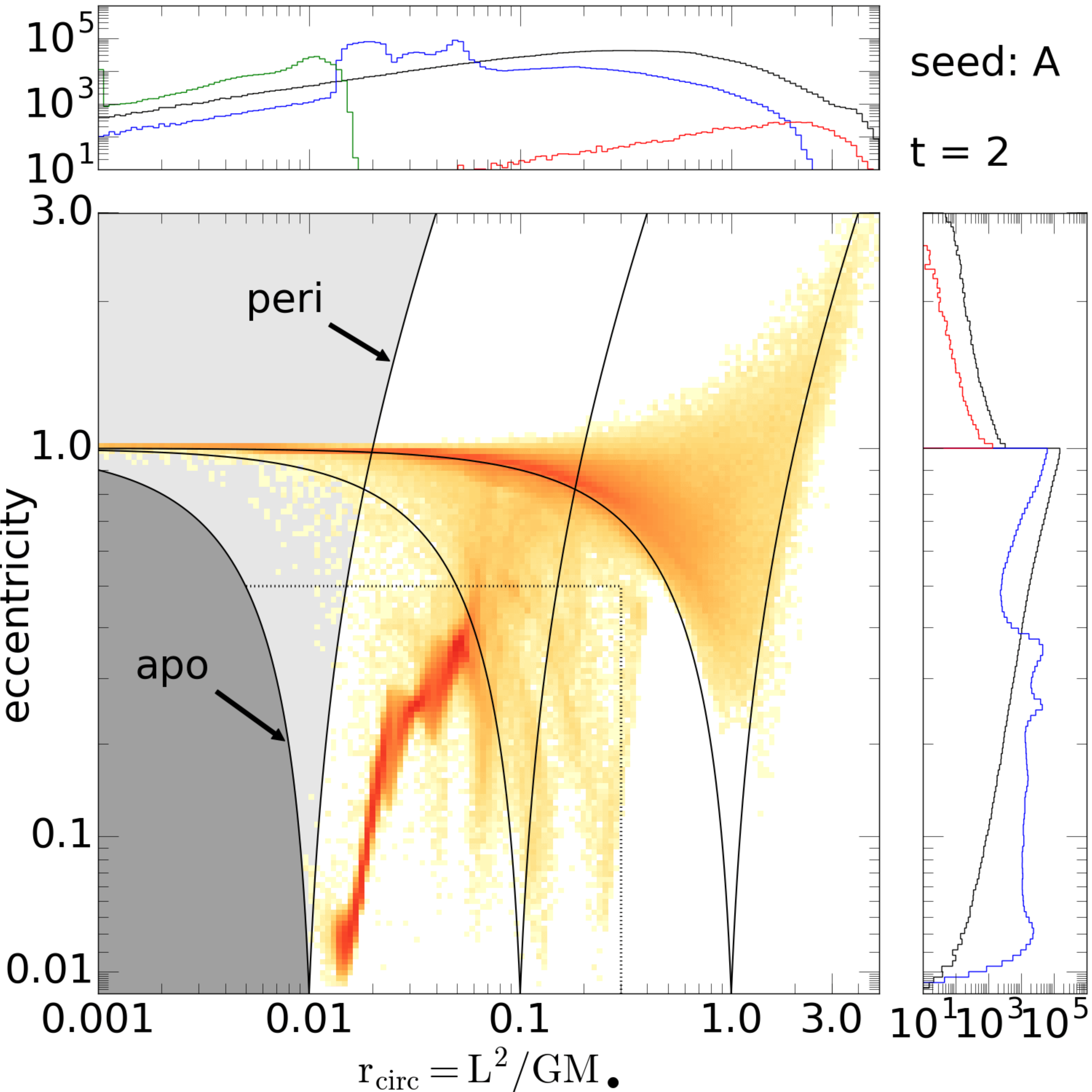}
	\hfill
	\includegraphics[width=43.3mm]{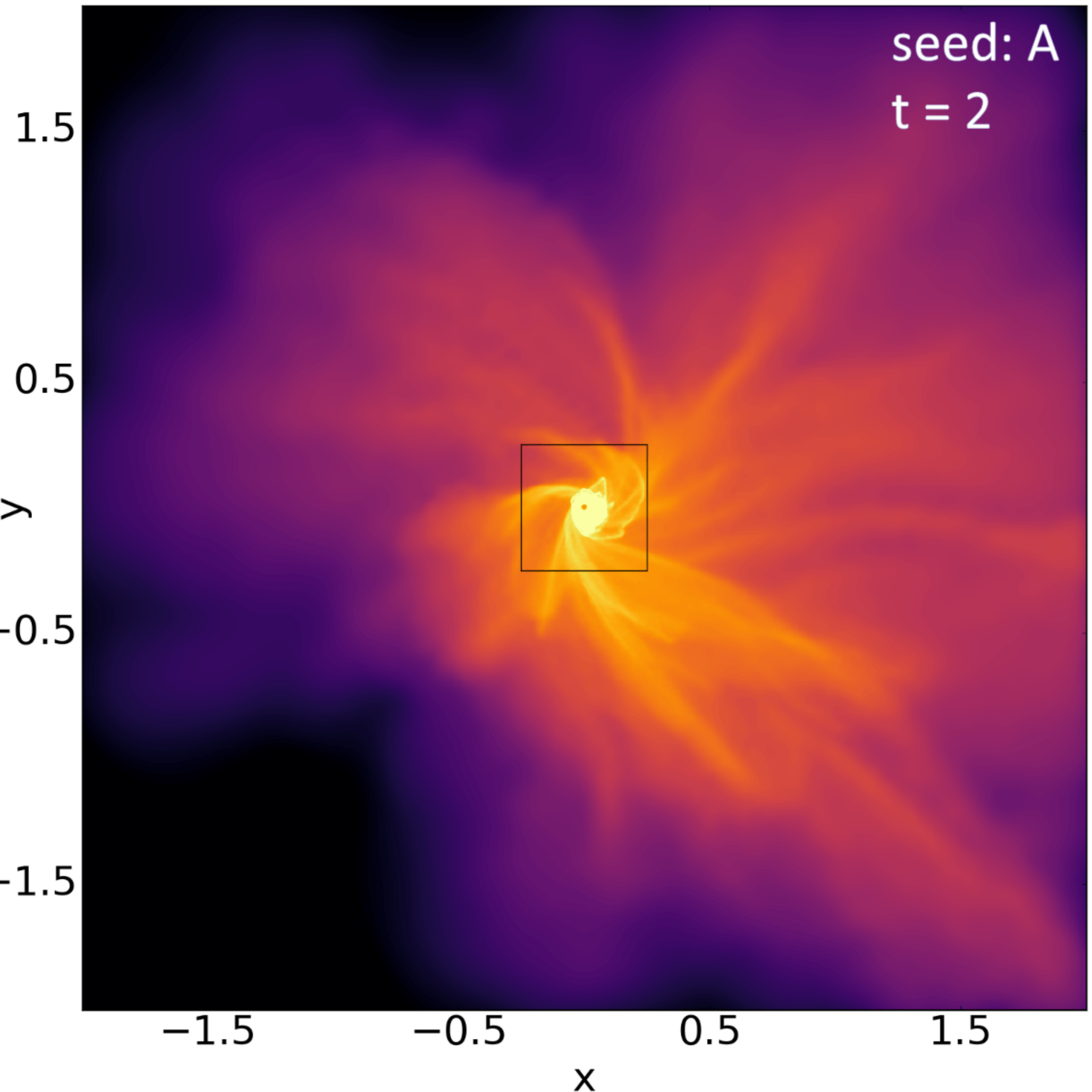}
	\includegraphics[width=43.3mm]{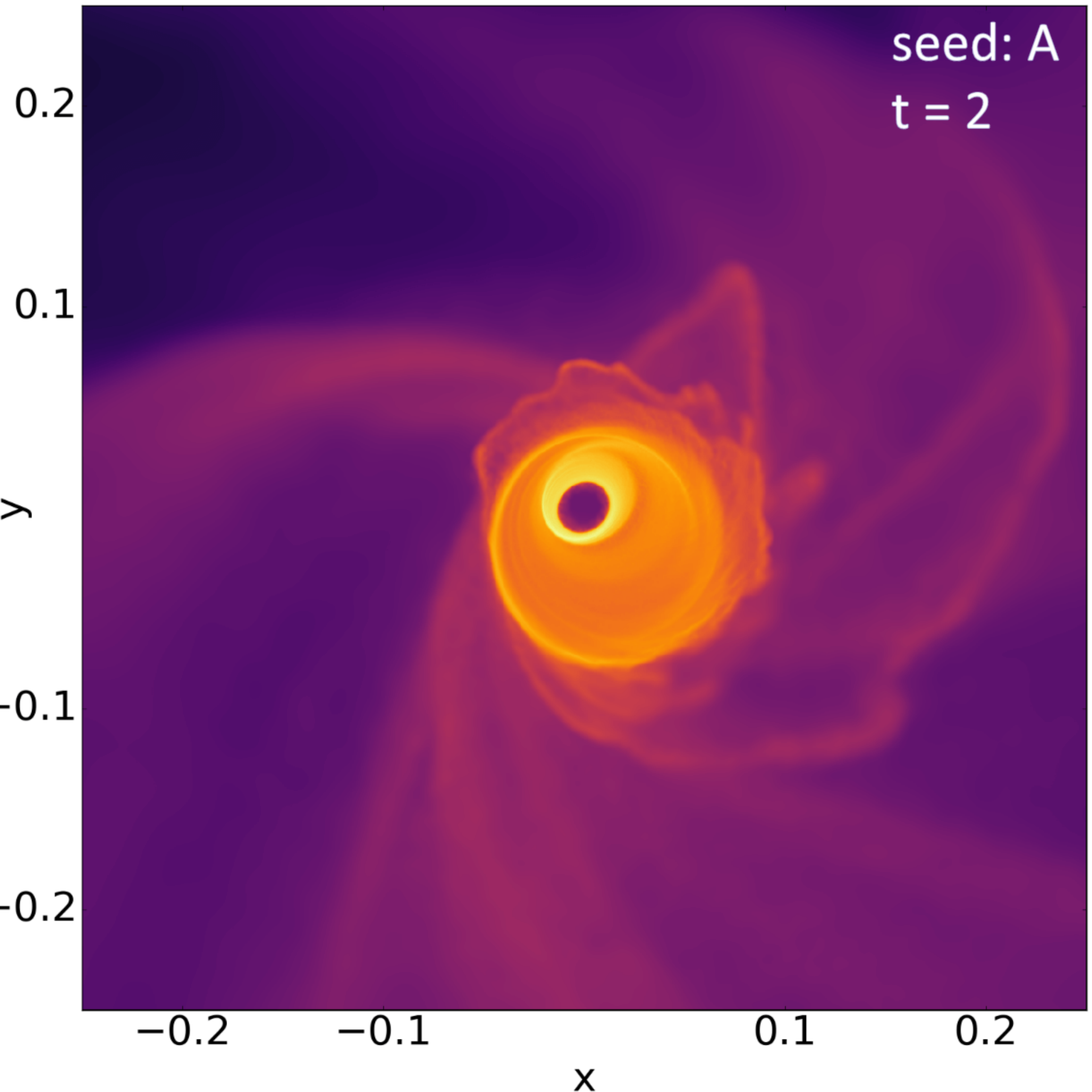}
	\includegraphics[width=43.3mm]{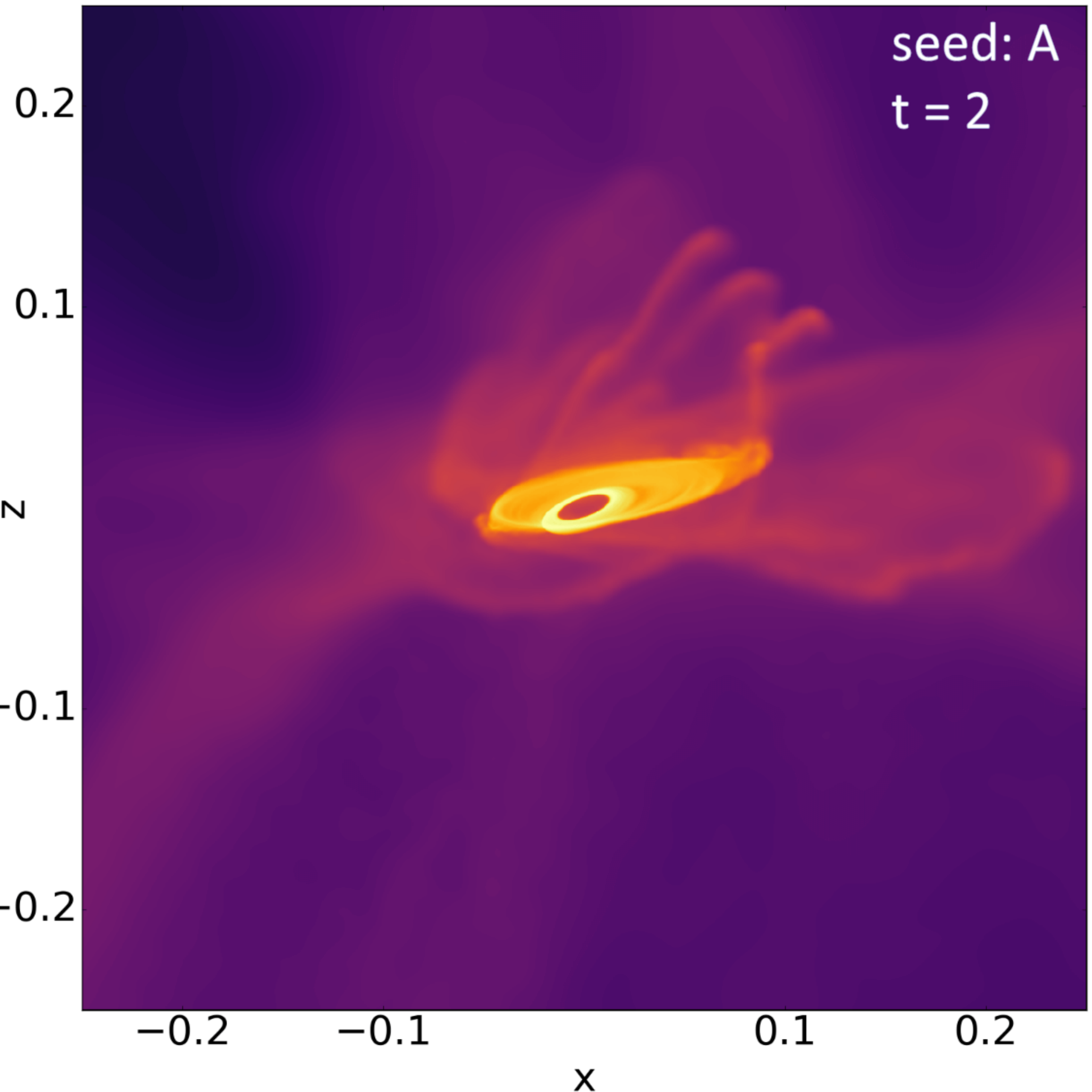}
    
	\includegraphics[width=43.3mm]{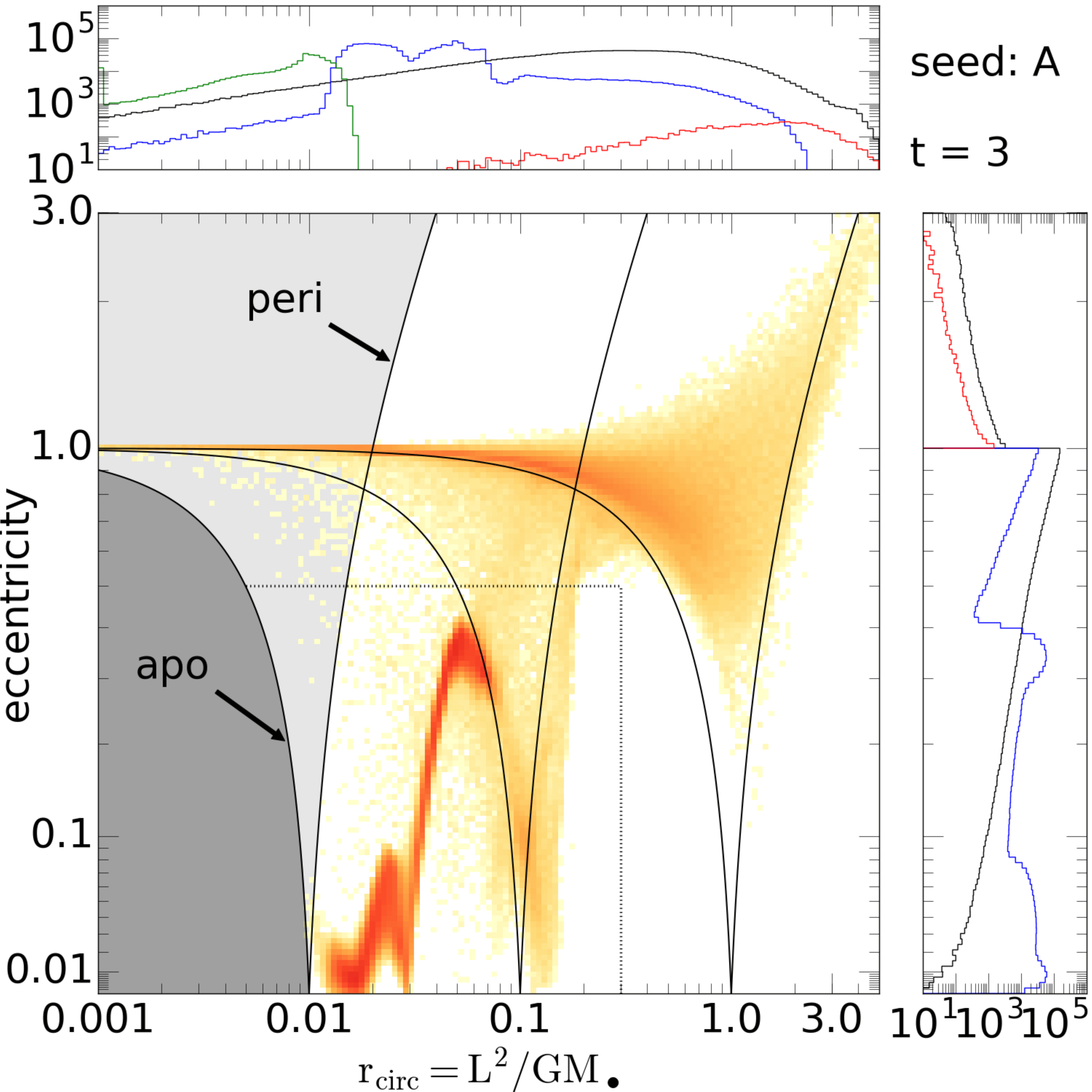}
	\hfill
	\includegraphics[width=43.3mm]{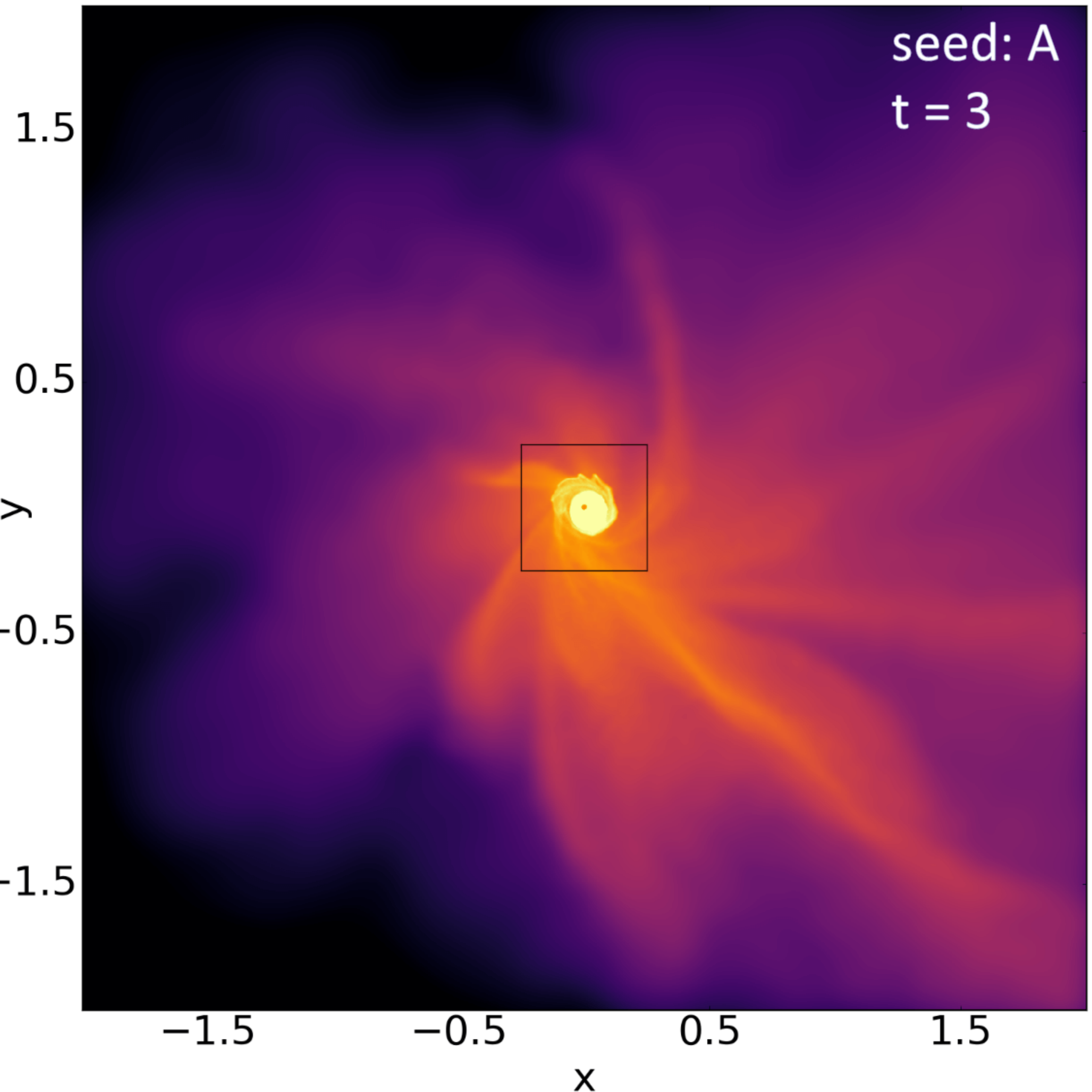}
	\includegraphics[width=43.3mm]{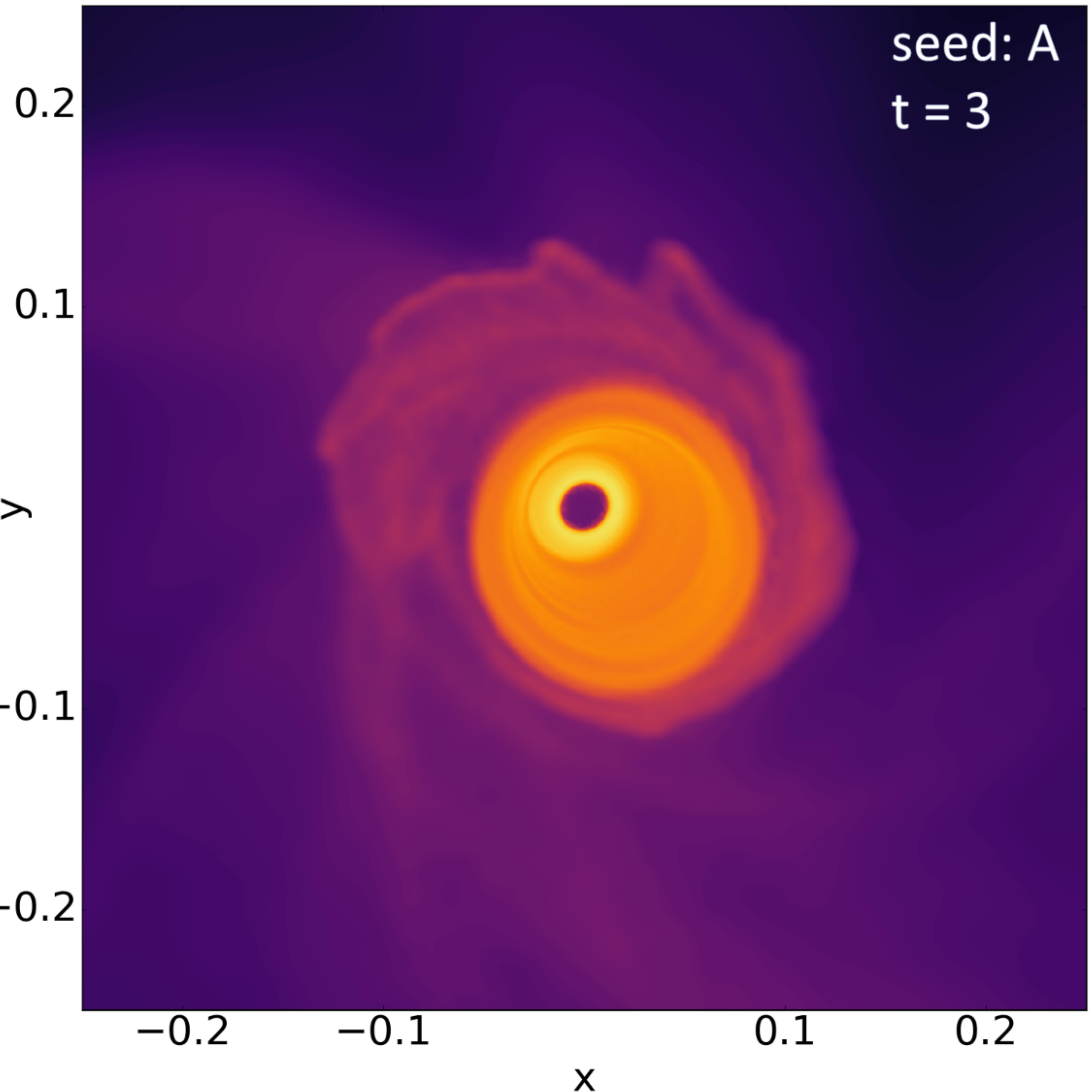}
	\includegraphics[width=43.3mm]{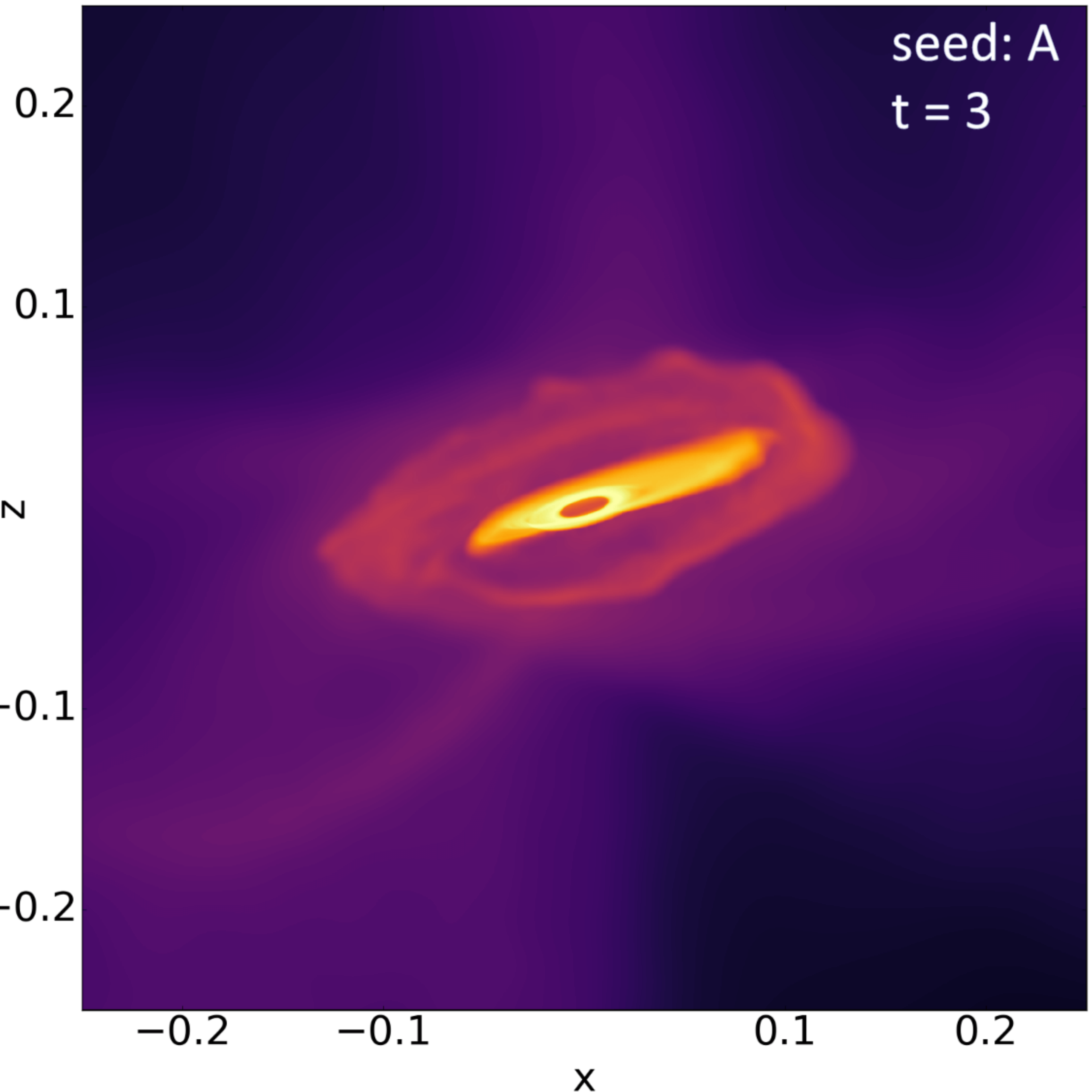}
	
	\vspace{-1.0mm}
	\caption{\label{fig:combined_ref-seed633}
	Snapshots for the reference simulation at (top to bottom) $t=0.5,\,1.0,\,1.5,\,2.0,\,3.0$. \emph{Column 1}: gas distributions over $e$ and $\rcirc$ (as in Fig.~\ref{fig:initial2Dhist_ref-seed633}). The black and green histograms in the top and right sub-panels refer, respectively, to the initial state and the absorbed gas (the bin at $\rcirc=0.001$ includes gas absorbed with $\rcirc\le0.001$). \emph{Column 2}: gas density (log scale with range of 10$^6$) over $(x,\,y$) for gas near $z=0$, centred on the sink particle. \emph{Column 3}: a zoom into the inner part of the plots in column 2. \emph{Column 4}: like column 3, but for gas near the $y=0$ plane. The central gap in the disc visible in columns 3\&4 is due to the inner boundary, where gas reaching $\sub{r}{sink}=0.01$ is absorbed into the sink. In column 1 this disc corresponds to the structure at $e\lesssim0.3$ and $\rcirc\lesssim0.1$.}
\end{figure*}

In view of the lack of detailed observational data and the huge parameter space, we chose to pick typical physical parameters to define a reference simulation. In this section we present the results from this reference simulation (and its random sisters) in detail, while in the Section~\ref{sec:parasweep} individual parameters are altered to investigate their importance. The physical parameters for the reference simulation have already been mentioned in the previous section and can also be found in the top row of Table~\ref{tab:paroverview}.

\subsection{A detailed look at a representative simulation}
\label{sec:ref:detail}
A representation of the initial conditions can be seen in Figure~\ref{fig:initial2Dhist_ref-seed633}. The central panel shows the distribution of the initial gas shell over eccentricity $e$ and the circularisation radius $\rcirc=\vec{L}^2 /GM_\bullet$. The majority of gas is initially bound ($e<1$) and, as $r\sim\sub{r}{shell}=1$, resides between the curves for $\sub{r}{peri}=\sub{r}{shell}$ and $\sub{r}{apo}=\sub{r}{shell}$. Most of the unbound gas will quickly become bound due to the initial interactions caused by the turbulence imposed on the initial velocity field. 

Material with $\rcirc\ll\sub{r}{shell}$ and $e\sim1$ has very small angular momentum, but resides near its apo-centre. As the simulation progresses, this tail of the distribution remains occupied by low-angular-momentum gas near its apo-centre (i.e.\ at $r\gg\rcirc$). Individual gas particles stay only briefly (typically much shorter than a local dynamical time) in this region, because their angular momentum is altered by local hydrodynamics and minute changes in the SMBH position and velocity.

The top and right panels plot the gas distributions over $\rcirc$ and $e$, respectively, using different colours for bound (blue) and unbound (red) gas. In later versions of this plot, we distinguish the initial distributions and gas absorbed into the central sink particle as well (black).

Figure~\ref{fig:combined_ref-seed633} shows snapshots at $t=0.5,\,1,\,1.5,\,2,$ and 3 (from top to bottom) utilising different representations. The left column shows the gas distribution over $e$ and  $\rcirc$ as in Figure~\ref{fig:initial2Dhist_ref-seed633}, while the remaining graphs display column density plots at different scales and projections. 

The time evolution of the gas happens roughly in two phases, which are approximately separated by the free-fall time
\begin{equation}
    \sub{t}{ff}=\frac{\pi}{\sqrt{8}}\sqrt{\sub{r}{shell}^3/GM_\bullet},
\end{equation}
when most of the gas reaches the inner regions around the SMBH and forms or feeds a nuclear disc. In the first $t\sim0.1\sub{t}{ff}$ (not represented in Figure~\ref{fig:initial2Dhist_ref-seed633}), the turbulent velocity field produces filaments and clumps, which are $\sim10$ times denser than the initial state resulting in a clumpy, weakly rotating shell of gas. During this very early phase, the amount of unbound gas decreases (as energy is dissipated), though the distribution over $e$ and $\rcirc$ hardly changes otherwise.

Subsequently, the filaments and clumps fall towards the central region and are stretched into extended streams by tidal forces. The resulting focusing of material drastically enhances the chance of interactions and results in angular-momentum cancellation. This in turn reduces the average $\rcirc$ and increases the average $e$. By $t\sim0.5$ (Figure~\ref{fig:combined_ref-seed633}, first row) the first simulated gas has reached the inner region with some being absorbed into the sink and some starting to circularise and to form a disc. The orientation of this early disc is roughly edge-on if viewed along the $z$-axis, i.e.\ not aligned with the overall initial angular momentum of the shell. Further infall changes this original tilt and by $t\sim\sub{t}{ff}$ (Figure~\ref{fig:combined_ref-seed633}, second row) a disc has formed, which is aligned with the overall angular momentum. This disc is clearly visible as a peak at $e\sim0.2$ and $\rcirc\sim0.025$ in the distribution over $e$ and $\rcirc$ (Figure~\ref{fig:combined_ref-seed633}, first column, second row).

In the following evolution, additional infalling filaments hit the disc and may cause angular-momentum cancellation followed by circularisation resulting in the growth of the disc out to $0.1\sub{r}{shell}$, though still with substantial eccentricities of $e\sim0.1-0.4$. This phase last roughly to $t=1.5\sub{t}{ff}$, when further infall of filaments and clumps onto the disc ceases sufficiently for the disc to settle. In particular in its inner parts, the eccentricity drops to $e\ll0.1$, while it remains quite eccentric in its outer parts. The disc suffers a small, but constant flow of gas into the (unresolved) inner region at $r<\sub{r}{sink}$. 

Since the focus of this study is the interaction of the filaments, the simulation is stopped at $t=3$ (Figure~\ref{fig:combined_ref-seed633}, bottom row). A comparison of the distributions over $\rcirc$ (Figure~\ref{fig:combined_ref-seed633}, left column: top sub-panels) between the initial conditions (black) with the gravitationally bound gas (blue) and gas absorbed (green) at the end of the simulations clearly shows a considerable shift towards lower $\rcirc$. At the end of the simulation $\sim1/5$ of the gas has reached $r<\sub{r}{sink}=0.01\sub{r}{shell}$ (at which moment gas particles are absorbed into the sink) and a further $\sim2/3$ are inside the disc region. The majority of the remaining gas ($\sim1/10$ of the initial amount) is bound, but on eccentric orbits.

\begin{figure*}
	\includegraphics[width=50.5mm]{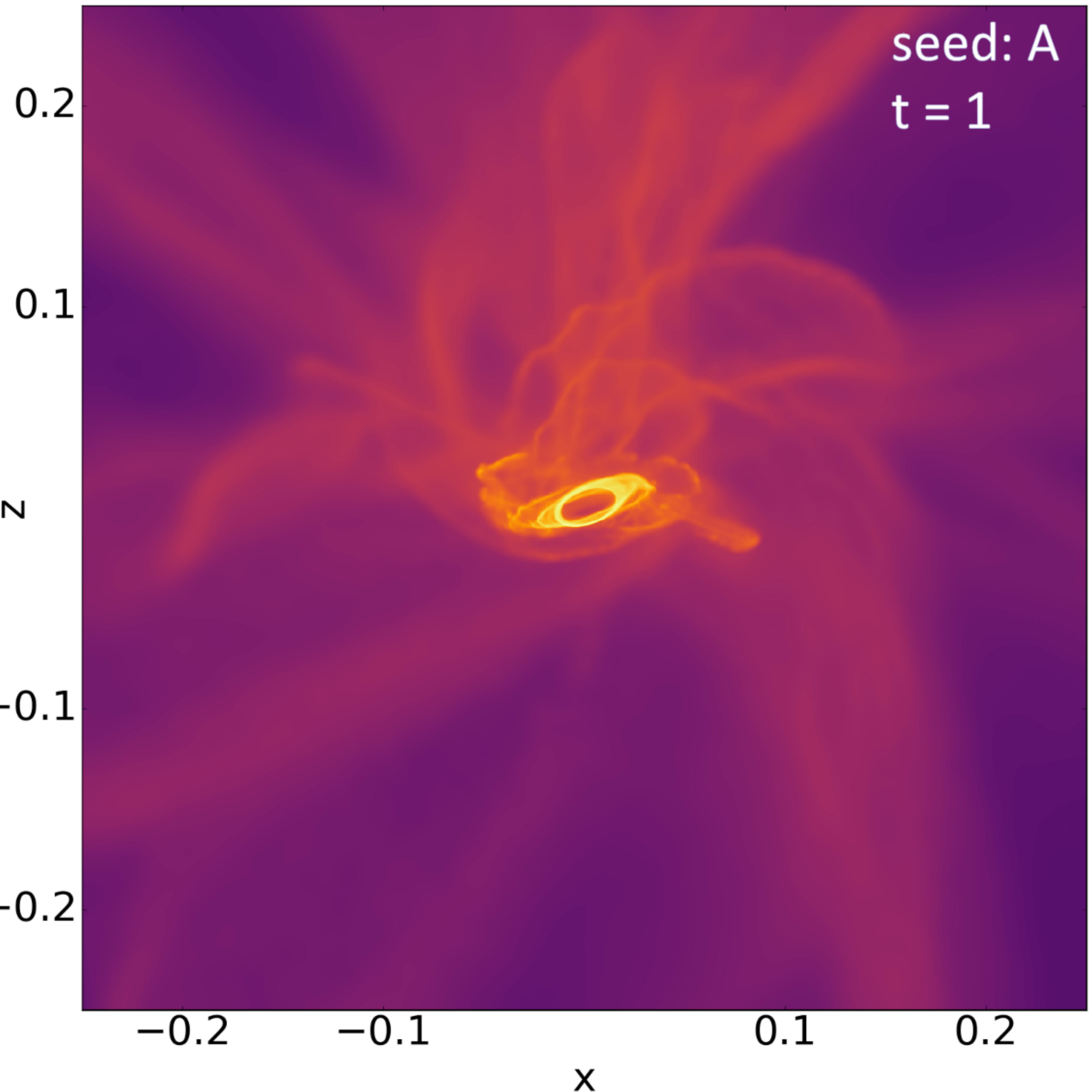}\hfil
	\includegraphics[width=50.5mm]{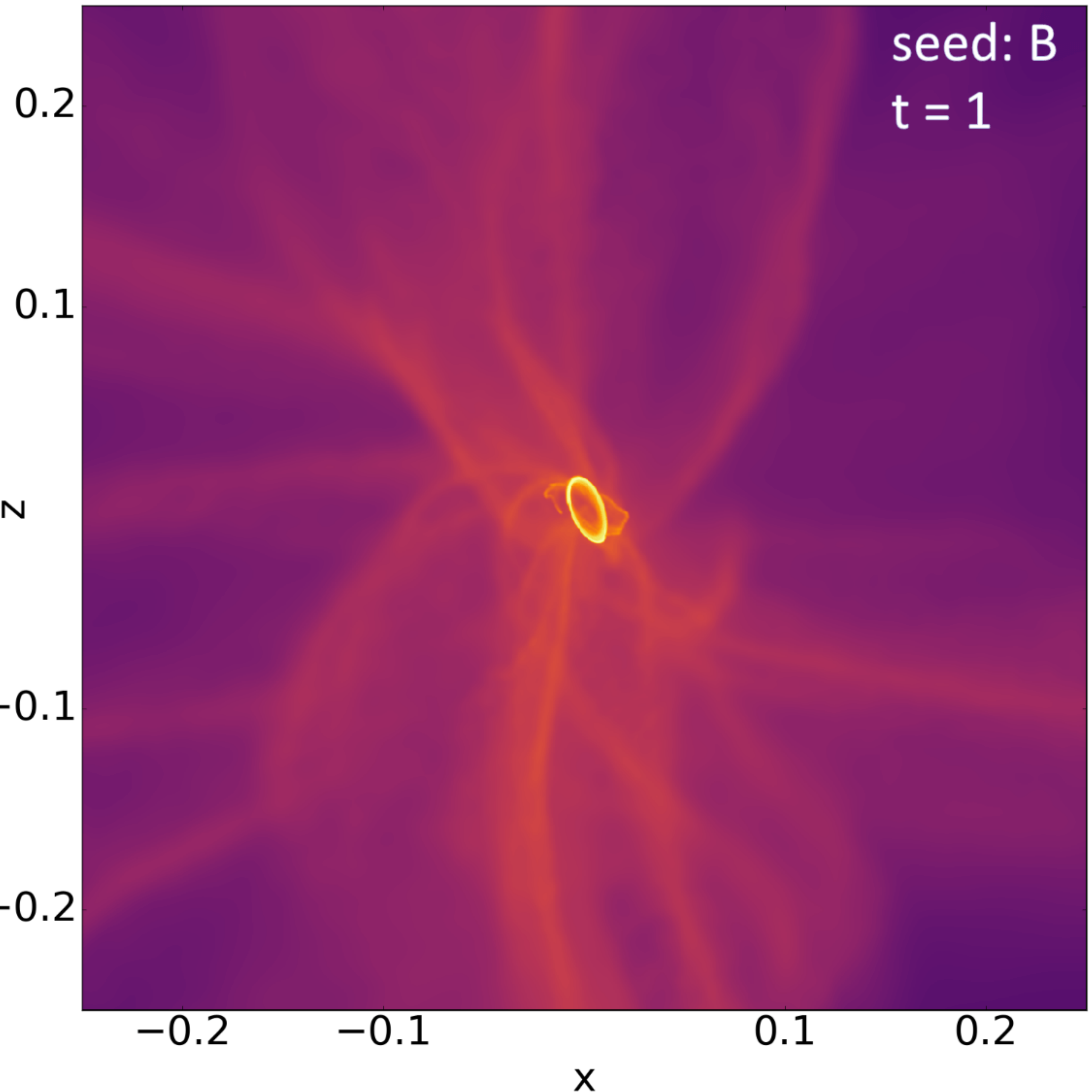}\hfil
	\includegraphics[width=50.5mm]{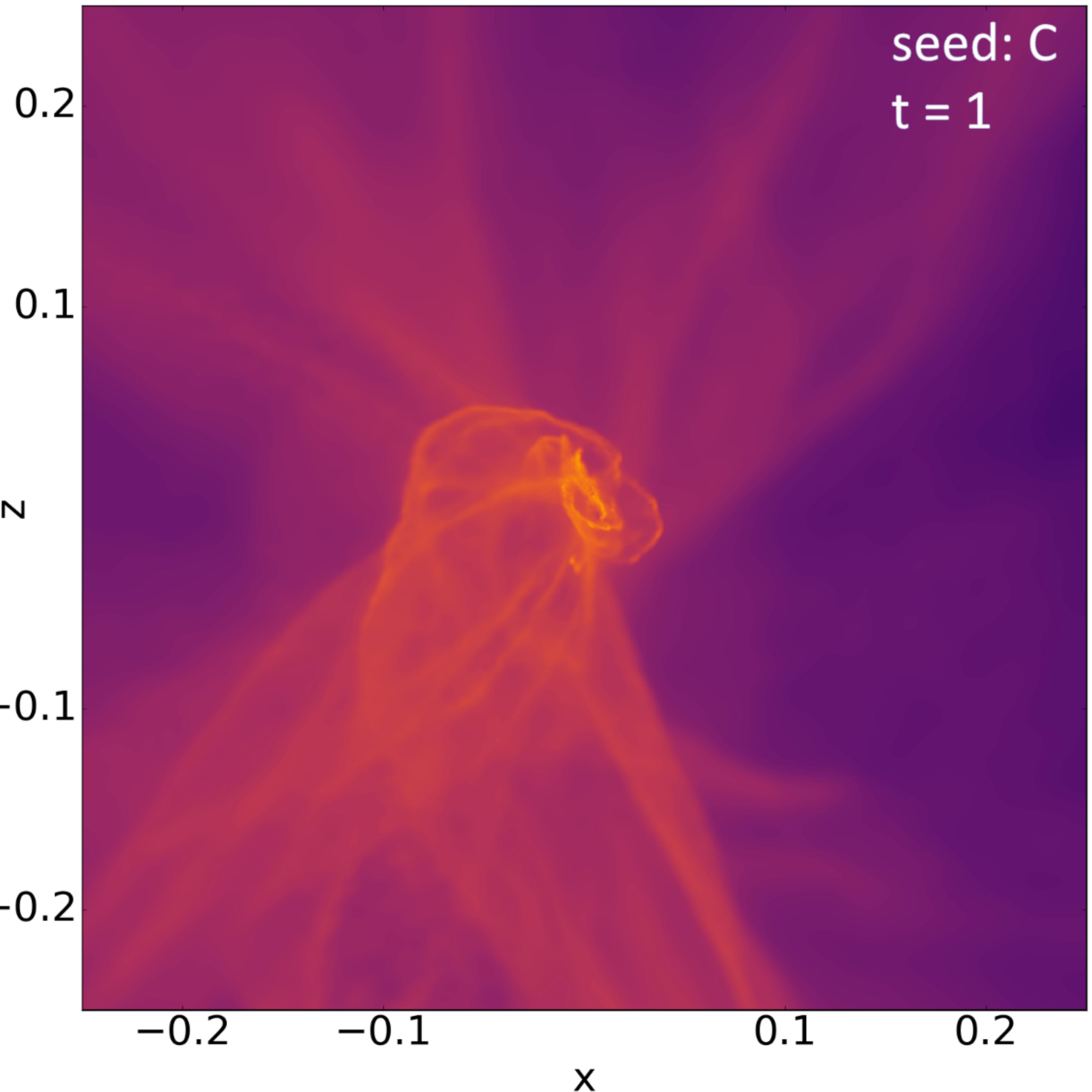}

	\includegraphics[width=50.5mm]{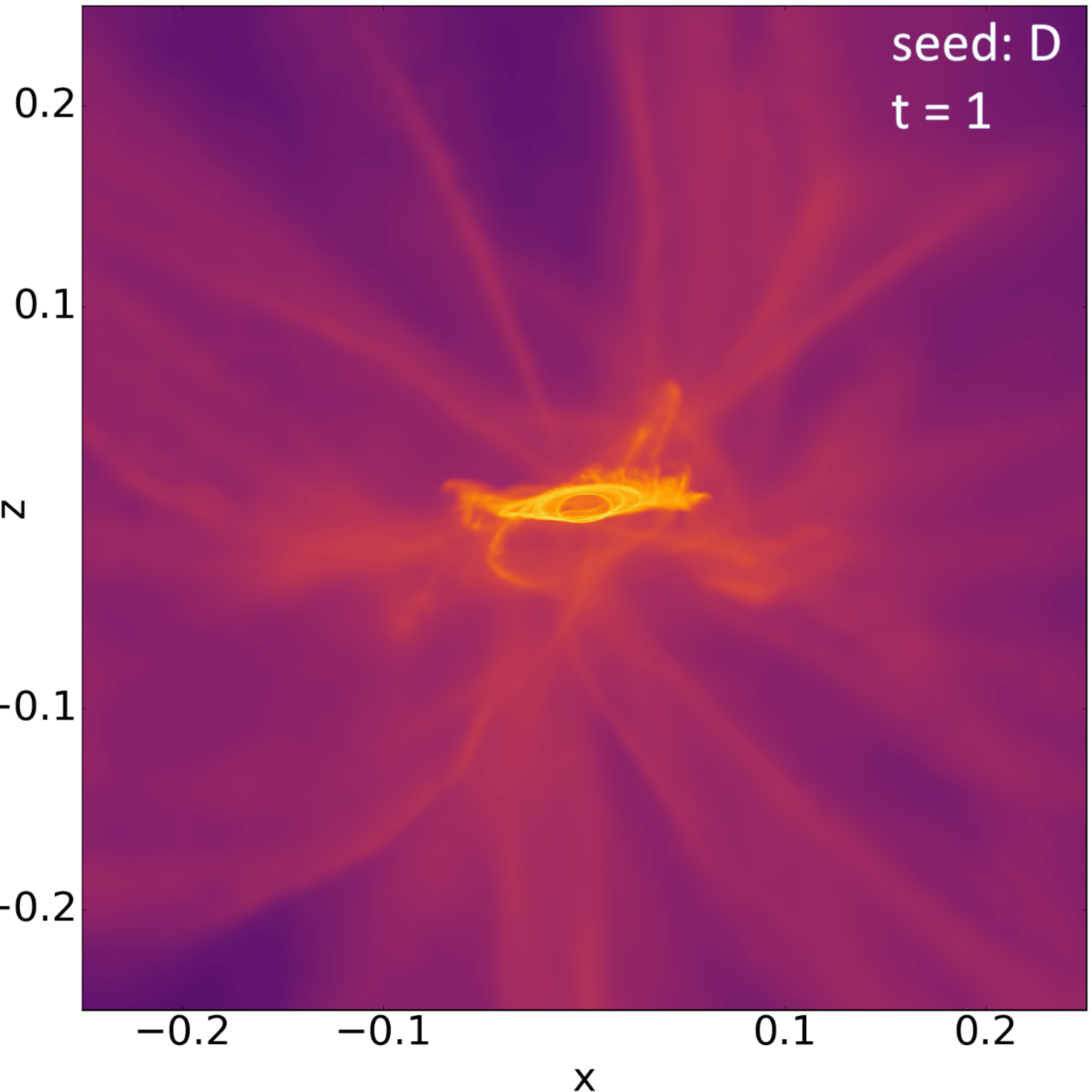}\hfil
	\includegraphics[width=50.5mm]{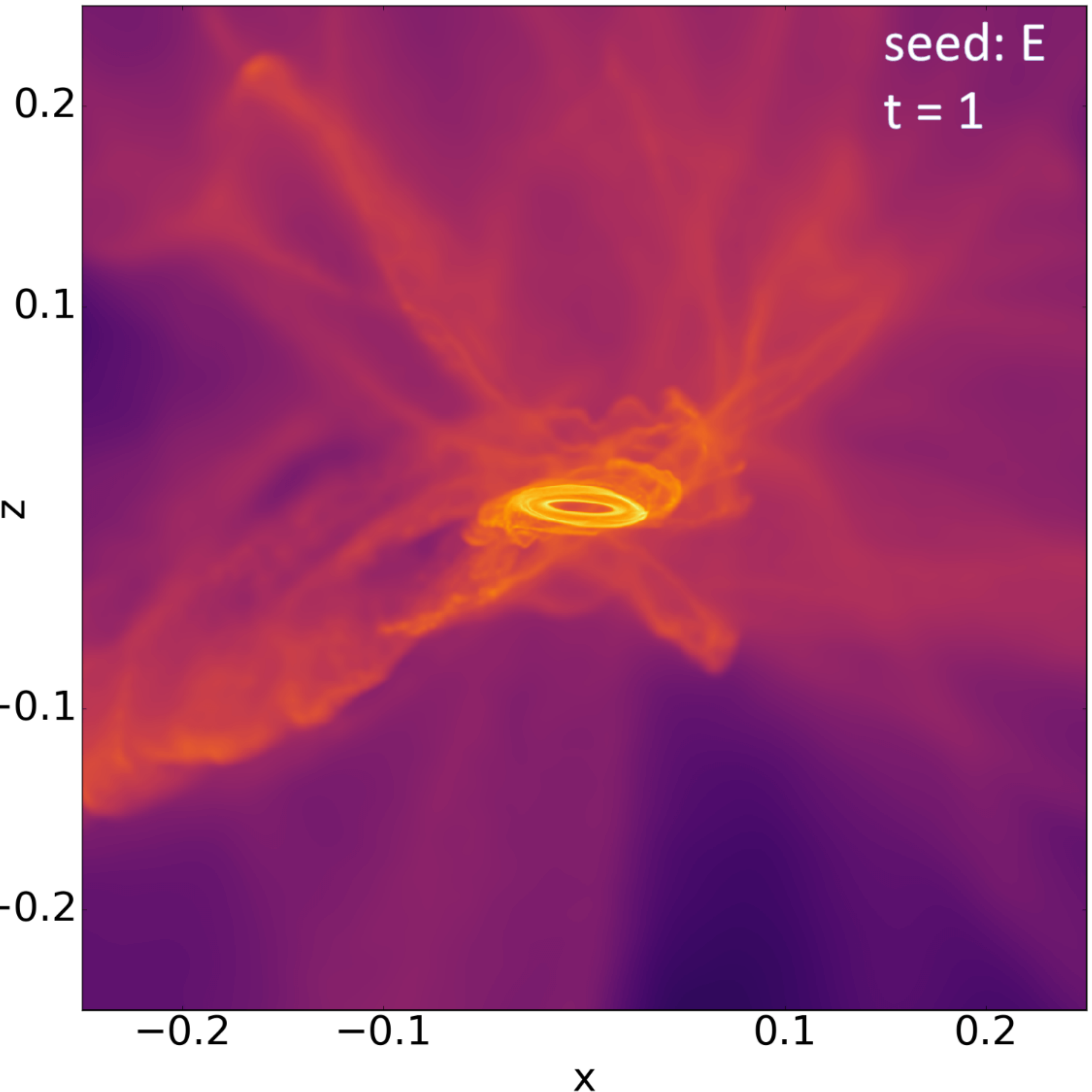}\hfil
	\includegraphics[width=50.5mm]{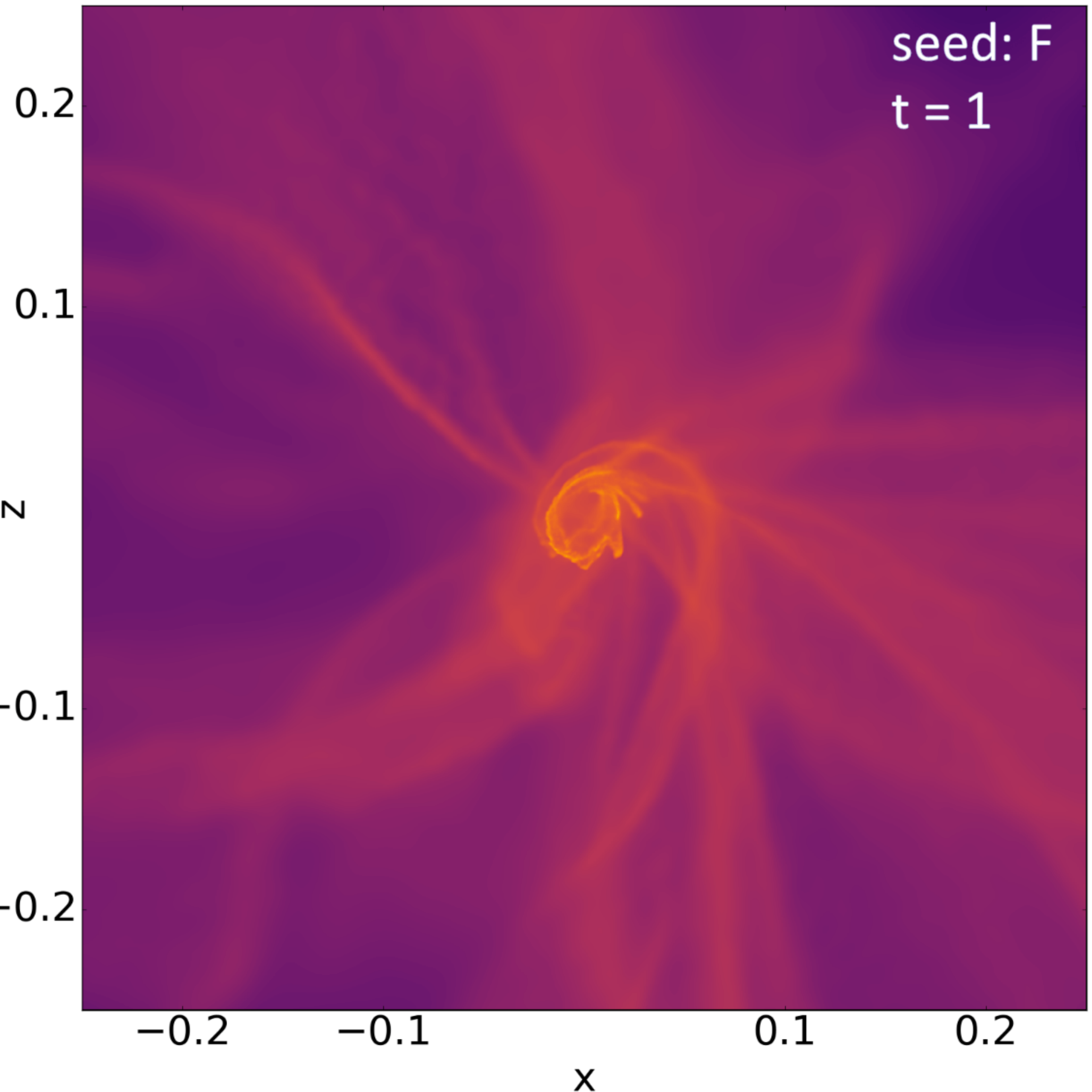}
	
    \caption{\label{fig:density_ref-seedx}
    Density plots (as in column 4 of Figure~\ref{fig:combined_ref-seed633}) at time $t=1$ for the sextet of reference simulations which differ only by the random seed used to generate the turbulent velocities and have the same parameters as the simulation presented in detail in Section~\ref{sec:ref:detail} and Figure~\ref{fig:combined_ref-seed633} (shown here in the top left). Discs orientated in accordance with the net angular momentum of the initial conditions would appear edge-on and horizontally aligned.}
\end{figure*}

\begin{figure*}
	\includegraphics[width=50.5mm]{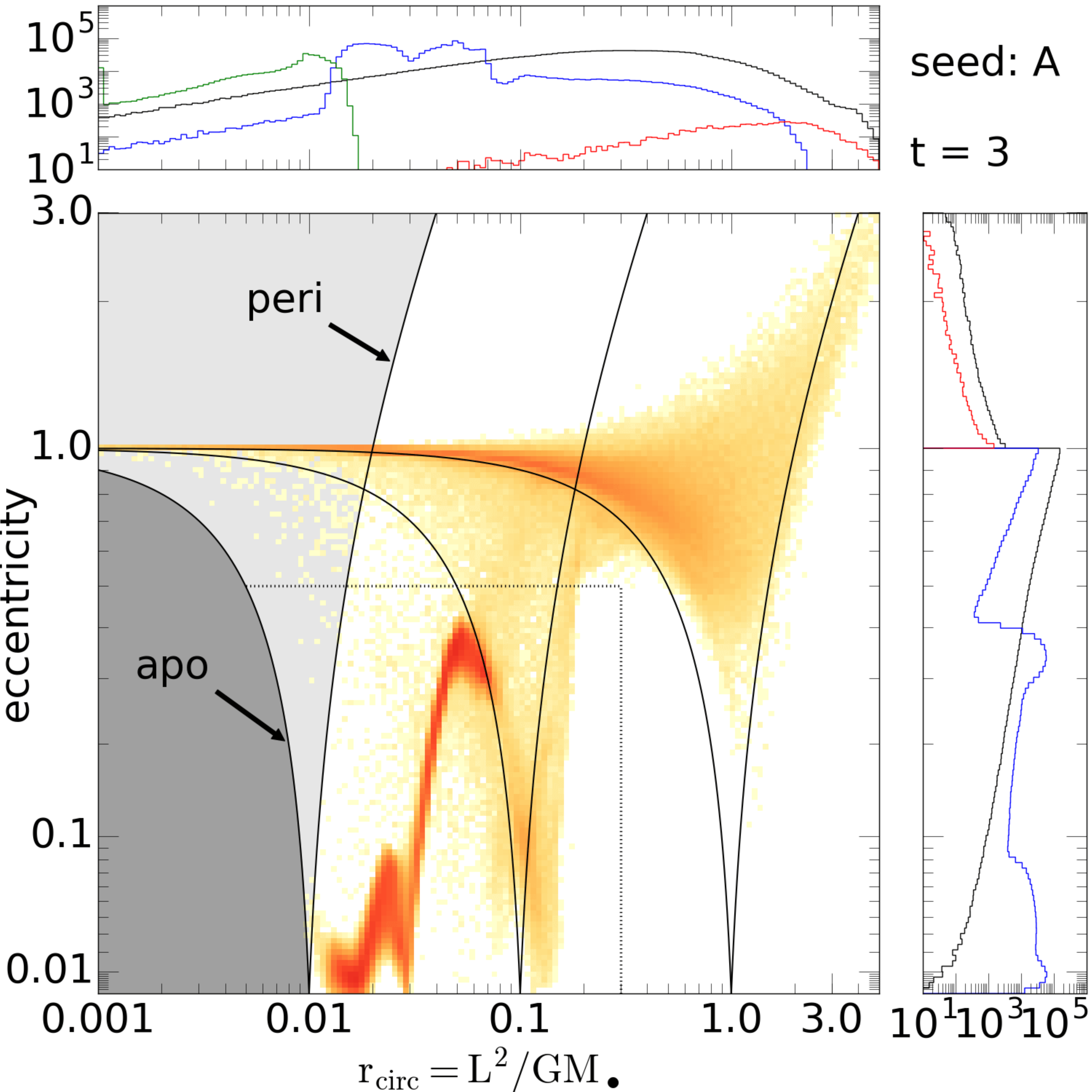}\hfil
	\includegraphics[width=50.5mm]{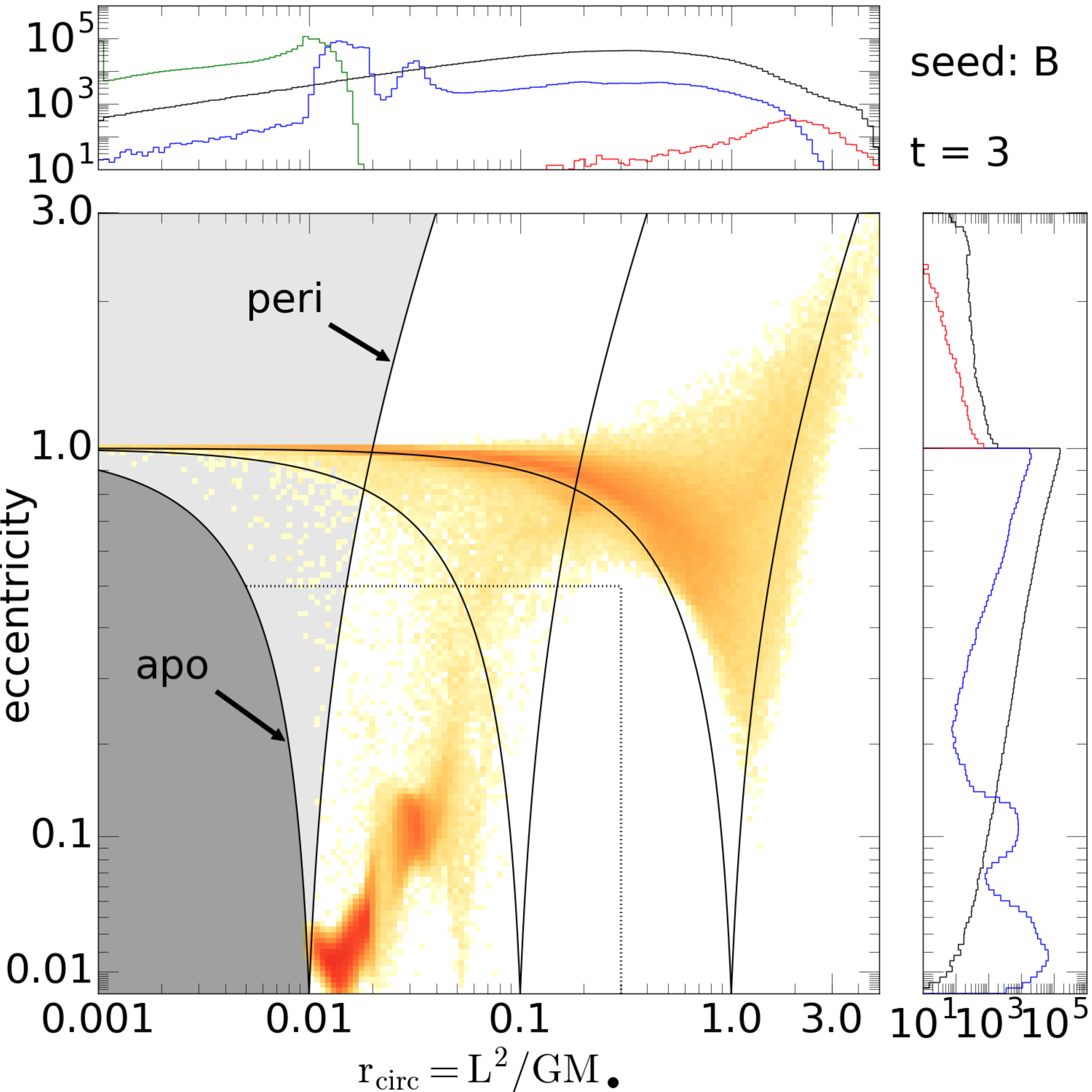}\hfil
	\includegraphics[width=50.5mm]{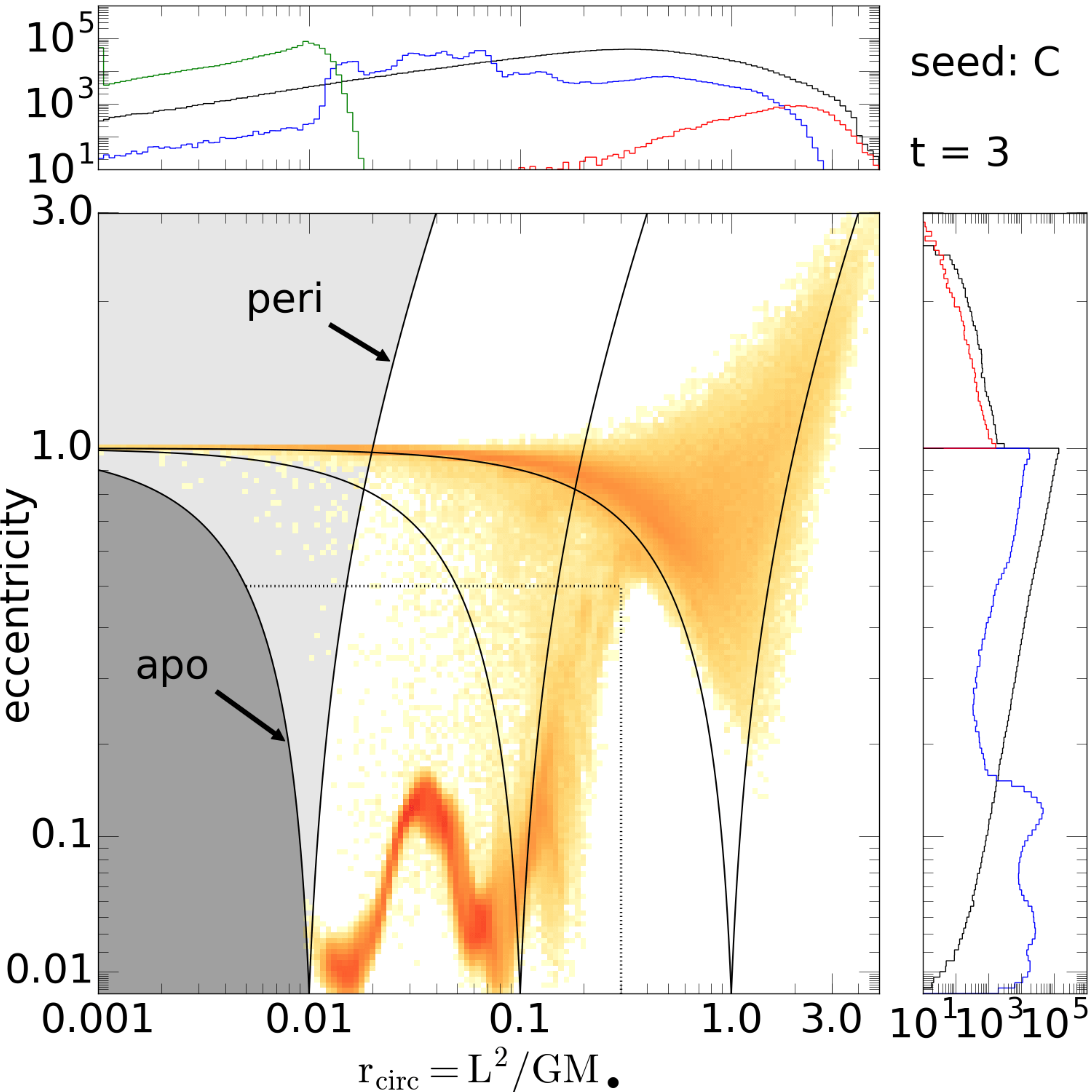}

	\includegraphics[width=50.5mm]{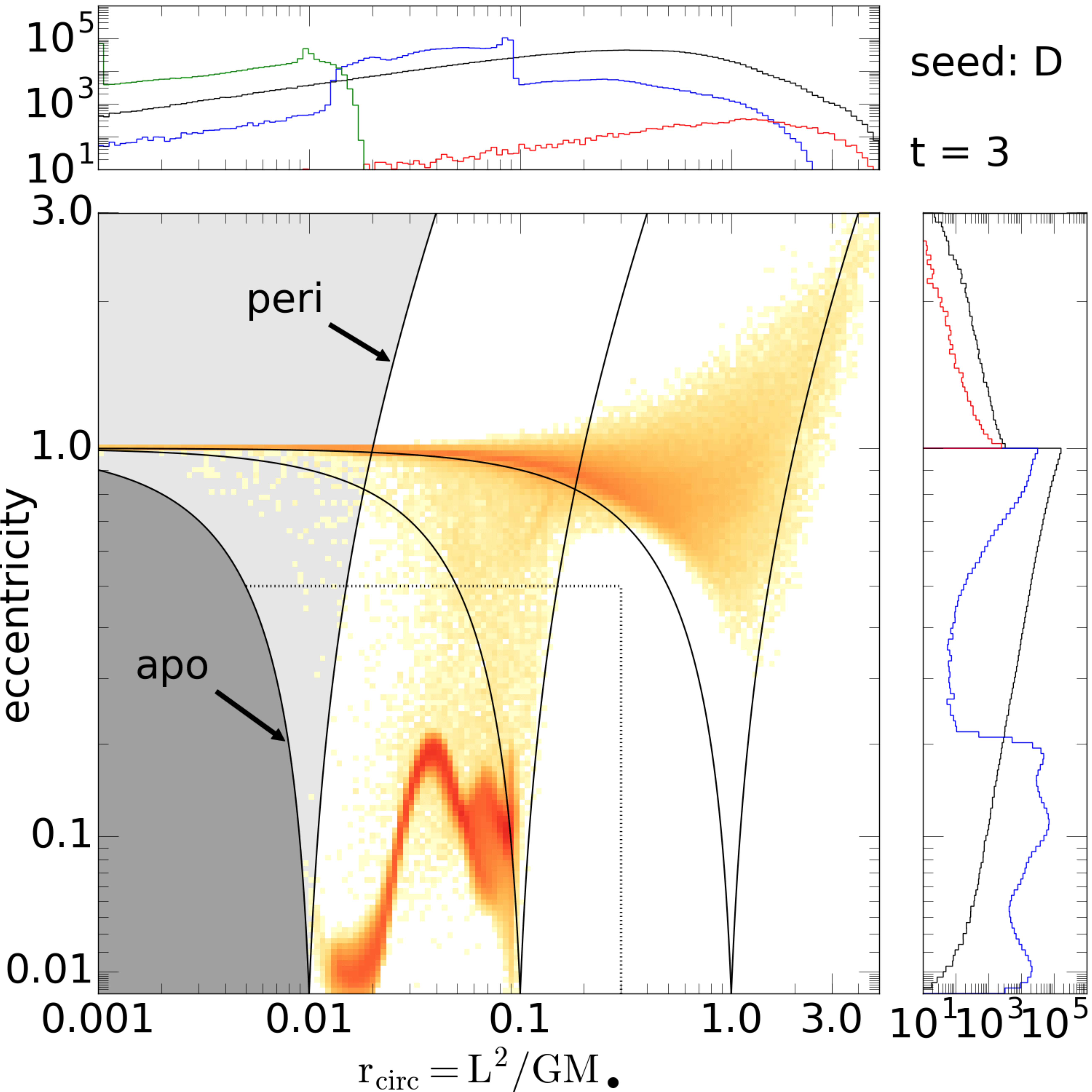}\hfil
	\includegraphics[width=50.5mm]{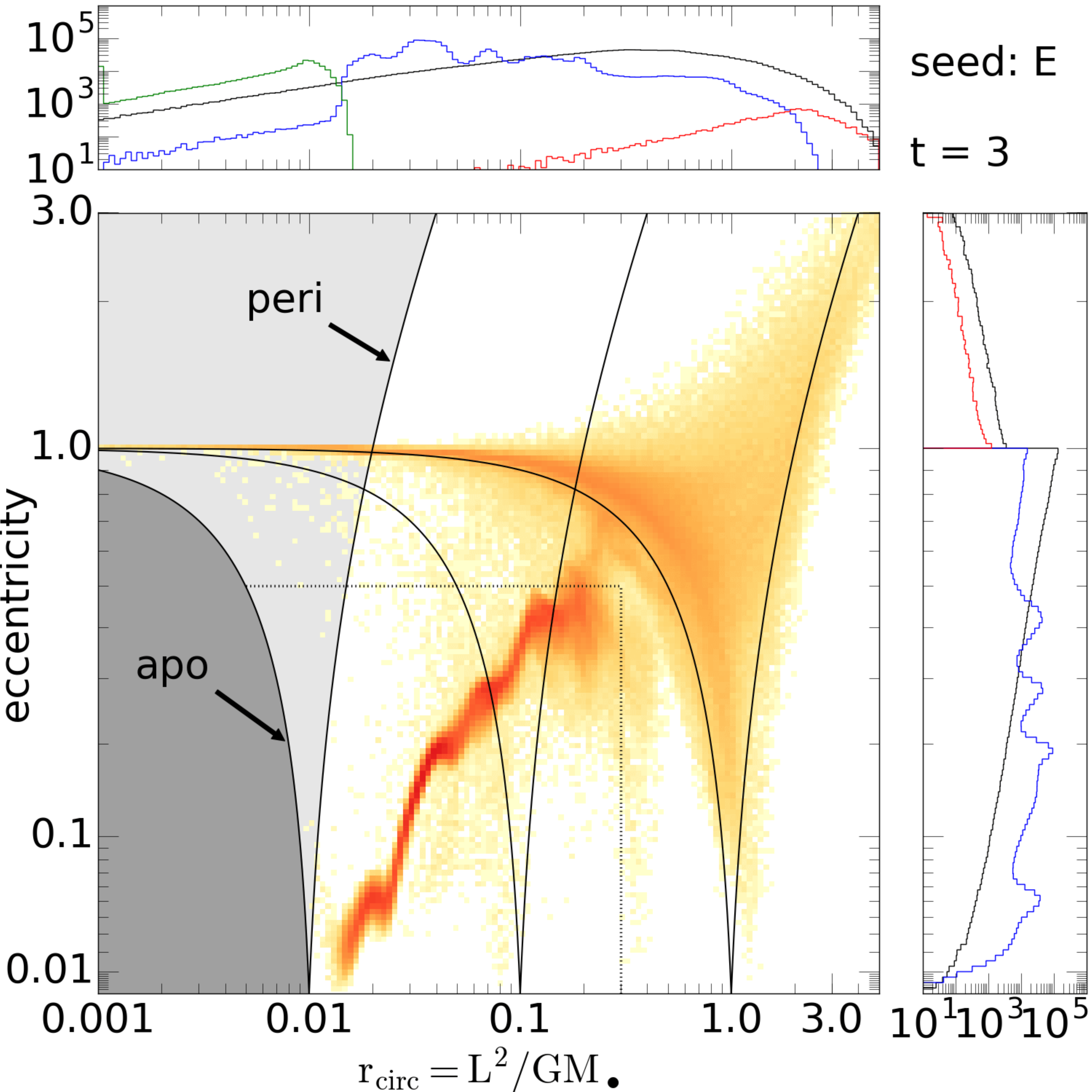}\hfil
	\includegraphics[width=50.5mm]{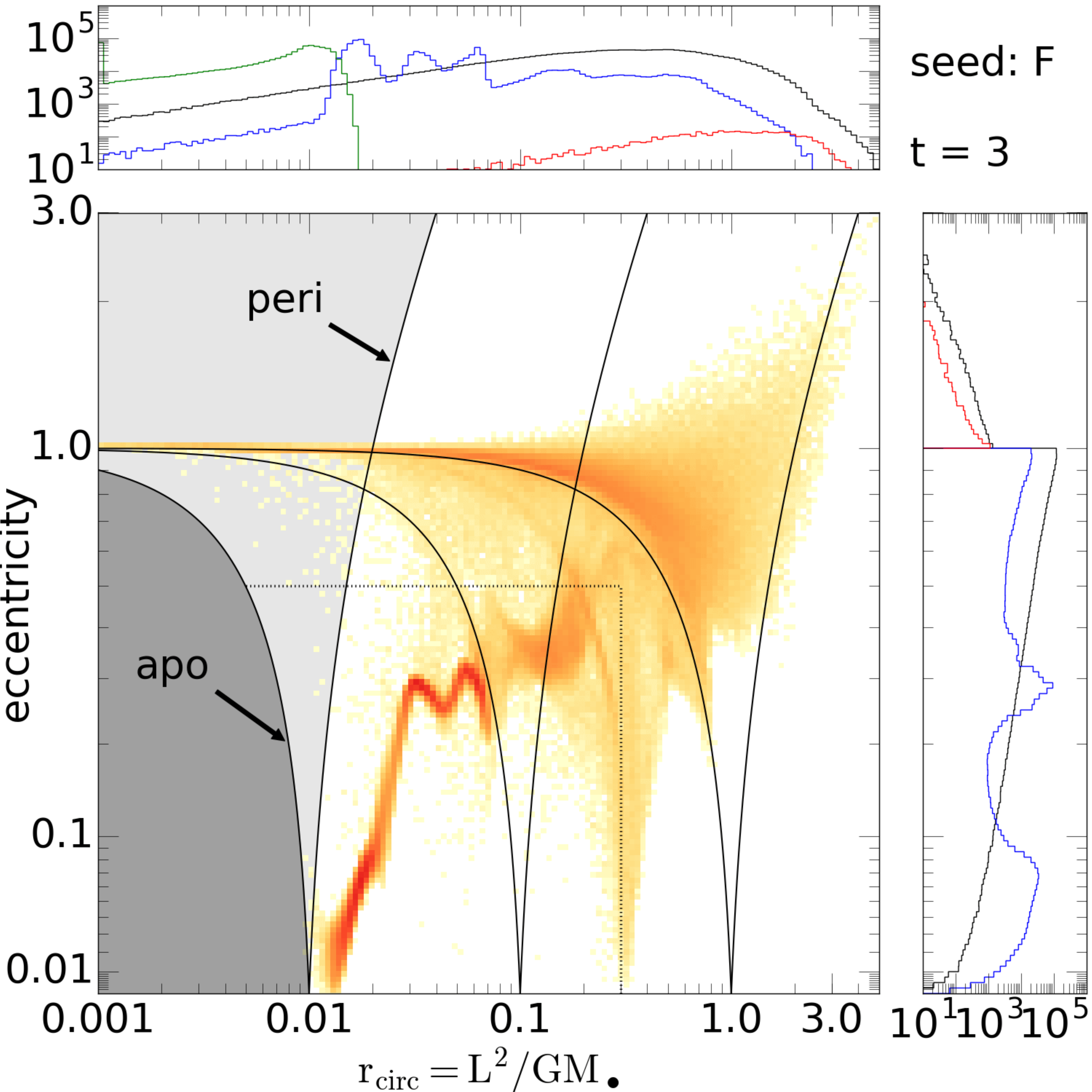}

	\caption{\label{fig:dist_ref-seedx}
	Distributions over $e$ and $\rcirc$ (as in Figure~\ref{fig:initial2Dhist_ref-seed633} or the left column of Figure~\ref{fig:combined_ref-seed633}) at the final simulation time $t=3$ for the sextet of reference simulations which differ only in the random components of their initial conditions (the same simulations for which Figure~\ref{fig:density_ref-seedx} shows snapshots at time $t=1$).
	While there is considerable variation of the disc structure and extent, the overall distributions of circularisation radii (top histograms) are similar.
	}

\end{figure*}

\subsection{Simulations differing only by the random seed}
\label{sec:ref-seeds}
The choice of the random number seed used to generate the turbulent velocity field does not affect the velocity power spectrum, but results in large local differences in the velocity and, consequently, in the emerging density distributions of the infalling gas as well. Therefore, any two simulations with different random seed may differ considerably in their details. In order to assess how much variation there is and which results are least variable, we conduct a suite of five additional simulations with the same parameters as those presented in the previous sub-section, but utilising different random seeds.

For the resulting simulation sextet, Figure~\ref{fig:density_ref-seedx} shows the density distributions near the $y=0$ plane at time $t=1$ (Figure~\ref{fig:combined_ref-seed633}, column 4). At this early time, only three of the six simulations (A, D, E) have formed a disc that is roughly aligned with the net angular momentum imposed on the initial conditions, while one (B) has a disc that is almost perpendicular to that orientation and two (C\&F) have hardly formed any well-defined disc structure at all. There is large variation in the tilt/warp and eccentricity of these discs, as well as in the filamentary structures. Over the course of these simulations, new infall can substantially disrupt any previously formed discs including re-orientation or the formation of nested, but mutually (strongly) inclined discs.

For the same six simulations, Figure~\ref{fig:dist_ref-seedx} shows the distributions over $e$ and $\rcirc$ at time $t=3$, the end of the simulations. All simulations have eventually formed some sort of disc corresponding to the structures at $e\lesssim0.5$ and $\rcirc\lesssim0.3$. Moreover, these discs all approach circularity ($e\lesssim0.02$) at their inner simulated edge at $\rcirc\sim0.01$. At $\rcirc\lesssim4 \sub{r}{sink}$, such circularisation is largely an artefact of the inner boundary condition, as simulations with smaller $\sub{r}{sink}$ demonstrate (see Section~\ref{sec:hstar}). However, the general trend that the disc is less eccentric at smaller radii, (also seen in the aforementioned simulations with smaller $\sub{r}{sink}$) and can be understood in terms of the faster evolution (shorter dynamical time) at smaller radii. Apart from this general trend, the details of the disc structures vary considerably between the six simulations including their sizes: typically the disc edge occurs at $\rcirc\sim0.1$, but shows a variation of a factor $\sim10$ (between $0.03$ and $0.3$ for simulations B and E, respectively).

\begin{figure*}
	\includegraphics[width=145mm]{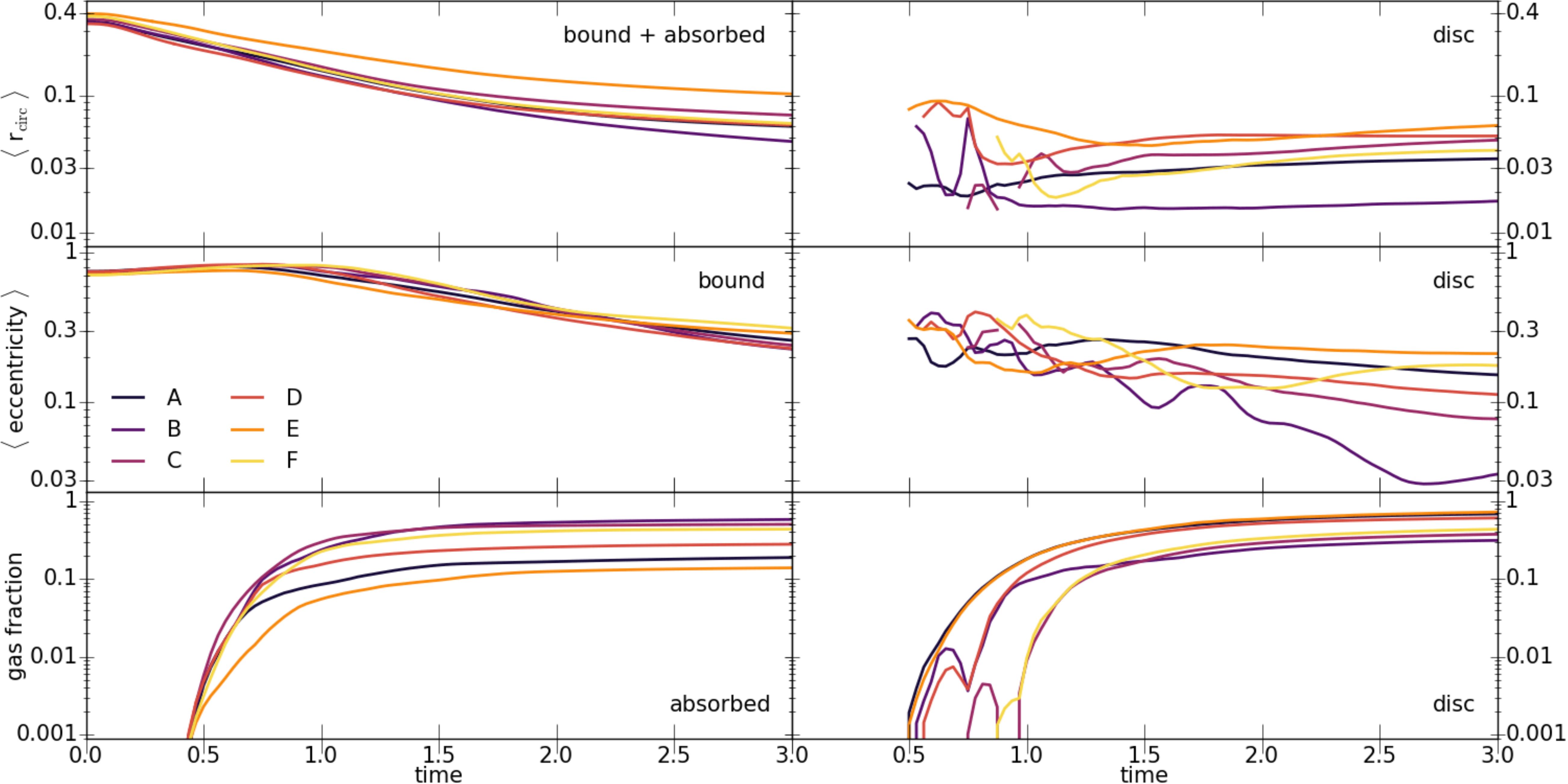}
	\caption{\label{fig:pathpar_ref}
	Time evolution of the average circularisation radius and eccentricity as well as the gas fraction for different portions of the gas and for each of the sextet of reference simulations already shown in Figs.~\ref{fig:density_ref-seedx} and ~\ref{fig:dist_ref-seedx}. `Absorbed' refers to simulated gas that at some point prior to $t$ has reached $r<\sub{r}{sink}=0.01$ and was absorbed into the central sink particle; 'bound' refers to all remaining gas that at the given time has $e<1$, 'disc' represents all bound gas at $e\le0.5$ and $\rcirc\le0.3$ indicated by a thin dotted rectangle in Figure~\ref{fig:initial2Dhist_ref-seed633} (which may also contain a small fraction of impacting filaments).
    }
\end{figure*}

Figure~\ref{fig:pathpar_ref} shows how the averages of $\rcirc$ and $e$ as well as the gas fractions evolves for certain gas components (see the figure caption for the precise definition of the categories `bound', `absorbed' and `disc'). Again, we can distinguish several phases of evolution. Before $t=0.5$ no gas has reached $r=\sub{r}{sink}$ and the formation of density filaments from the turbulent velocity field resulted in some reduction of the mean $\rcirc$ owing to angular-momentum cancellation between impacting gas. Between $t=0.5$ and $t=1$ infalling material is either directly absorbed (because it reached $r=\sub{r}{sink}$) or contributes to a forming disc. The early phases or disc formation may even involve the complete disruption of an earlier disc (simulations C \& F). After $t=1$, the disc formation consolidates, when further infalling filaments contribute to the growing disc, which prevents any significant further direct infall to $r<\sub{r}{sink}$.

The largest variations occur in the disc properties, whose mean eccentricity and circularisation radius varies by factors of $\sim5$, while the total amount of gas in the disc varies by a factor of 2-3 with less massive discs forming later. The amount of gas at $r<\sub{r}{sink}$ (particles absorbed into the sink), i.e.\ material that may ultimately reach the central SMBH varies by a factor $\sim4$.

\begin{figure*}
	\includegraphics[width=145mm]{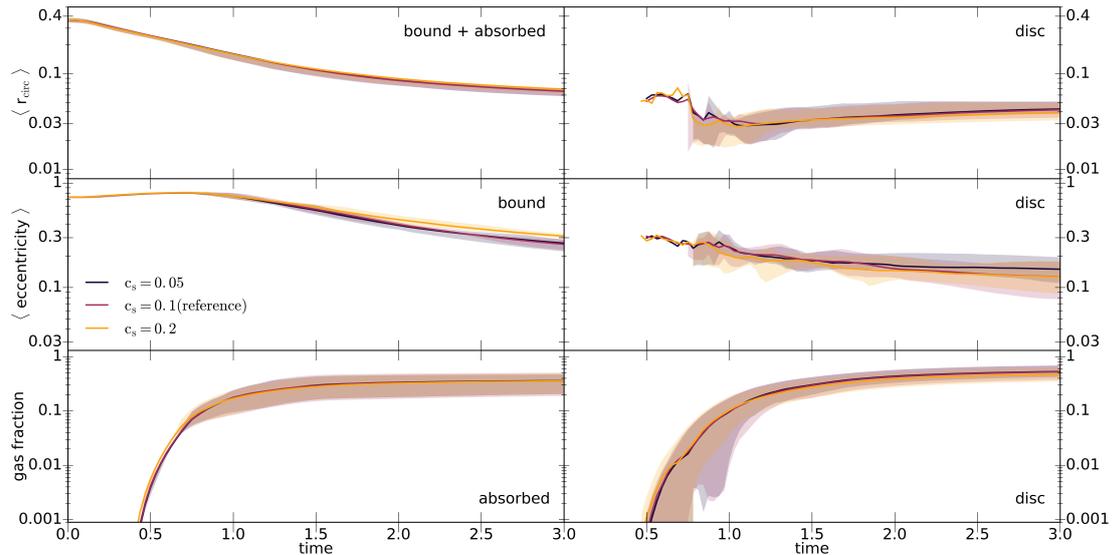}
	\caption{\label{fig:compare_cs}
	Similar to Figure~\ref{fig:pathpar_ref}, except that we compare three sextets of simulations, each with a different value of the gas temperature (or sound speed as indicated). Each sextet is indicated by a curve for its mean and a band covering the 17 and 83 percentiles (excluding the two most extreme simulations).}
\end{figure*}

\section{Effects of varying the parameters}
\label{sec:parasweep}
In this section, we vary (usually) one of the parameters of the simulations, but keep the initial conditions otherwise identical (as much as possible) to those used for the six simulations presented in the previous section. In this way, the effect of the parameter considered can be isolated in the clearest way, while at the same time ensuring that we capture any variation across the different random realisations of the same physics.

\begin{figure*}
	\includegraphics[width=145mm]{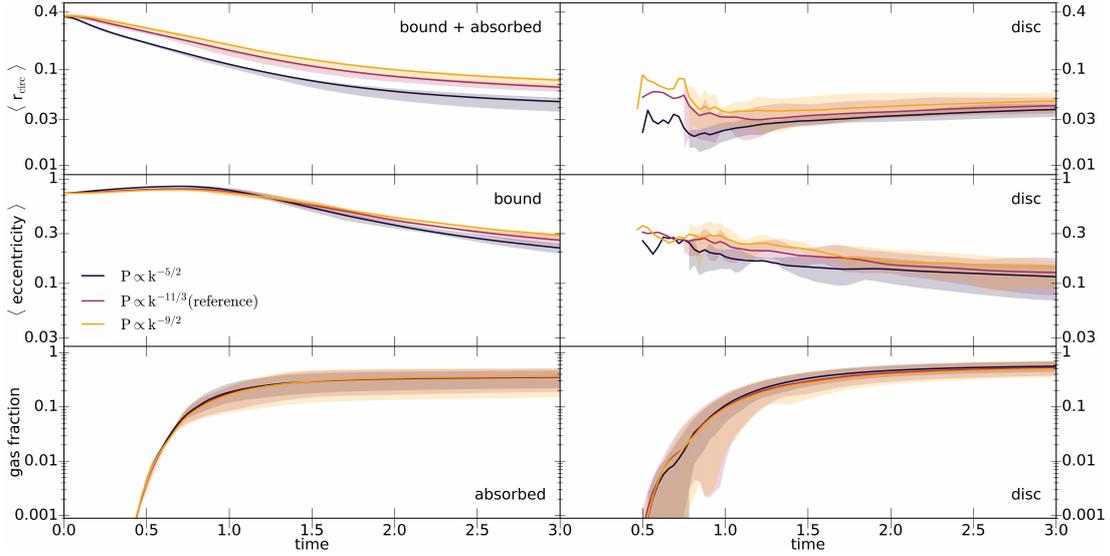}
	\caption{\label{fig:compare_pwr}
	As Figure~\ref{fig:compare_cs}, but for simulations with different power spectrum of the turbulent velocity field.}
\end{figure*}

\subsection{Sound speed}
\label{sec:cs}
The assumed overall gas temperature or, equivalently sound speed, influences the nature of the filaments: small $c_s$ (low temperature) leads to denser filaments with smaller cross sections for collision, which may reduce the efficiency of angular-momentum cancellations. Large $c_s$ (higher temperature) smooths out small-scale modes of the turbulent shell and results in a lower density in the filaments. Additionally the disc structure is directly influenced by the value of $c_s$ through the conditions for a vertical hydrostatic equilibrium with smaller values resulting in a thinner discs.

The value of the sound speed used for the reference simulation is $0.1$, i.e.\ a tenth of the circular speed at $r=\sub{r}{shell}$. Two suites of simulations with $c_s=0.05$ and $0.2$ with the latter corresponding to an increase in temperature by a factor $4$ are summarised in Figure~\ref{fig:compare_cs}.

Simulations $c_s=0.05$ behave very similarly to the reference simulations. However, the slightly denser infalling filaments are more likely to form rings, which can be up to 10 times denser than the more uniform disc structure in the reference simulations. The disc is on average more eccentric than for larger $c_s$ and the latest infall remains often visible as a ring separated by a low-density gap from the original disc.

Simulations with $c_s=0.2$ form wider and less dense filaments, which have a higher collision cross section and form a disc earlier than the reference simulation. A larger pressure is likely to better erase the small-scale fluctuations of the initial velocity field and, consequently, simulations with $c_s=0.2$ share some characteristics with simulations starting from velocities with a steeper power spectrum ($P\propto k^{-9/2}$ in Figure~\ref{fig:compare_pwr}): larger mean eccentricities and circularisation radii in the bound gas and fewer material in the disc.

In general, however, there is very little variation in the gross statistical properties between simulations the differ by a factor 4 in the sound speed (or a factor 16 in temperature).

\subsection{Power spectrum of the turbulent velocity field}
\label{sec:pwr}
Changing the slope of the power spectrum has a profound impact on the evolution of the simulation due to its influence on the distribution of the energy in the initial turbulent velocity field. For the reference simulation $P\propto k^{-11/3}$, bracketed by the two additional sets of simulations with $P\propto k^{-5/2}$ (shallower) and $P\propto k^{-9/2}$ (steeper), respectively, the results of these simulations are summarised in Figure~\ref{fig:compare_pwr}.

For $P\propto k^{-5/2}$ there are more small-scale fluctuations in the emerging density field shortly after the start of the simulations, when compared to the reference simulations. This is reflected in more substructure in the filaments, while the  maximal density remains similar. The main effect in terms of the average properties of the resulting gas flows is a stronger reduction of the average angular momentum, or $\rcirc$, in the early phases. Conversely, for $P\propto k^{-9/2}$ fewer, but more pronounced large scale filaments emerge at the start of the simulations and the average $\rcirc$ declines less; again with the main difference being in the early phase (before $t=1$).

There is no significant difference in the rate of gas inflow onto the inner simulation boundary (particle absorbed onto the sink particle) nor in the mass and formation rate of the gas disc. However, there is a difference in the structure of the disc: with more power on smaller scales the disc forms at slightly smaller radii. Furthermore it is slightly less eccentric, but this may simply be a consequence of the faster circularisation rate at smaller radii.

\subsection{Balance between rotation and turbulence}
\label{sec:ratio}
\begin{figure*}
	\includegraphics[width=145mm]{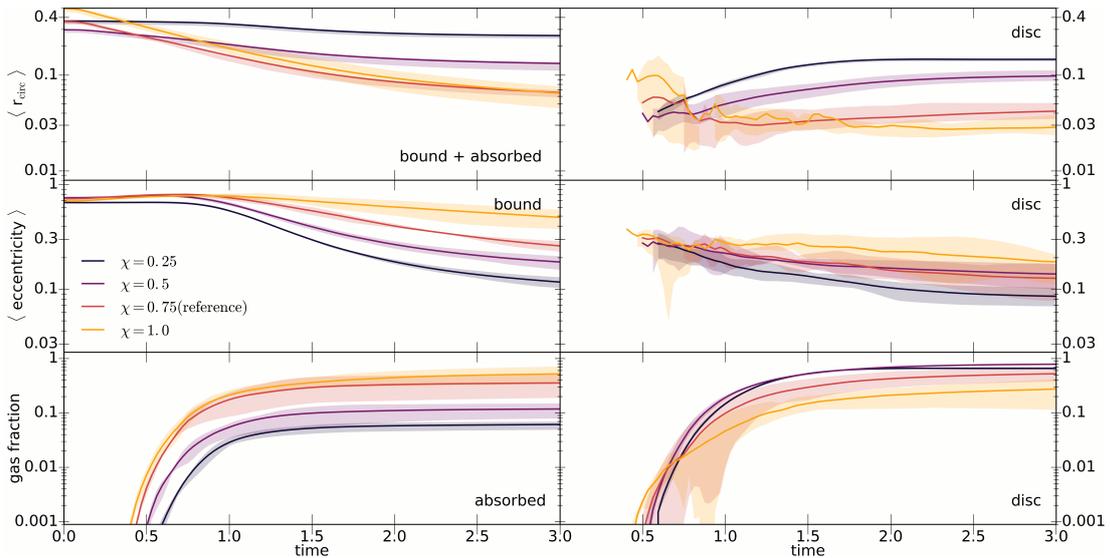}
	\caption{\label{fig:compare_ratio}
	As Figure~\ref{fig:compare_cs}, but for simulations with different balance between rotation and turbulence in the initial velocity field according to equation~\eqref{eq:v:initial}. For $\chi=1$ no rotation has been added to the turbulent velocities, but any random rotational component of the latter has not been removed.}
\end{figure*}
The parameter $\chi$ determines the relative contributions of the turbulent velocities and solid-body rotation to the initial velocity field according to equation~\eqref{eq:v:initial}. For $\chi=0$ the velocities field only contains solid-body rotation and no random (turbulent) component resulting in rather unrealistic situation, while for $\chi=1$ there is no rotational component apart from the residual rotation of the random velocity field. For four values $\chi>0$, Figure~\ref{fig:compare_ratio} summarises the time evolution of the mean properties for the emerging gas flows\footnote{The initial mean circularisation radius is minimal near $\chi=0.5$, when the rotational and turbulent velocities are comparable. A simple analytic estimate for our initial model suggests $\langle\rcirc\rangle\approx\tfrac{2}{3}\eta(\chi^2+[1-\chi]^2\eta)\sub{r}{shell}$, which shares this property.}. There are clear trends with $\chi$: the more rotational supported the initial velocity fields, the larger, more massive, and more circular the forming gas discs\footnote{It appears from Figure~\ref{fig:compare_ratio} that for $\chi=0.25$ the discs stop growing at $t=2$, but this is an artefact of our definition of the disc region in ($e$, $\rcirc$), which excludes disc material at $\rcirc>0.3$, when in fact the discs for these simulations grow larger.} and the less gas is `absorbed' into the sink particle.

Of particular interest are the simulations with $\chi=1$, when the only rotational component of the initial velocity field is the small residual rotation of the random velocity component. The six simulations for this choice of $\chi$ show large variety in their disc properties, often including the complete destruction of an early disc by later infall. However, by the end all six of these simulations have formed some gas disc (with random orientation in line with the expectation of stochastic accretion described in Section \ref{sec:introduction}), often exhibiting large eccentricities and warps or gaps (rings). These discs are typically much smaller than those formed in simulations with $\chi\le0.75$, but still significantly larger than the (artificial) sink absorption radius of $0.01$. When reducing this radius to half, i.e.\ $\sub{r}{sink}=0.005$, we found no significant change in the evolution, in particular the outer structure of these small discs remains hardly affected, although the inner parts are, of course, is altered by the change of $\sub{r}{sink}$.

\subsection{Solenoidal velocity field}
\label{sec:sol}
\begin{figure*}
	\includegraphics[width=145mm]{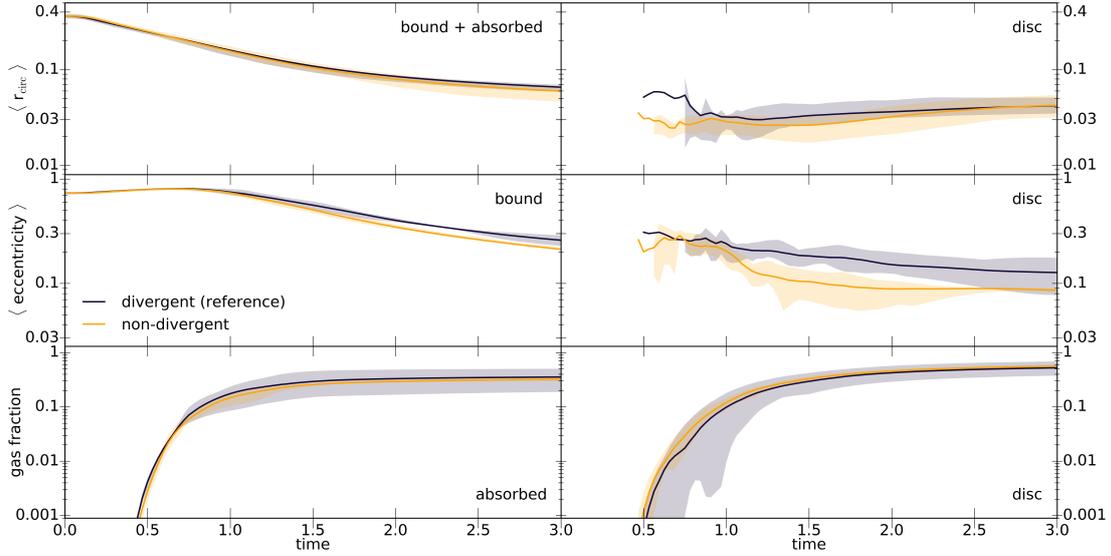}
	\caption{\label{fig:compare_sol}
	As Figure~\ref{fig:compare_cs}, but for simulations with initial velocities satisfying $\vec{\nabla}\cdot\vec{v}=0$ (neither diverging nor converging).}
\end{figure*}
In Figure~\ref{fig:compare_sol} we summarise the results of six simulations, which differ from the reference simulations only by the choice of the velocity field such that $\vec{\nabla}\cdot\vec{v}=0$ everywhere. It is important to note that our method for creating such initial conditions (see Appendix~\ref{sec:turb}) produces a different field instead of transforming a general field into a divergent-free one. Therefore the same seed numbers will produce a different asymmetric initial condition, such that a one-to-one comparison with the simulations in the reference set is not sensible, but only a comparison between either set of simulations in a statistical sense.

The overall evolution of the system is very similar to the set of reference simulations up to $t\sim1$. Thereafter, the discs formed from divergent-free initial conditions are initially slightly smaller (but eventually equally large) and distinctly less eccentric, but equally massive than for the reference case.

\begin{figure*}
	\includegraphics[width=145mm]{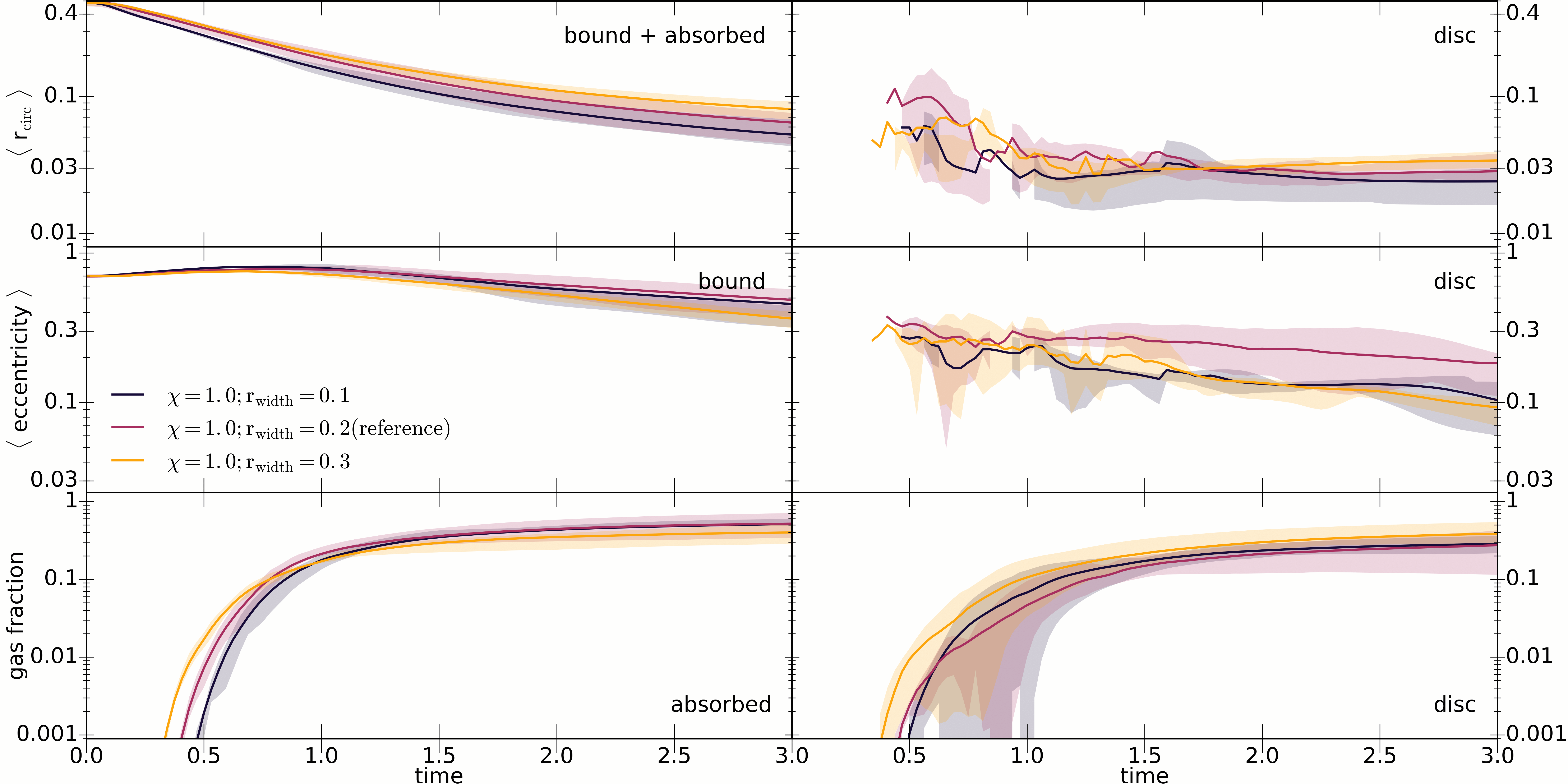}
	\caption{\label{fig:compare_width}
	As Figure~\ref{fig:compare_cs}, but for simulations with different width $\sub{r}{width}$ of the initial shell of gas.}
\end{figure*}

\subsection{Width of the initial gas shell}\label{sec:addpar:width}
The parameter $\sub{r}{width}$ controls the Gaussian width of the initial shell and defaults to $0.2\sub{r}{shell}$. We have also run two sets of simulations for $\sub{r}{width}=0.1$ and $0.3$, see Fig.~\ref{fig:compare_width}. A smaller (larger) width implies a smaller (larger) spread of peri-centre arrival timings across the shell. This in turn increases (decreases) the chance of collisions and hence angular-momentum cancellation and reduction of $\rcirc$. We find indeed that the final averaged $\rcirc$ for $\sub{r}{width}=0.1$ is almost twice as large as for $\sub{r}{width}=0.3$, which  between them bracket the result for our default simulations with $\sub{r}{width}=0.2$.

\begin{figure*}
	\includegraphics[width=58.5mm]{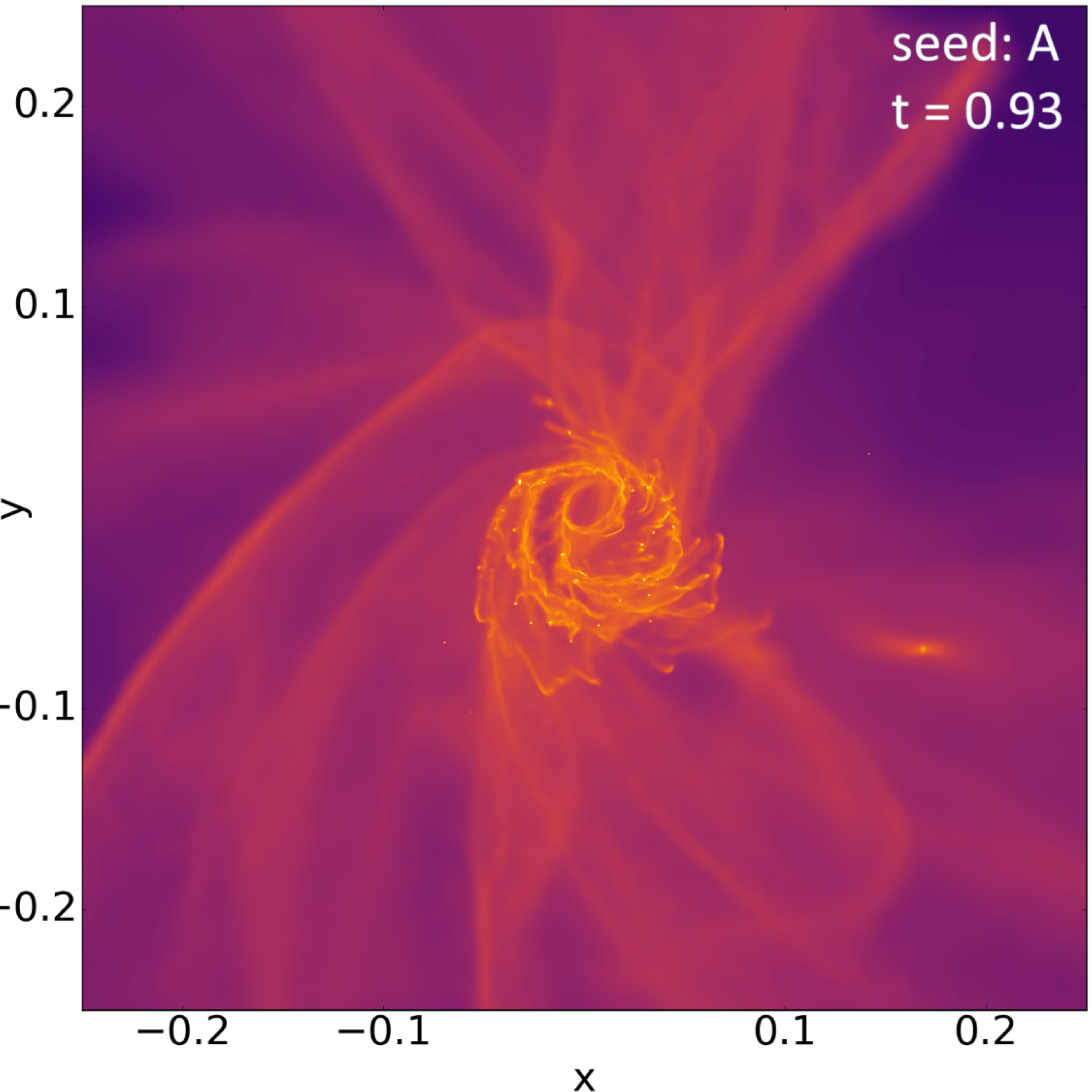}
	\includegraphics[width=58.5mm]{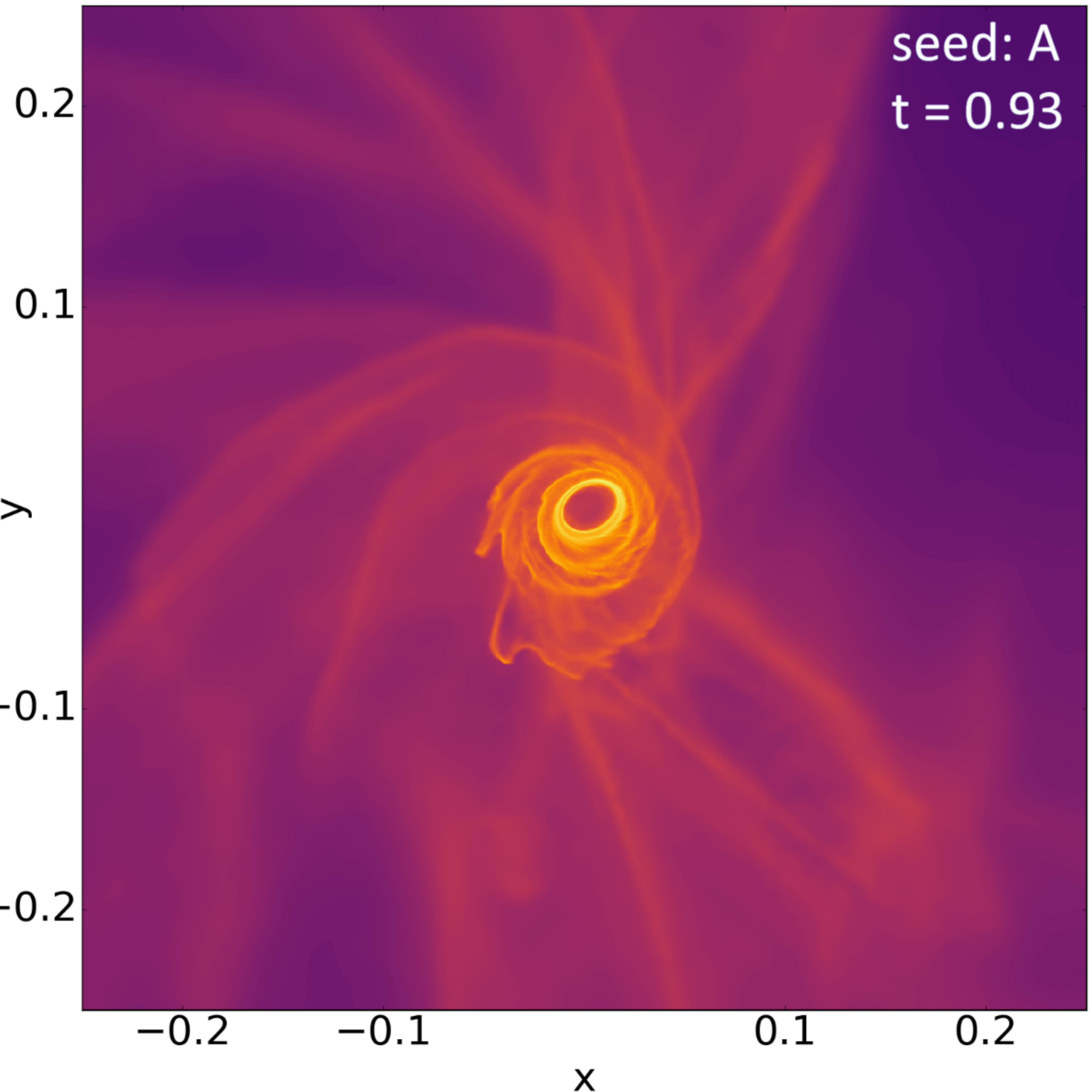}
	\includegraphics[width=58.5mm]{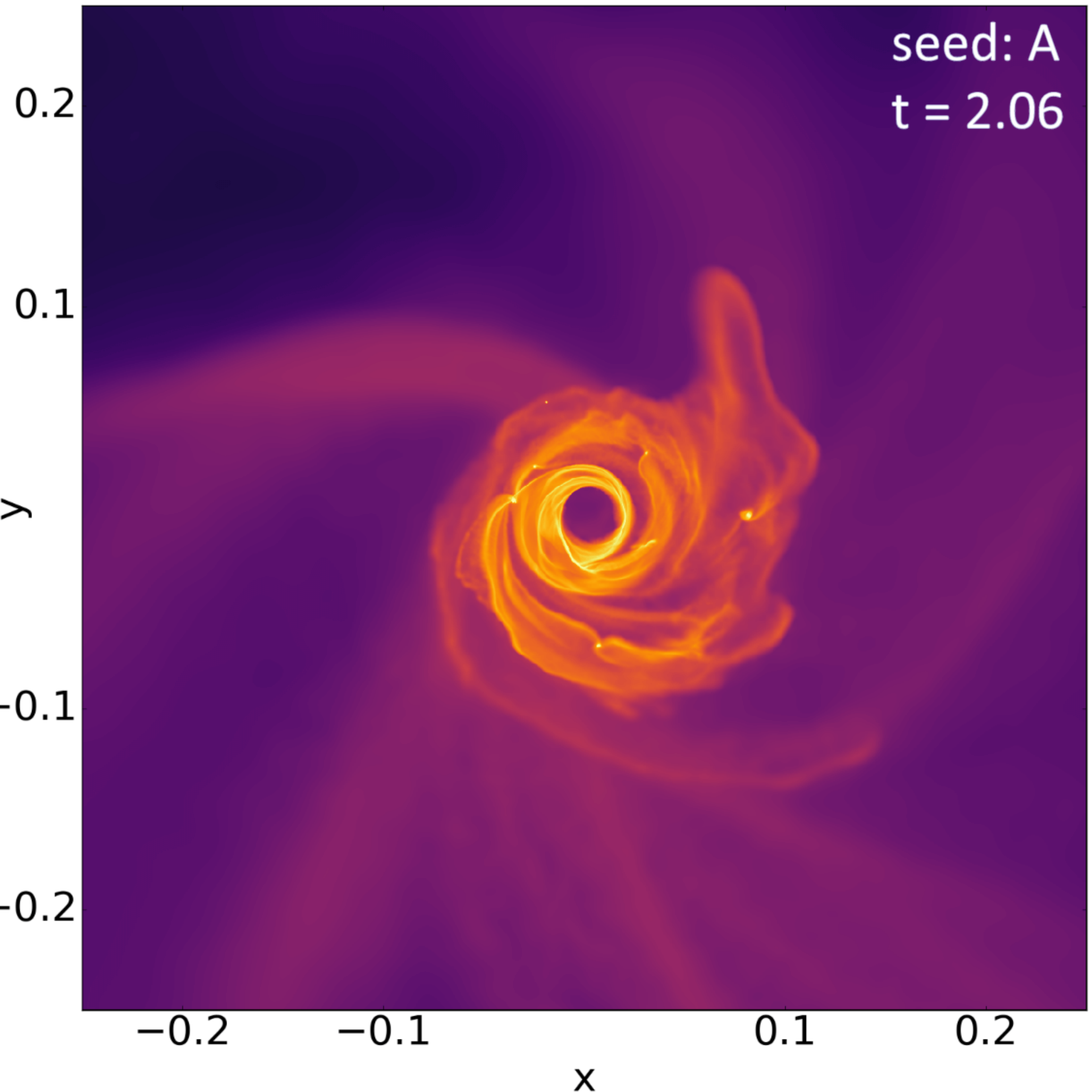}
	\caption{\label{fig:xz_grav}
	Density plots, similar to Figure~\ref{fig:density_ref-seedx}, for simulations starting from the same initial conditions as those in Fig.~\ref{fig:combined_ref-seed633}, but with gas self-gravity and $\sub{M}{shell}=1$ at $t=0.93$ (\emph{left}) or $\sub{M}{shell}=0.1$ at $t=0.93$ (\emph{middle}) and $t=2.06$ (\emph{right}).}
\end{figure*}

\subsection{Mass of the initial gas shell}
\label{sec:addpar:mass}
In all simulations presented so far, the total amount of gas equals 1\% of that of the SMBH, $\sub{M}{shell}=0.01M_\bullet$. We also considered the cases of $\sub{M}{shell}/M_\bullet=0.1$ and 1. The resulting simulated gas flows (not shown) behave very similar to the reference simulations. One difference is a slightly higher fraction of gas `absorbed' onto the sink particle, i.e. removed from the simulation, because it reached $r<\sub{r}{sink}=0.01\sub{r}{shell}$. This can be explained by the increased amplitude of the sink particle's random walk (`Brownian motion') owed to the relatively larger momentum it absorbs due to the increased mass of the flow\footnote{The initial random gas velocities contain some small centre-of-mass motion with respect to the hole. When removing this small momentum, simulations show hardly differences.}. For high $\sub{M}{shell}$ the picture is quite different, if the gas self-gravity is accounted for (see Section~\ref{sec:addpar:grav}).

\subsection{The initial kinetic energy of the gas}
\label{sec:addpar:eta}
The parameters $\sub{\eta}{rot}$ and $\sub{\eta}{turb}$ control the relative amount of kinetic energy and hence the deviation of the initial conditions from virial equilibrium. In all simulations so far, both were set equal to $\eta=0.9$, when the overall virial ratio for the reference simulations is $\approx0.74$ (according to equation~\ref{eq:virial:ratio}). Here, we report simulations with $\eta=0.5$ and 1.1, when the virial ratio becomes $\approx0.35$ and 0.97, respectively. 

This change in the initial velocity amplitudes directly affects the initial distribution of angular momenta such that the average initial $\rcirc$ is roughly proportional to $\eta$. With smaller velocities, the gas streams are on average more plunging and collide at smaller radii where collisions are more likely (due to the smaller volume). This results in more cancellation of angular momenta and hence reduction of $\rcirc$ for simulations with smaller $\eta$. The size and mass of any gas discs formed in the later stages of the simulations increase with the initial velocity amplitude, as expected.

Arguably, our simulations with $\eta=0.5$ are somewhat under resolved and would have benefited from a smaller value for $\sub{r}{sink}$, the radius at which particles are absorbed into the central sink particle. With the value $\sub{r}{sink}=0.01$ used, these simulations struggle to form continuously existing discs as most of the inflowing material is lost to absorption onto the sink particle (about twice as much as for the reference simulations).

\subsection{Gas self-gravity}
\label{sec:addpar:grav}
We ran several simulations with self-gravity of the gas particles turned on. In this case, we limit the Jeans mass (such that it is always resolved) by limiting the gas softening lengths to $\epsilon\ge 10^{-3}$ (while otherwise $\epsilon_i$ is proportional to the SPH smoothing length). We experimented with self-gravity for the default gas mass $\sub{M}{shell}/M_\bullet=0.01$, but also for the larger values of $0.1$ and $1$. 

The inclusion of the gas self-gravity has little effect on the gas inflow in the early phases of the simulations prior to $t\sim\sub{t}{ff}$, in particular the re-distribution of angular momenta (as reflected in the distribution of $\rcirc$) and absorption onto the sink particle. The main effect at later stages is the formation of dense clumps within the gas discs, where the gas is densest and most prone to the Jeans instability, see Figure~\ref{fig:xz_grav}. In particular, in simulations with $\sub{M}{shell}=M_\bullet$ a multitude of such clumps form, each surrounded by its own mini-disc, rendering the smooth disc seen without self-gravity into a rather messy arrangement. For $\sub{M}{shell}/M_\bullet=0.1$, the effect is much milder and virtually absent at the reference value $\sub{M}{shell}/M_\bullet=0.01$.

\begin{figure*}
	\includegraphics[width=43.3mm]{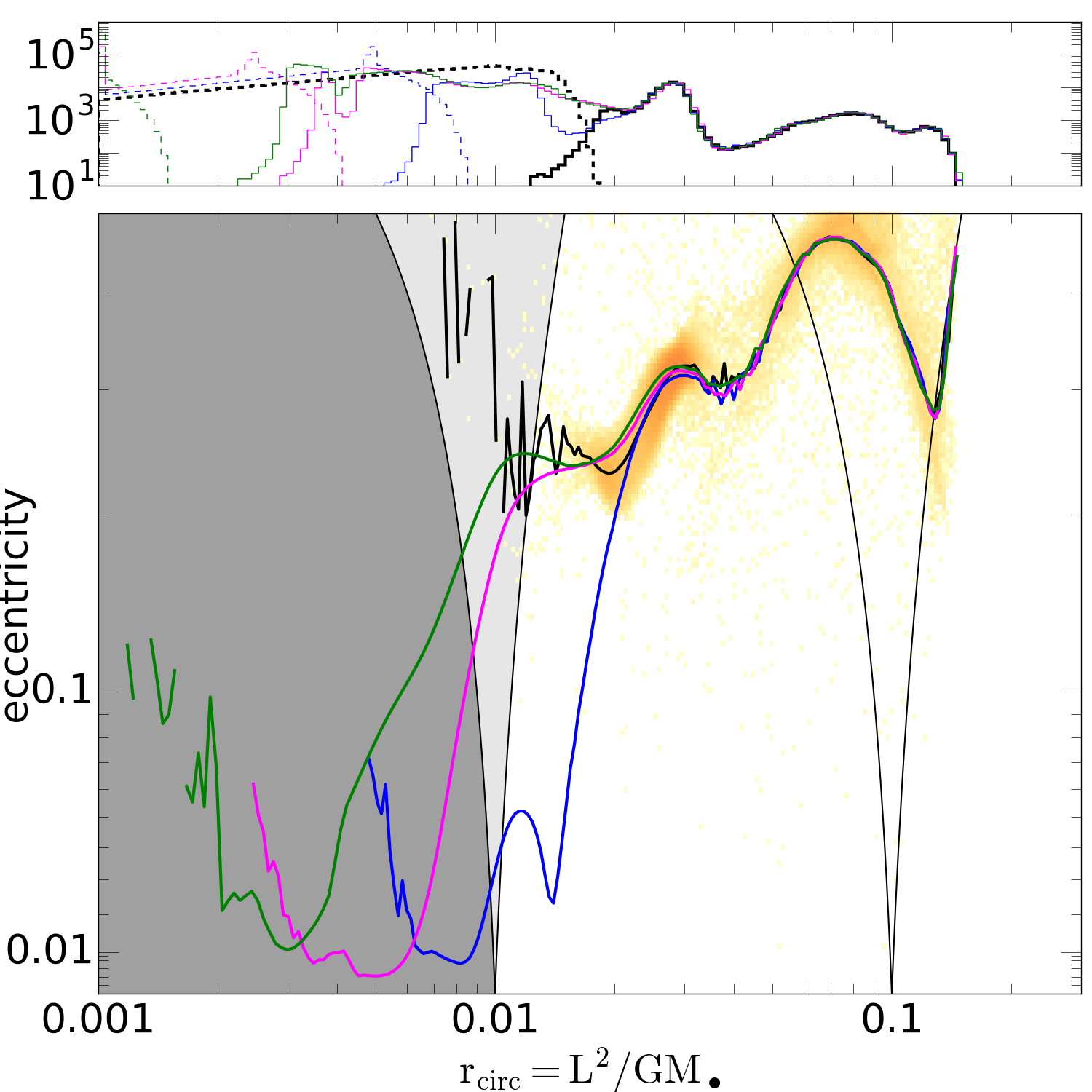}\hfil
	\includegraphics[width=43.3mm]{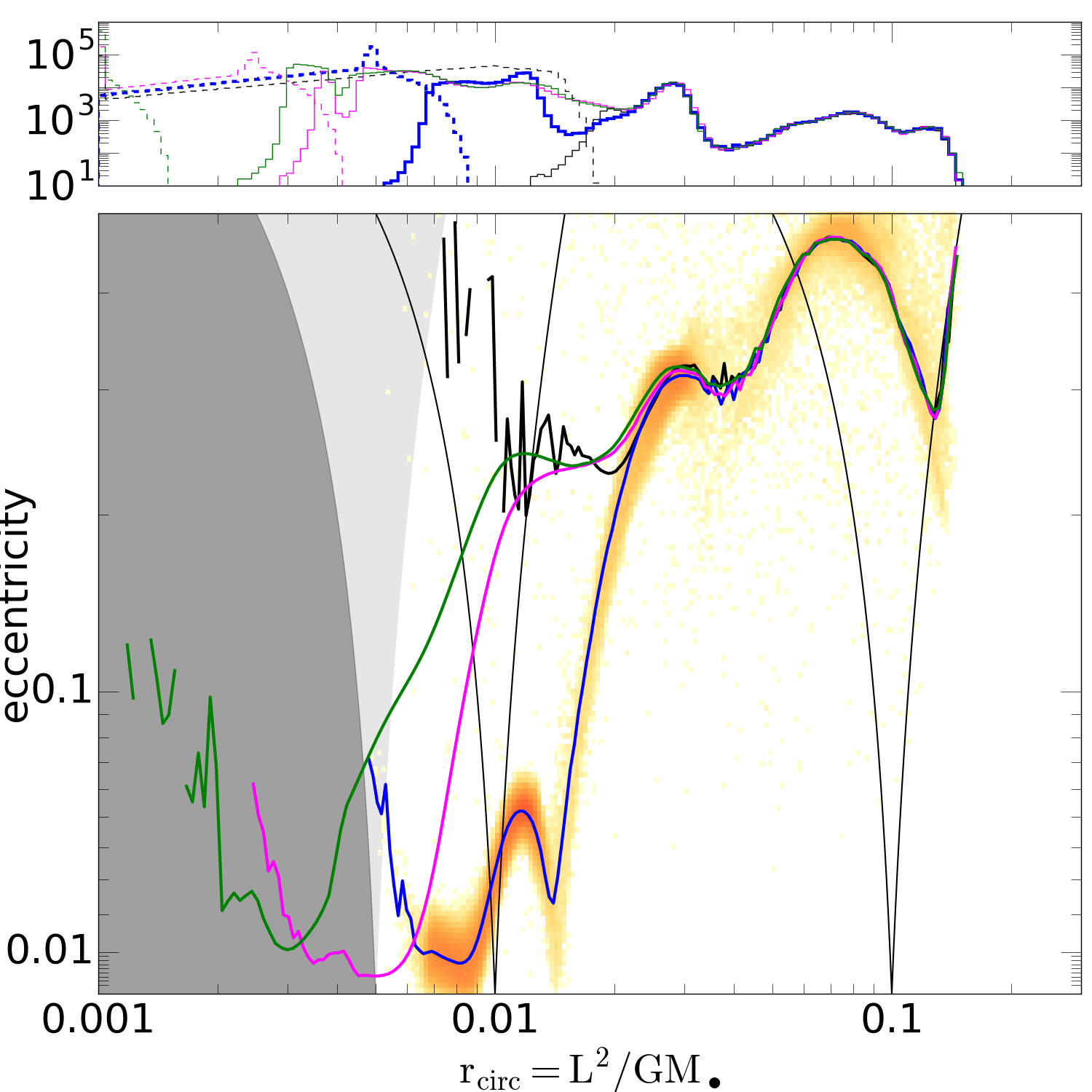}\hfil
	\includegraphics[width=43.3mm]{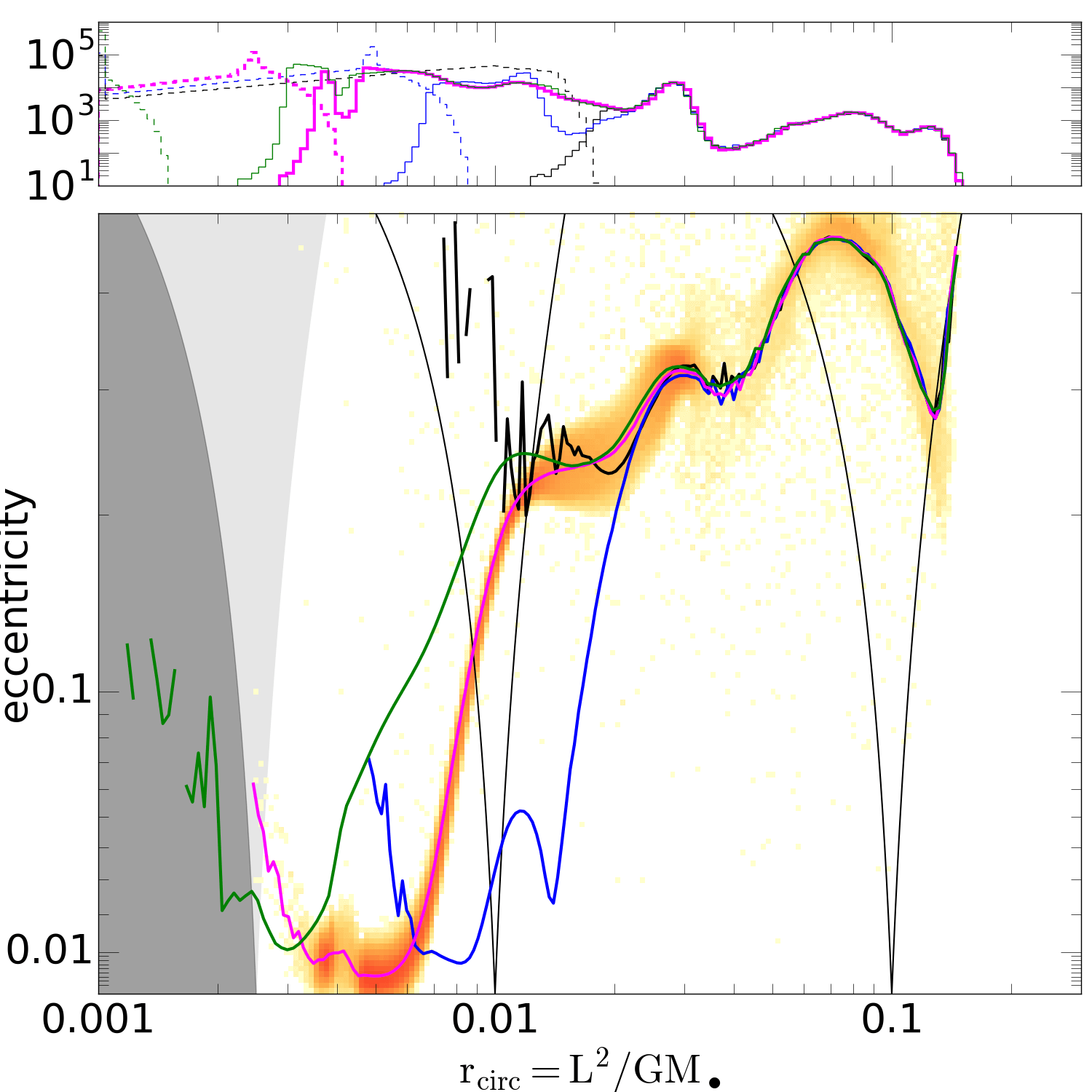}\hfil
	\includegraphics[width=43.3mm]{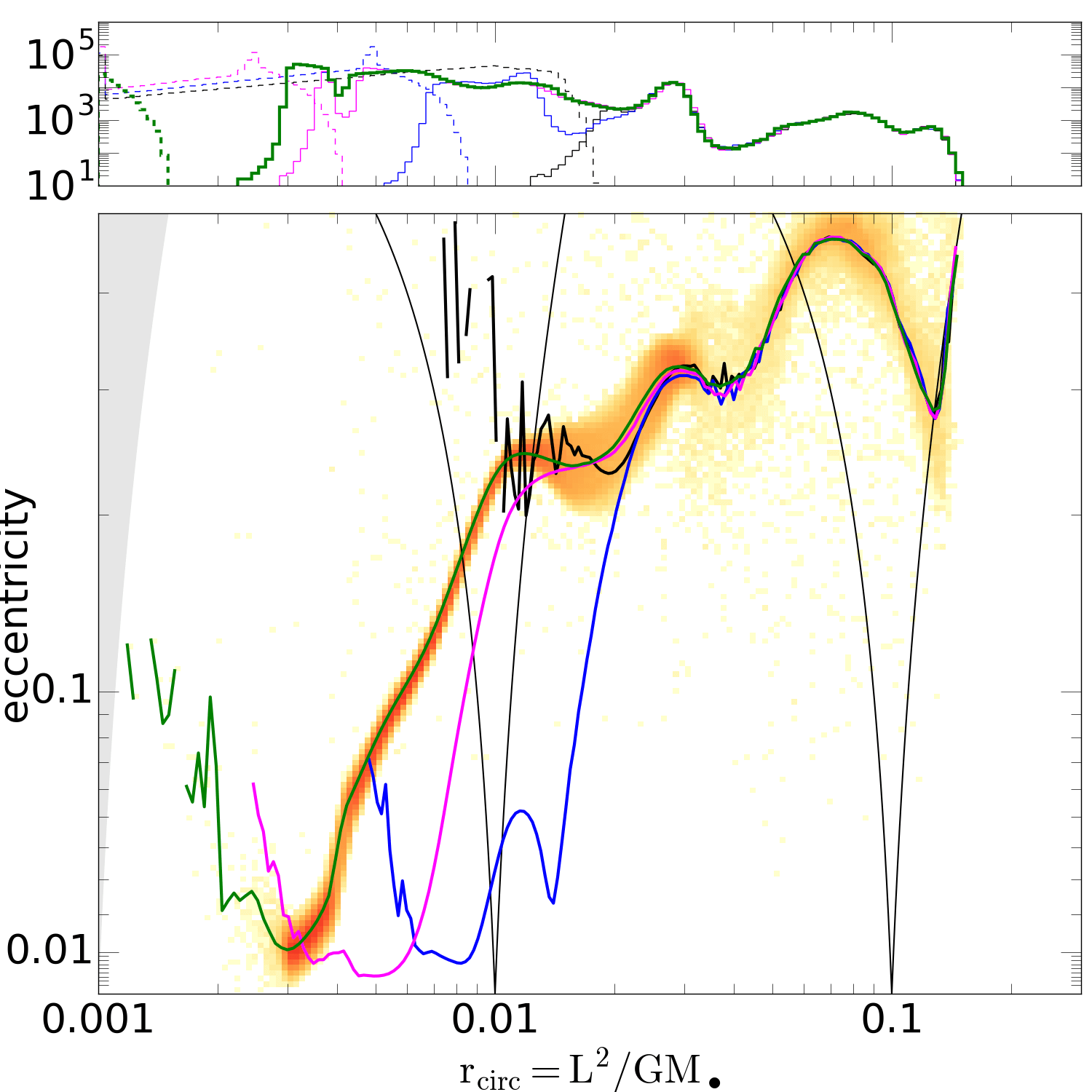}

	\caption{\label{fig:scatter_hstar}
	The distributions over $\rcirc$ and $e$ in the disc region for four simulations which differ only in the choice for the inner boundary radius: $\sub{r}{sink}=0.01$, 0.005, 0.0025, and 0.001 from left to right. For better comparison, the thin curves indicate the mean eccentricity within the disc regions of these four simulations. Similarly, the top panels show the $\rcirc$ distributions for all four simulations.
	}
\end{figure*}

\begin{figure*}
	\includegraphics[width=43.3mm]{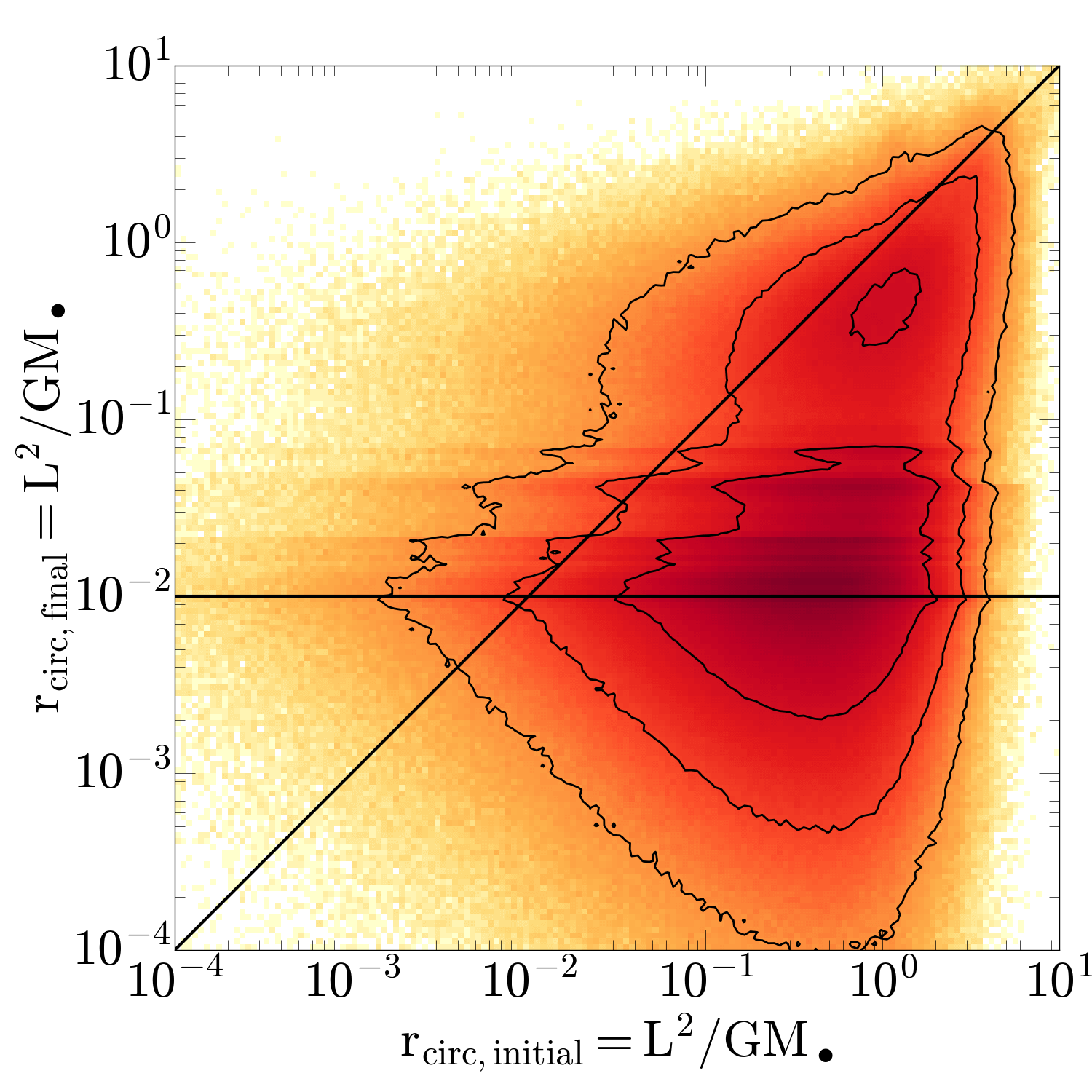}\hfil
	\includegraphics[width=43.3mm]{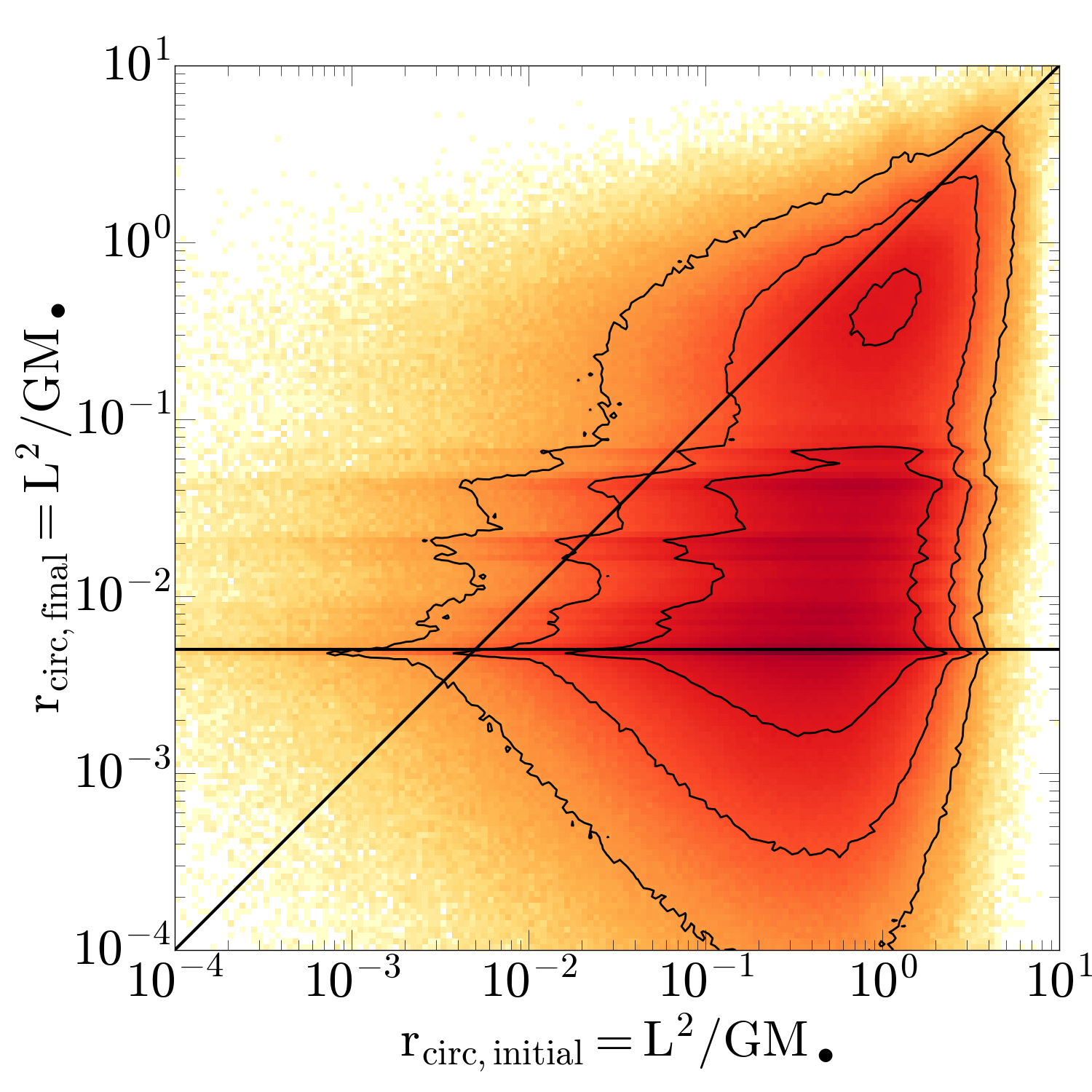}\hfil
	\includegraphics[width=43.3mm]{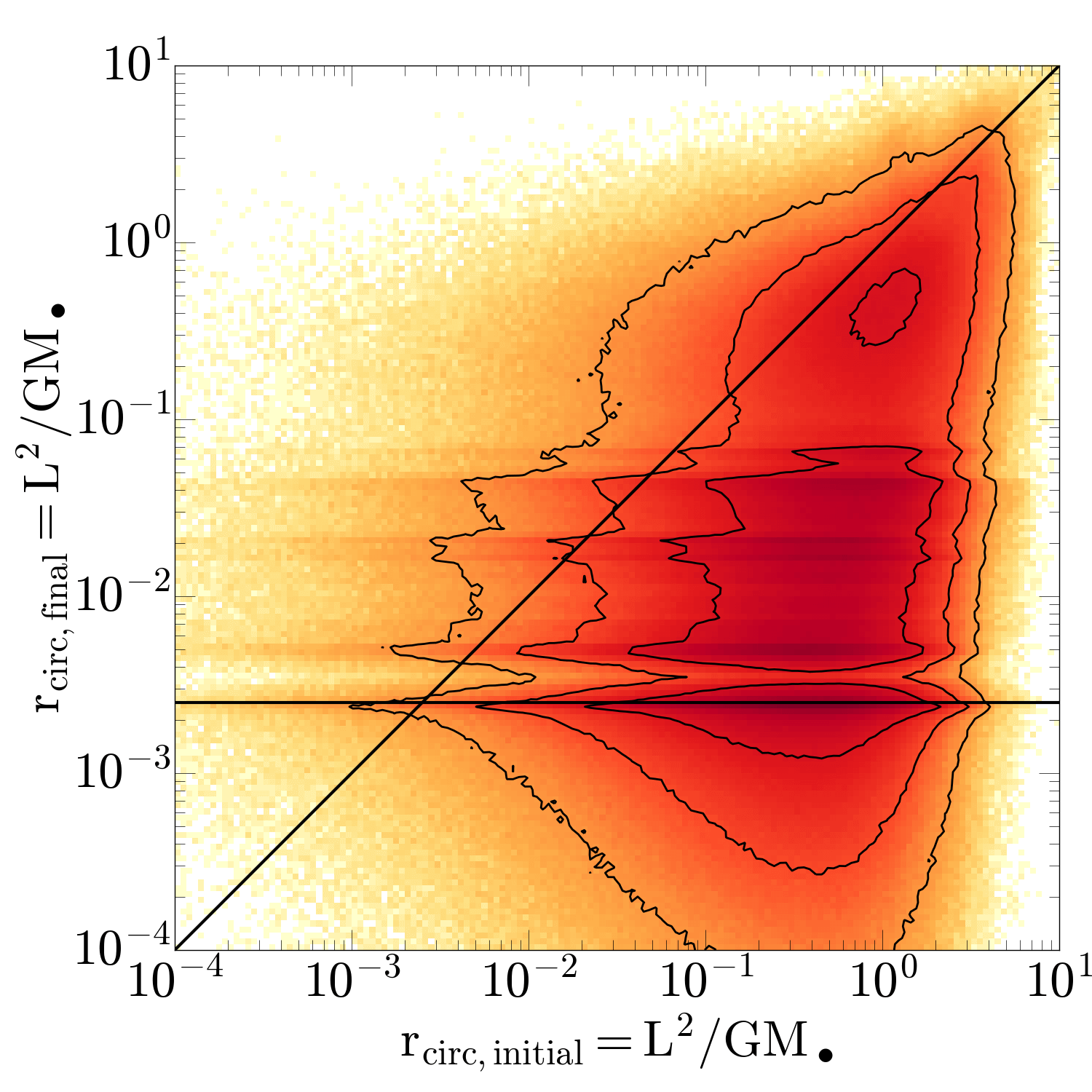}\hfil
	\includegraphics[width=43.3mm]{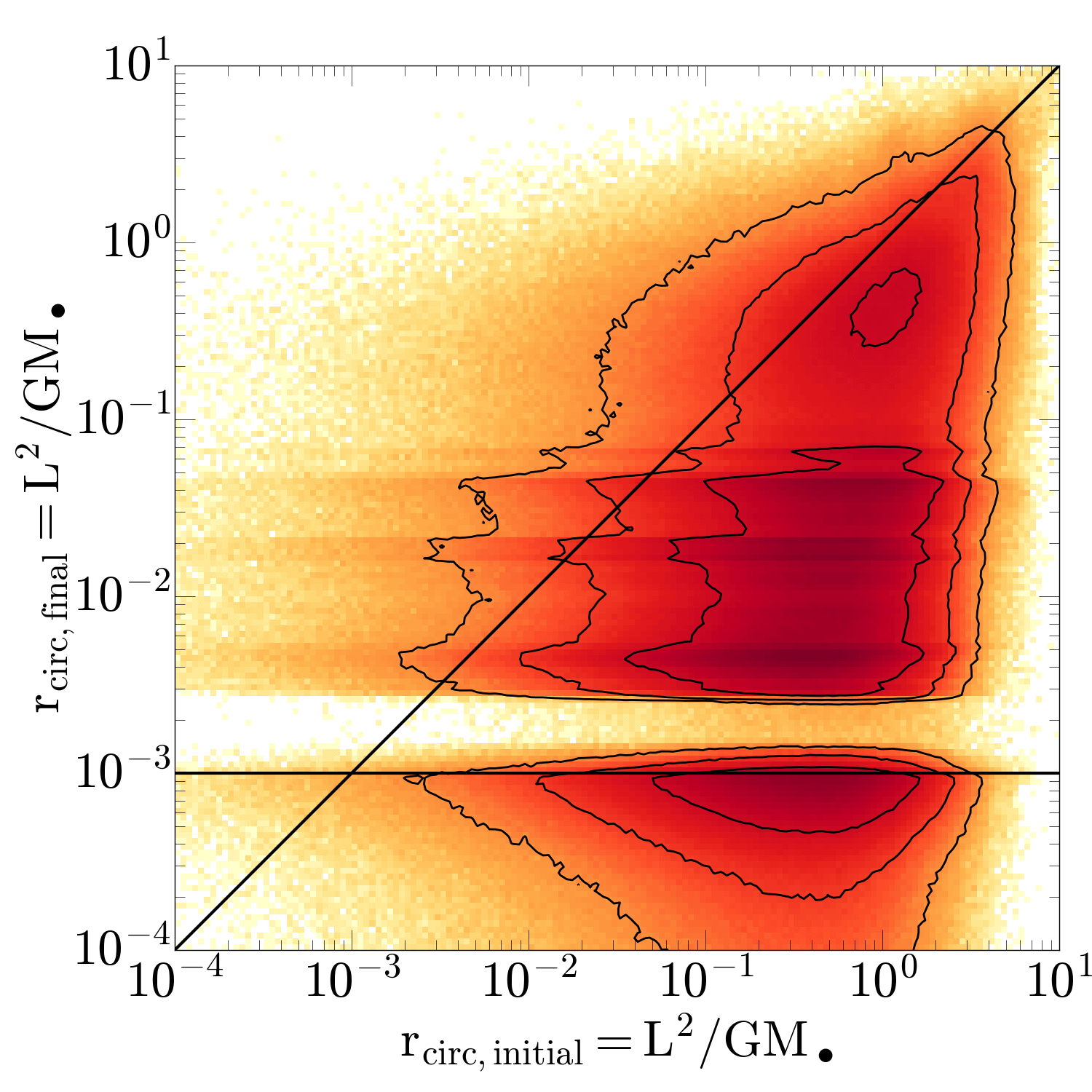}

	\caption{\label{fig:dist_rcirc_init_final}
	The distributions over initial and final $\rcirc$ for four sets of simulations which differ only in the choice for the inner boundary radius $\sub{r}{sink}$ as indicated by the horizontal lines. The horizontal features are gas rings formed in the later stages of the simulations\addtocounter{footnote}{1}\protect\footnotemark\addtocounter{footnote}{-2}.}
\end{figure*}

\subsection{The inner boundary and convergence}
\label{sec:hstar}
All our simulations must employ an inner boundary condition to avoid simulating the gas flow very close to the SMBH, where the dynamical time scales are too short for efficient modelling. We have implemented this in the usual way by a sink radius $\sub{r}{sink}$ around the SMBH particle, such that any gas particle found at distance $r<\sub{r}{sink}$ from the SMBH is `absorbed' into the latter and removed from the simulation. Of course, this is somewhat artificial and unphysical. 

The main adverse effect of this boundary is that gas particles which orbit the SMBH on an elliptic orbit with apo-centre outside of $\sub{r}{sink}$ but peri-centre inside are removed and thus prevented from interacting with and affecting the gas flow further out\footnote{Even with sophisticated criteria for particle absorption, this problem cannot be avoided, only alleviated at higher computational costs, which is equivalent to reducing $\sub{r}{sink}$.}. 
\addtocounter{footnote}{1}
	
In order to assess the effect of the inner boundary more quantitatively, we ran three additional sets of six simulations each with $\chi=1$ and $\sub{r}{sink}=0.005$, 0.0025, and 0.001 (the default value was 0.01). Fig.~\ref{fig:scatter_hstar} presents the resulting distributions over circularisation radius and eccentricity at the end ($t=3$) of these simulations for seed A. The gas flows are very similar, if not identical, at $\rcirc\gtrsim 4\,\sub{r}{sink}$, but differ sometimes significantly at small angular momenta. This implies that the details of the flows at these small radii, for instance the presence and characteristics of the inner disc formed at later stages of the simulations are not reliable.

As a direct consequence, our simulations cannot directly give the fraction of gas which lost sufficient angular momentum for viscous accretion to become efficient, i.e. reaching $\rcirc\sim10^{-4}\sub{r}{shell}$. However, we can attempt to estimate this by extrapolating to smaller $\sub{r}{sink}$. To this end, we consider the distribution of gas over initial and final $\rcirc$, which we can directly interpret as the conditional probability for gas to reach some final value for $\rcirc$ given its initial value. For the four sets of simulations with different $\sub{r}{sink}$, Fig.~\ref{fig:dist_rcirc_init_final} plots the combined distributions from the six simulations for each set.

These distributions are fairly wide: the range of $\sub{r}{circ,final}$ for a given $\sub{r}{circ,initial}$ spans at least four orders of magnitude and vice versa. Obviously, $\sub{r}{circ,final} < \sub{r}{circ,initial}$ for the bulk of the distribution: most gas has lost angular momentum. Within each simulation, material with large initial $\rcirc$ underwent most angular-momentum reduction. This is simply a consequence of the inner boundary: particles with larger initial $\rcirc$ typically require more interactions and angular-momentum reduction before they are taken out of the simulation at that boundary. Similarly, with a larger dynamic range (decreasing $\sub{r}{sink}$) the typical reduction of $\rcirc$ increases, which can be attributed to the same cause.

\begin{figure}
	\includegraphics[width=78mm]{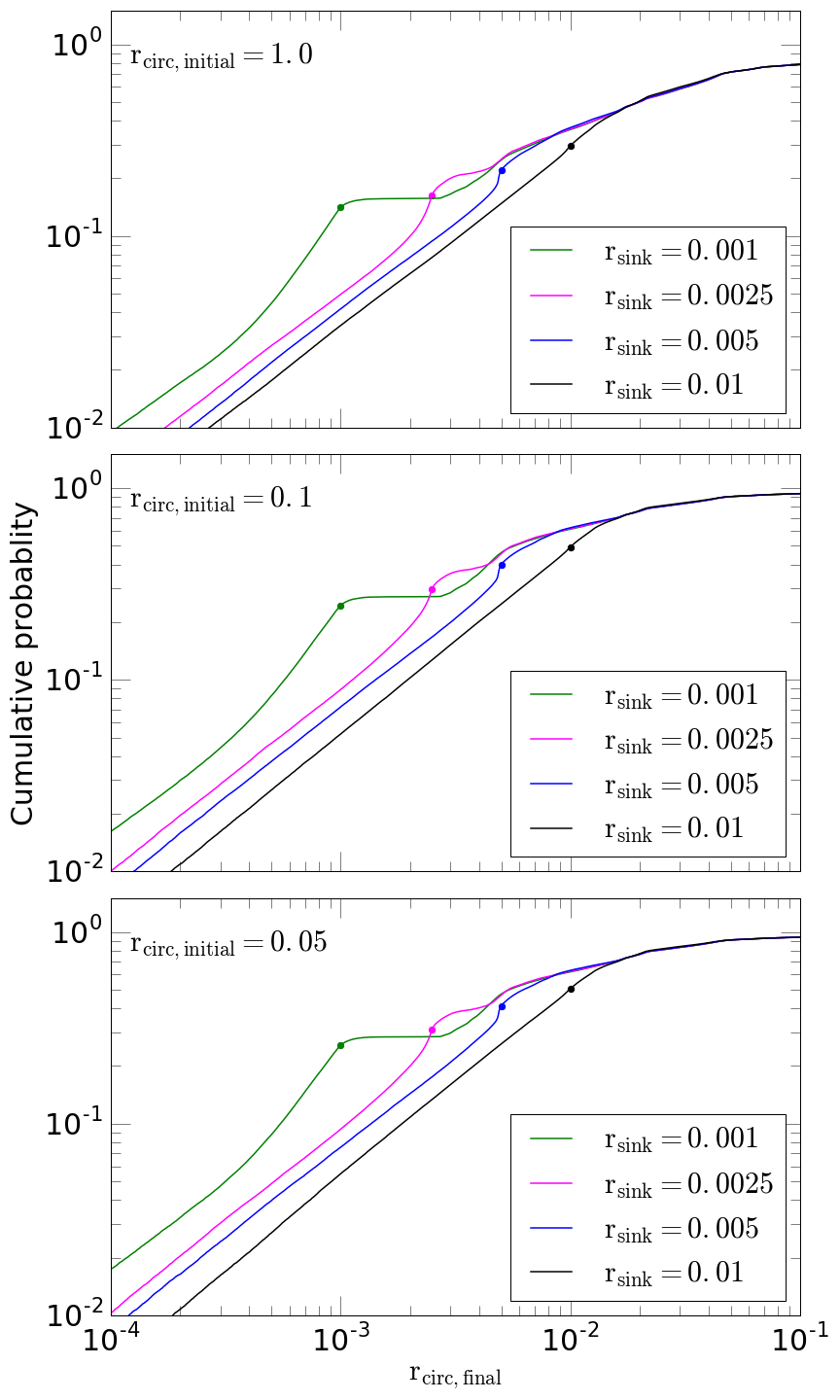}
	\caption{\label{fig:dist_rcirc_ratio}
	The cumulative distributions over the final circularisation radius $\rcirc=L^2/GM$ for gas with three different initial values as indicated (gas particles within $0.1$ dex of this value are included in the plots). The four distributions per panel refer to sextets of simulations with different dynamic range or $\sub{r}{sink}$ as indicated (the solid dots are at $\sub{r}{sink,final}=\sub{r}{sink}$).}
\end{figure}

\footnotetext{The simulations for $\sub{r}{sink}=0.001$ show a gap at $\rcirc \sim0.002$, which is also present, to a lesser degree, at $\sub{r}{sink}=0.0025$ or in simulations with $\sub{N}{gas}\lesssim10^6$ and $\sub{r}{sink}=0.01$ (not shown). This effect is an artefact of low resolution: the innermost particles' smoothing length becomes comparable to $\sub{r}{sink}$ and we cannot expect an adequate model of the gas flow.}

In order to extrapolate these trends to the yet smaller values of $\rcirc\sim10^{-4} \sub{r}{shell}$ required for efficient viscous accretion, we plot in Fig.~\ref{fig:dist_rcirc_ratio} the cumulative distributions over the final circularisation radius for gas with the same initial value of $\rcirc=0.05$, 0.1, or 1 (top to bottom), but obtained from simulations with different $\sub{r}{sink}$. Simulations with smaller $\sub{r}{sink}$ always obtain more particles at smaller $\sub{r}{circ,final}$. In the converged part, the cumulative distributions approximately follow a power law with index $\gtrsim-0.5$. Extrapolating, we estimate that at least 10 percent of the gas will reach $\rcirc=10^{-4}\sub{r}{shell}$.

\section{Discussion}
\label{sec:discuss}
Our simulations do not model the feedback and sweeping-up of gas from the vicinity of the SMBH, but start from a gaseous shell of radius $\sub{r}{shell}$ with non-uniform velocity distribution and a combination of rotational and turbulent motion. Shortly after the start of each simulation, the non-uniformity of the velocity field generates non-uniformity in the gas density, forming clumps, which subsequently fall into the cavity where tidal forces transform them into radially elongated streams. As modelling the gas flow close to the SMBH is prohibited by ever shorter dynamical time scales, we impose an inner boundary at $\sub{r}{sink}=0.01\sub{r}{shell}$ representing the sub-parsec volume around a SMBH. This is still much larger than $\sub{r}{visc}$ if we take $\sub{r}{shell}\sim10$\,pc, the typical radius of a momentum-driven outflow corresponding roughly to the size of the SMBH's sphere of influence for $M_\bullet \approx 10^8 M_\odot$.

As the streams approach their orbital peri-centre, there is a good chance of collisions, which typically diminish the circularisation radius $\rcirc$ of the gas, with an average reduction by a factor of a few. This allows a higher portion of the gas to cross $\sub{r}{sink}$, while the most of the remaining gas circularises and forms disc-like structures at $\sim0.05\sub{r}{shell}$. The earlier phases of these discs are typically randomly oriented and can be completely destroyed and subsequently reformed from impacting streams. After roughly one free-fall time, the disc dominates the further evolution of the distribution of the simulated gas over angular momentum (i.e.\ $\rcirc$) and eccentricity $e$. Continued gas infall often has larger $\rcirc$ and extends the disc rather than destroying it. At the end of our simulations, the disc is still far from settled but possesses significant eccentricity and often warps or gaps.

Of course, the details of these gas flows are highly dependent on the details of the initial conditions. Therefore, we perform for each set of physical parameters six simulations which differ only by the random seed used to generate the turbulent velocities. Amongst simulations from such sextets no are identical or even similar in their gas flows, but differ in details like the shape of the disc, its formation time, the fraction of gas crossing $\sub{r}{sink}$ etc.. Despite the differences, the overall behaviours are similar and follow closely that outlined above.

In order to assess the robustness of our results, we varied the physical parameters of our simulations, such as the gas temperature (assumed to be constant), total gas mass, power-spectrum of the initial turbulent velocities, total kinetic energy, rotational support of the gas, and the width of the initial gas shell. Most of these variations have little effect on the overall reduction of the gas circularisation radii (but see Section~\ref{sec:parasweep}). The one exception is the initial rotational support of the gas as compared to the turbulence. In case of significant rotation, the streams fall in a more orderly fashion, reducing the number of collisions such that only interactions with the forming disc cancel some angular momentum. Conversely, simulations without systematic rotation show the strongest average reduction of $\rcirc$ and the largest amount of gas crossed into $r<\sub{r}{sink}$. This result agrees with simulations presented by \cite{Hobbs11} who used a similar setup, but focused on the gas transportation in the galactic bulge. They argue that the formation of dense material caused by the turbulence leads to 'ballistic accretion' whereby the angular momentum of those filaments barely mixes with the ISM and therefore the gas can reach smaller radii directly.

While the interactions of the streams with the disc are not the focus of this study, it is worth commenting on the ability of some discs to resist destruction or at least major disturbance caused by the gaseous infall. One might expect that low-angular-momentum gas on plunging orbits can fall to small scales, but a number of simulations (see Sections~\ref{sec:pwr} and~\ref{sec:sol}) indicate that dense disc-like structures are effective at preventing gas infall to smaller scales. 

The realism of our simulations is limited in various ways. One such limitation is our simple treatment of the thermodynamics, where we assumed a constant gas temperature (isothermal equation of state). While the gas may cool rapidly after stream-stream collisions, when our approach is reasonably accurate, it is insufficient for the high densities in the disc, when a detailed thermodynamics model including cooling would be desirable. However, the disc formation is not our main focus here.

A second limitation is the omission of gas self-gravity in most simulations. Even when including self-gravity, we had to suppress small-scale clumping and hence star formation, which may well occur in reality. The main effect of star formation for the purpose of our simulation is (a) a reduction of the amount of gas available for feeding the SMBH and (b) the heating of gas by stellar feedback (driven by supernovae and winds). Both should reduce the efficiency of stream collision and disrupt the formation of the disc. The latter is indeed what we find in simulations with gas self-gravity and a larger total gas mass.

Another limitation is introduced by the (unavoidable) inner boundary, represented by a radius $\sub{r}{sink}$ from the black hole, inside of which particles are removed from the simulation (and their mass, momentum, and angular momentum added to the SMBH particle). This inevitably introduces some artefacts near the boundary. Simulations run with up to ten times smaller $\sub{r}{sink}$ show a larger reduction of $\rcirc$ and more gas in the low-angular-momentum tail of the distribution, since the collision cascade can penetrate to smaller radii. Extrapolation of these results suggest that $\gtrsim10\%$ of the gas can circularise at $r\lesssim10^{-3}\,$pc, which is the scale of efficient viscosity-driven accretion discs.

Finally, our simulations ignore for simplicity the gravity of the host galaxy. A more accurate picture might be obtained by a static potential or even an $N$-body model for the stellar cusp. However, as the SMBH dominates gravity within the simulated volume, such treatment is unlikely to improve the realism of the simulations in view of the other aforementioned limitations.

\section{Conclusion}
\label{sec:conclude}
We have presented smoothed particle hydrodynamic (SPH) simulations investigating the scenario proposed by \cite{Dehnen13} to overcome the angular momentum problem impeding SMBH feeding. In this scenario, gas swept-up by momentum-driven feedback from a previous accretion event falls back towards the hole. As this occurs near-simultaneously for most of the gas, the chances for collisions are enhanced. Such collisions promote the (partial) cancellation of angular momentum and increase the amount of material at circularisation radius $\rcirc\lesssim\sub{r}{visc}$, the radius at which classical disc-driven viscous accretion becomes efficient. The goal of this study was to assess the efficiency of this process more quantitatively.

Our suite of simulations provides strong support for this scenario of SMBH feeding, as they demonstrate a reduction of $\rcirc$ by a factor of a few on average and by much larger factors for a small fraction of the gas. These reductions are caused predominantly by stream-stream collisions but also interactions with discs that form from the infalling material. The details of each simulation depend both on the random velocity field and the physical parameters (such as gas temperature, velocity power spectrum, or velocity amplitudes), but the reduction of the specific angular momentum for most of the gas flow is hardly affected by changing these parameters. Our simulations confirm the suggestions by \citeauthor{Dehnen13} of the formation and maintenance of a near-toroidal dynamical gas structure caused by the continuous circularisation of infalling gaseous streams; the creation and destruction of randomly orientated discs; and high rates of gas passing through the inner numerical boundary, which potentially drive growth of the SMBH.

The scenario of angular-momentum reduction via a collisional cascade requires the near-simultaneous infall of gas from different trajectories. In our simulations this was provided by the fallback of a shell of gas assumed to have previously been swept up by AGN feedback. However, other initial situations are also possible, for example the collision of two massive clouds/streams of gas, resulting in a near-cancellation of their angular momenta and the subsequent infall of their shreds.

When adding star formation to this picture, one expects stars to form both from gas in the newly formed disc and from gas on plunging streams, possibly triggered by stream collisions. Of the latter some may come close enough to the SMBH to suffer from tidal disruption of binaries \citep{Hills1988} and subsequent capture of one binary component into an eccentric orbit around the SMBH. This fits with the observational situation in our own Galaxy\footnote{The radius of the sphere of influence for Sgr\,A$^{\!\star}$ is 2-3\,pc, which should be scaled to $\sub{r}{shell}$ of our simulations.}, where young stars (4-6\,Myr old) are found in a disc at $\sim 0.1$\,pc \citep{PaumardEtAl2006}, while the so-called S-stars on eccentric isotropic orbits at much smaller ($\sim0.01$\,pc) distances from Sgr\,A$^{\!\star}$ have a similar age \citep[3-10\,Myr,][]{HabibiEtAl2017}, which appears to coincide with the driving of the Fermi bubbles and the likely associated AGN activity \citep{Zubovas11,Zubovas12}. However, this time span appears too short to change the initially very eccentric orbits of tidally captured stars into a thermal distribution, as observed for the S-stars \citep{Gillessen2017}, solely by stellar dynamics \citep[in particular scalar resonant relaxation,][see also the review by \citealt{Alexander2017}]{Perets2009}. But within the gas-rich environment during  accretion-disc formation gravitational interactions with the gas may play an important role.

\section*{Acknowledgements}
We thank the anonymous referee for their helpful comments. We also thank Sergei Nayakshin, Chris Nixon, and Andrew King for helpful discussions as well as Martin Bourne and Hossam Aly.
Research in Theoretical Astrophysics at Leicester is supported by STFC grant ST/M503605/1. Some calculations presented in this paper were performed using the ALICE High Performance Computing Facility at the University of Leicester. Some resources on ALICE form part of the DiRAC Facility jointly funded by STFC and the Large Facilities Capital Fund of BIS. This work used the DiRAC Data Analytic system at the University of Cambridge, operated by the University of Cambridge High Performance Computing Service on behalf of the STFC DiRAC HPC Facility (\url{www.dirac.ac.uk}). This equipment was funded by BIS National E-infrastructure capital grant (ST/K001590/1), STFC capital grants ST/H008861/1 and ST/H00887X/1, and STFC DiRAC Operations grant ST/K00333X/1. DiRAC is part of the National E-Infrastructure.

\bibliographystyle{mnras}
\bibliography{CFbib}

\appendix
\section{Generating turbulent initial velocities}
\label{sec:turb}
An isotropic Gaussian random vector field $\vec{v}(\vec{x})$ with a given power spectrum 
\begin{equation}
    P(k) = \left\langle|\hat{\vec{v}}(\vec{k})|^2\right\rangle_{|\vec{k}|=k}
\end{equation}
can be created simply as three independent Gaussian random scalar fields \citep[e.g. see][]{Efstathiou85} with the same power spectrum divided by a factor 3. In order to generate a Gaussian random scalar field $f(\vec{x})$, its Fourier transform $\hat{f}(\vec{k})$ is sampled as complex random variable with uniform phase and normally distributed amplitude with zero mean and variance equal to the power $P(k)$.

In practice, we taper the velocity power spectrum at some maximum wave length by replacing $P(k)\propto k^{-n}$ with
\begin{equation}
    P(k) \propto (k^2+\sub{k}{min}^2)^{-n/2},
\end{equation}
where $\sub{k}{min}=2\pi/\sub{\lambda}{max}$ \citep[as suggested by][]{Dubinski95}. In our simulations $\sub{\lambda}{max}$ is set equal to the initial radius $\sub{r}{shell}$ of the gas shell.

In order to obtain a velocity field satisfying $\vec{\nabla}\cdot\vec{v}=0$, there are two obvious routes. First, one may obtain $\vec{v}$ as the curl of a Gaussian random field $\vec{u}(\vec{x})$ with power spectrum steeper by a factor $k^2$ or, equivalently set $\hat{\vec{v}}=i\vec{k}\times\hat{\vec{u}}$ where the variance of $\hat{\vec{u}}(\vec{k})$ equals $k^{-2}P(k)$. Second, one may simply project a general Gaussian random vector field onto its divergent free part, which is most easily achieved in Fourier space by replacing $\hat{\vec{v}}$ with 
\begin{equation}
    \hat{\vec{v}} - \frac{\vec{k}\vec{k}\cdot\hat{\vec{v}}}{\vec{k}^2}.
\end{equation}
In our simulations, we used the first of these methods.

\label{lastpage}
\end{document}